\documentclass{mn2e}
\usepackage{graphics}

\begin{document}

\title[The Stellar Mass Density at $z\approx 6$]{The Stellar Mass Density at $z\approx 6$ from {\em Spitzer} Imaging of $i'$-drop Galaxies}

\author[Eyles et al.\ ]{Laurence P.\ Eyles$^{1}$,
Andrew J.\ Bunker\,$^{1}$, Richard S.\ Ellis\,$^{2}$,\newauthor
Mark Lacy\,$^{3}$,
Elizabeth R.\ Stanway\,$^{4}$, Daniel P.\ Stark$^{2}$, Kuenley Chiu$^{1}$ \\
$^{1}$\,School of Physics, University of Exeter, Stocker Road, Exeter, EX4 4QL,
U.K. {\tt email:
eyles@astro.ex.ac.uk}\\
$^{2}$\,California Institute of
Technology, Mail Stop 169-327, Pasadena, CA~91109, U.S.A. \\
$^{3}$\,Spitzer Science Center,California Institute of Technology, Mail Code 220
-6, 1200 E.\ California Blvd., Pasadena, CA~91125, U.S.A.\\
$^{4}$\,Astronomy Department, University of Wisconsin-Madison, 475 N.\ Charter Street, Madison, WI~53706, U.S.A. \\
}

\date{Submitted for publication in MNRAS, 2006}

\maketitle

\begin{abstract}
We measure the ages, stellar masses, and star
  formation histories of $z\sim 6$ galaxies, observed within
  1\,Gyr of the Big Bang. We use imaging from the {\em Hubble Space
    Telescope (HST)} and the {\em Spitzer Space Telescope}
    from the public ``Great Observatories Origins Deep
  Survey'' (GOODS), coupled
  with ground-based near-infrared imaging, to measure their spectral
  energy distributions (SEDs) from $0.8-5\,\mu$m, spanning the
  rest-frame UV and optical. From our sample of $\approx 50$
  `$i'$-drop' Lyman-break star-forming galaxies in GOODS-South
  with $z'_{AB}<27$, we focus on $\approx 30$ with reliable
 photometric or spectroscopic redshifts. Half of these are confused with
  foreground sources at {\em Spitzer} resolution, but from
  the 16 with clean photometry we find that
  a surprisingly large fraction (40\%) have
 evidence for substantial Balmer/4000\,\AA\ spectral
  breaks. This indicates the presence of old underlying stellar
  populations that dominate the stellar masses.  For these objects, we
  find ages of $\sim 200 - 700$\,Myr, implying formation redshifts of
  $7\le z_f \le 18$, and large stellar masses in the range $\sim 1 -
  3\times 10^{10}\,M_{\odot}$.  Analysis of seven $i'$-drops that are
  undetected at 3.6\,$\mu$m indicates that these are younger,
  considerably less massive systems.
We calculate that emission line contamination should not severely
  affect our photometry or derived results.  Using SED fits out to $8\,\mu$m, we find 
  little evidence for substantial intrinsic dust reddening in our sources.
 We use our individual galaxy results to obtain an estimate of the
  global stellar mass density at $z\sim 6$.  Correcting for
  incompleteness in our sample, we find the $z\sim 6$ comoving stellar
  mass density to be $2.5\times 10^{6}M_{\odot}\,{\rm Mpc}^{-3}$.
  This is a lower limit, as post-starburst and dust-obscured objects,
  and also galaxies below our selection thresholds, are not accounted
  for.  From our results, we are able to explore the star formation
  histories of our selected galaxies,
and we suggest that the past global 
star formation rate may have been much higher than that observed at the $z\sim 6$ epoch.
The associated UV flux we infer at $z>7$ could have played a major role in reionizing the universe.
\end{abstract}

\section{Introduction}
\label{sec:INTRO}

Studying the stellar populations in the most distant objects known
could provide a key insight into galaxy formation, potentially
revealing the star formation history at even earlier epochs. The current
frontier for spectroscopically-confirmed galaxies is $z\sim 6$, with
unconfirmed candidates based on photometric redshifts at perhaps even
higher redshifts. The $i'$-drop technique is based on the Lyman-break technique
(Steidel et al.\ 1996) and robustly selects star-forming galaxies at
$z\sim 6$ (Stanway, Bunker \& McMahon
2003; Bunker et al.\ 2004; Bouwens et al.\ 2004a; Yan \& Windhorst 2004;
Giavalisco et al.\ 2004), when the universe was less than 1\,Gyr old.
If some of these $i'$-drop galaxies can be shown to harbour stellar
populations with ages of a few hundred Myr, then this pushes their
formation epoch to $z\sim 10$. Measurements of the stellar masses of
individual $z\sim 6$ galaxies can also constrain structure formation
paradigms; in a simple hierarchical model, massive galaxies assemble
at later times through merging, so it might be expected that in this
scenario the number density of massive evolved galaxies in the first
Gyr would be low.

In Eyles et al.\ (2005), we presented {\em Spitzer}/IRAC imaging of
$z\sim 6$ $i'$-drop galaxies with known spectroscopic redshifts,
sampling the rest-frame optical at $3.6-8\,\mu$m. Our previous work
with optical {\em HST}/ACS data and near-infrared imaging from
VLT/ISAAC and {\em HST}/NICMOS (Stanway, McMahon \& Bunker 2005)
explored the rest-frame ultraviolet (UV) in these galaxies, which is
dominated by recent or ongoing star formation.  The addition of {\em
  Spitzer}/IRAC imaging allowed us to fit spectral energy
distributions from the multi-wavelength broad-band photometry
to stellar population synthesis models. We were able to constrain the 
stellar masses and ages, and hence
explore the preceding star formation history and the formation epochs.
We concentrated primarily on two bright, well-detected objects
(SBM03\#1\,\&\,\#3 from Stanway, Bunker \& McMahon 2003) with
spectroscopic redshifts of $z\approx 5.8$ (Stanway et al.\ 2004a,b;
Bunker et al.\ 2003).  For these two sources, we found evidence for the
presence of Balmer/4000\,\AA\ spectral
breaks, indicating significant old stellar populations, with ages of the order of a few hundred million
years (see Section~\ref{sec:DETECTIONS}). From this we inferred formation redshifts of $z_{f}\sim 7.5 - 13.5$, and
that even more vigorous star formation had taken place prior to the
time of observation. Our work was confirmed independently
by Yan et al.\ (2005) who, in their selection of galaxies, studied one
of the same objects, SBM03\#1 (see also Finlator, Dav\'e \& Oppenheimer 2006).  The $z\sim 6$ epoch marks a pivotal
point in the history of the Universe -- the end of the reionisation
era (Becker et al.\ 2001; Kogut et al.\ 2003; Spergel et al.\ 2006).
Hence we suggested that if objects such as these were commonplace at
$z\ge 6$, the UV flux from their initial vigorous starbursts may have played a key role in the reionisation of the
Universe, supporting earlier work of Bunker et al.\ (2004) and Egami
et al.\ (2005).

The evidence found for the presence of significant Balmer/4000\,\AA\
breaks in our two well-detected $i'$-drops raises the question: are
these breaks rare in $z\sim 6$ objects, or are they commonplace?
In this paper, we now look to expand on our case studies of a few individual $z\sim
6$ sources, by considering the observed optical-infrared properties of
a larger population of $i'$-drop galaxies.
If Balmer/4000\,\AA\  breaks are found to be rare occurrences a possible scenario is
that most star-forming $i'$-drops could be young ``protogalaxies''
undergoing their first bout of star formation at $z\sim 6$.  On the
other hand, if these breaks are commonplace at $z\sim 6$, it could
be inferred that there is a significant population of
well-established objects in place 1\,Gyr after the Big Bang.
In these systems, vigorous star formation would be required at
$z\gg 6$ in order to assemble the bulk of the stellar mass.

Thus our primary goal is to obtain a robust estimate of the stellar
mass density at $z\sim 6$ from our sample of $i'$-drop galaxies;
coupled with age estimates (also derived
from our photometry), we may be able to uncover the preceding star
formation history.  This is immensely important when considering 
galaxy assembly scenarios and also the
reionisation of the Universe.  
The contribution of star-forming galaxies at $z\ge 6$ to the UV ionizing
background, and hence the reionisation of the Universe, is still
debated. Bunker et al.\ (2004) discovered $\sim 50$ $i'$-drop galaxies in the {\em Hubble}
Ultra Deep Field (HUDF) with $z'_{AB}<28.5$ ($10\,\sigma$). 
Star formation rates of these galaxies extend down to $1\,M_{\odot}\,{\rm yr}^{-1}$,
derived from the rest-frame UV continuum assuming a standard Salpeter
initial mass function (IMF). Bunker et al.\ concluded
that the star formation density from these observed sources would be
insufficient to reionize the Universe at $z\sim 6$, even with large escape fractions
for the Lyman continuum photons.
Yan \& Windhorst (2004) independently confirmed the Bunker et al.\ $i'$-drops
in the HUDF (see Stanway \& Bunker 2004 for a comparison), but suggested that unobserved galaxies below the detection
limit could contribute much of the flux if the faint-end slope of the rest-UV
luminosity function was much
steeper than $\alpha\sim -1.6$ seen for Lyman break galaxies at
lower redshifts ($z\sim 3-4$; Steidel et al.\ 1996). Stiavelli, Fall \& Panagia (2004) suggested
that a warmer intergalactic
medium (IGM), a top-heavy IMF and substantially lower metallicity (e.g., Population III)
at $z\sim 6$ might just provide sufficient ionizing flux if the escape
fraction was $f_{esc}\sim 0.5$, much higher than observed at  $z=0-4$
($f_{esc}=0.01-0.1$; Inoue, Iwata \& Deharveng 2006).
Whilst the slumping star formation rate
density from the observed $i'$-drops (Bunker et al.\ 2004, Bouwens et al.\ 2004a) might be
insufficient to account for reionisation at $z\sim 6$, it is possible that earlier more 
intense star formation played
a significant role in achieving reionisation at higher redshifts.

Using public imaging taken as part of the ``Great Observatories
Origins Deep Survey'' (GOODS; Dickinson \& Giavalisco 2003; Dickinson
et al.\ {\em in prep}), our group has 
explored the stellar mass density and ages of $v$-drop galaxies
at $z\sim 5$ (Stark et al.\ 2006). Comparing the inferred previous star 
formation
histories of these $v$-drops with observations of
the star formation rate density at higher redshifts, Stark et al.\
concluded that 
perhaps as much as half of the star formation occurring at $z > 5$ goes 
unobserved. Potential reasons for this include a high contribution
from low-luminosity sources, dust obscuration and/or a yet to be observed
phase of star formation at very high redshift ($z\gg 6$).  Recently, Yan et 
al.\ (2006) have provided an estimate to
the stellar mass density at $z\sim 6$, and concluded that the bulk of
reionising photons must have been provided by other sources, perhaps
by objects that are below current detection limits.  

It should be noted 
that such studies of Lyman break galaxies (LBGs) place {\em lower} limits on
the stellar mass density at the epoch of observation. The LBG  selection
technique is reliant upon an objects' detection in the rest-frame UV
(e.g., the $z'$-band for $i'$-drops at $z\sim 6$), so there must be at 
least {\em some} ongoing star formation
at the epoch of observation for the galaxy to be selected.  Attempts to find 
post-starburst (Balmer-break) objects at $z\sim 6$ are extremely difficult 
and uncertain due to a large population of lower redshift interloping 
galaxies.  For example, Mobasher et al.\ (2005) have recently suggested the 
presence of a massive post-starburst galaxy, HUDF-JD2, with a photometric 
redshift of $z_{phot}\approx 6.5$ (high-redshift solutions with $z>5$ being preferred
85 per cent of the time). This comes from the re-analysis of an 
IRAC-detected Extremely Red Object (IERO) identified in the HUDF by
Yan et al.\  (2004), who originally derived a photometric redshift of 
$z_{phot}\sim 3.4$
(see also Chen \& Marzke 2004). If
the higher redshift from Mobasher et al.\ (2005) is correct, then the 
SED suggests a
remarkably large stellar mass of $6\times
10^{11}\,M_{\odot}$ at $z\approx 6.5$ -- a factor of ten greater than the 
masses we presented in Eyles et al.\ (2005). However, deep spectroscopy has
yet to yield a redshift for this source, and the high photometric redshift 
estimate of Mobasher et al.\ has been disputed by Dunlop et al.\ (2006), who 
suggest that $z_{phot}=2.2$ is more plausible.

Rather than study all galaxies in the GOODS fields with potential
photometric redshifts at $z\sim 6$,  in this paper we restrict ourselves to the $i'$-drop
selection, which has proven to be a reliable technique in isolating $z\approx 6$
star-forming galaxies (see Bunker et al.\ 2005 for a review) and so should minimize low redshift contaminant sources.
Hence by investigating the stellar masses of our $z\sim 6$ Lyman-break galaxies,
we should be able to derive lower limits on the stellar mass density at this epoch.

In this paper we examine the SEDs of $i'$-drop galaxies in the
GOODS-South field, using observations from {\em HST}/ACS in the
optical, VLT/ISAAC in the near-IR, and {\em Spitzer}/IRAC to span the
rest-frame UV/optical. The work presented in this paper
is an independent analysis of the $i'$-drop 
population, and differs from the recent work of Yan et al.\ (2006) in several ways.  In 
addition to photometric data gathered in {\em HST}/ACS \& {\em Spitzer}/IRAC 
wavebands, we also use ground-based near-IR imaging of the GOODS-South
field to better constrain the SEDs and the stellar population fitting (see 
Sections~\ref{sec:PHOT}\,\&\,\ref{sec:SEDS}).  Rather than assign a common 
redshift of $z=6$ to all galaxies in our sample, we choose to use $i'$-drops with 
either spectroscopically confirmed or robust photometric redshifts, and 
provide photometry
for each source and details of the best-fit stellar population models for individual galaxies
(Sections~\ref{sec:REDUCT}\,\&\,\ref{sec:ANALYSIS}).  For those objects which suffer from 
confusion with 
neighbouring sources, we attempt to subtract the contaminating objects in order to obtain accurate 
aperture photometry, rather than simply discarding these $i'$-drops from our 
sample (see Section~\ref{sec:GALFIT}).  We look to build on our previous work
in Eyles et al.\ (2005)
by now considering a full sample of $i'$-drop candidates, and exploiting new, 
improved imaging datasets provided by the GOODS team (see 
Section~\ref{sec:IMAGING}), including both epochs of the {\em Spitzer}/IRAC imaging.  

 An outline of this paper is as follows:
Section~\ref{sec:OBSERVATIONS} provides a summary of the imaging
datasets used in this study and our selection of $i'$-drop galaxies.
In Section~\ref{sec:REDUCT} we describe the photometry and the
removal of contaminating sources, and also the fitting of stellar
population synthesis models to the observed spectral energy
distributions.  We discuss our results and their implications in
Section~\ref{sec:ANALYSIS}, and our conclusions are presented in
Section~\ref{sec:CONCLUSIONS}.  Throughout we adopt the standard
``concordance'' cosmology of $\Omega_M=0.3$, $\Omega_{\Lambda}=0.7$,
and use $H_0\,=\,70\,{\rm km\,s^{-1}\,Mpc^{-1}}$, which is within $2\,\sigma$
of the latest WMAP determination (Spergel et al.\ 2006) -- in our
adopted cosmology,
the Universe today is 13.67\,Gyr old, and at $z=6$ its age was
914\,Myr.  All quoted magnitudes are on the $AB$ system (Oke \& Gunn
1983).

\section{Observations and Object Selection}
\label{sec:OBSERVATIONS}

\subsection{Imaging Data}
\label{sec:IMAGING}

In this paper, we use multi-waveband data of the GOODS-South field
which is centred on the {\em Chandra} Deep Field South (Giacconi et
al.\ 2002) and also contains the {\em Hubble} Ultra Deep Field
(Beckwith et al. 2003).

The Advanced Camera for Surveys (ACS; Ford et al. 2003) onboard the
{\em HST} has provided deep optical imaging of the GOODS-South field,
as part of {\em HST} Treasury Programs \#9425\,\&\,9583 (Giavalisco
et al.\ 2003; 2004), using the F435W ($B$), F606W ($v$), F775W
(SDSS-$i'$) and F850LP (SDSS-$z'$) broad-band filters, with a pixel
scale of $0\farcs05$.  In our analysis we make use of the publicly
available version-1.0 release\footnote{available from \newline
  {\tt ftp://archive.stsci.edu/pub/hlsp/goods/}} of the reduced data
  from the GOODS team.  These data, taken over 
5 observing epochs, had
been `drizzled' using a `multidrizzle' technique (Koekemoer et al.\
2002), producing combined images with a pixel scale of $0\farcs03$,
mosaicked in a $10'\times 15'$ area.  
The $3\sigma$ limiting
magnitudes,  measured in $0\farcs5$ diameter
apertures, are $B_{AB}=29.4$, $V_{AB}=29.5$, $i'_{AB}=28.8$ \& $z'_{AB}=27.9$
For the purpose of this
study, we are primarily concerned with the $i'$ \& $z'$-band images.

Deep ground-based near-infrared data were obtained using the Infrared
Spectrometer and Array Camera (ISAAC) on the Very Large Telescope
(VLT), as part of the ESO Large Programme LP168.A-0485(A) (PI: C.\
Cesarsky). We utilise the $J$-band ($\lambda_{\rm cent}\approx
1.25\,\mu$m) and $K_{s}$-band ($\lambda_{\rm cent}\approx 2.15\,\mu$m)
data which were released as part of the publicly available v1.5 reduced
dataset\footnote{available from \newline {\tt
    http://www.eso.org/science/goods/releases/20050930/}} (Vandame et
al. {\em in prep}).  We note that the $J$ and $K_s$ data release is an
updated version of the v1.0 release used in Eyles et al.\ (2005) and
Stanway et al.\ (2003; 2004a,b).  The mosaicked $J$ and $K_s$ images
each cover $\approx 159$\,arcmin$^{2}$, and comprise of 24 tiles taken over
many nights. These data have a pixel scale of $0\farcs15$, five times
larger than that of the `drizzled' ACS pixels, and each individual
tile has undergone a rescaling in order to provide an homogeneous
zeropoint AB-magnitude of 26.0 in both the $J$ \& $K_{s}$-bands.  The $J$
and $K_{s}$-band $3\sigma$ limiting magnitudes, measured in $1''$-diameter
apertures are $J\sim 25.6$\,\&\,$K_s\sim 26.0$ (see Section~\ref{sec:PHOT}).

Imaging of the GOODS-South field using the Infrared Array Camera
(IRAC; Fazio et al.\ 2004) onboard {\em Spitzer} was conducted as part
of the GOODS Legacy programme (PID 194, Dickinson et al.\
{\em in prep}).  IRAC uses four broad-band filters with central
wavelengths at approximately 3.6\,$\mu$m, 4.5\,$\mu$m, 5.8\,$\mu$m \&
8.0\,$\mu$m (channels 1-4).  The data were taken over two observing
epochs, with the telescope roll angle differing by $180^{\circ}$. Note
that at the time of our previous study (Eyles et al.\ 2005) only the
epoch 1 data
were available; as the area covered by channels 1 \& 3 is offset by
6.7 arcmin from that covered by channels 2 \& 4, IRAC imaging of the
entire GOODS-South field was not complete in any of the four
channels.  With the release of the epoch 2 data\footnote{Second Data
  Release (DR2); see \newline {\tt
    http://data.spitzer.caltech.edu/popular/goods/ \newline
    Documents/goods$\_$dr2.html}}, 
the full $10'\times 16.5'$ field is now covered in all four wavebands.
We use the publicly-released reductions of the GOODS-South images
produced by the GOODS team (the updated DR3 for epoch 1 rather
than the original DR1, and DR2 for epoch 2).
As with the {\em HST}/ACS GOODS data, these reduced IRAC images had been `drizzled' by the GOODS team,
resulting in combined images
with a pixel scale of $0\farcs6$ (approximately half the original pixel size), a factor of twenty greater than that
of the `drizzled' ACS data.  
The $3\sigma$
limiting magnitudes are 26.5 \& 26.1 for IRAC channels 1 \& 2
respectively, measured in $2\farcs4$ diameter apertures, and 23.8 \&
23.5 for IRAC channels 3 \& 4 measured in $3\farcs 0$ \& $3\farcs 7$ diameter apertures
respectively (these limits include aperture corrections appropriate for
unresolved sources, see Section~\ref{sec:PHOT}). In this work, we 
concentrate on the two shorter wavelength IRAC filters which have
the greatest sensitivity.

\subsection{The $i'$-band Dropouts}
\label{sec:IDROPS}

A catalog of $z\sim 6$ galaxy candidates present in the GOODS-South
field was constructed (Stanway 2004), with objects selected via the
Lyman-break technique (Steidel, Pettini \& Hamilton 1995; Steidel et
al. 1996; 1999). For a $z\sim 6$ Lyman-break galaxy, the flux decrement due to
absorption by the intervening Lyman-$\alpha$ forest is redshifted
to fall between the $i'$-band and $z'$-band filters, causing it
to ``drop-out'' in the $i'$-band. Previous studies (Stanway et
al.\ 2004b; Dickinson et al.\ 2004) have shown that a colour-cut of
$(i'-z')_{AB} >1.3$\,mag reliably selects star-forming galaxies at $z
\sim 6$, with some contamination from lower redshift, passively
evolving galaxies (the ``Extremely Red Object" ERO population,
e.g., Cimatti et al.\ 2002, Doherty et al.\ 2005), and also from low-mass
stars (e.g., Hawley et al.\ 2002).  In this paper we apply our
$i'$-drop selection criteria to the total
GOODS-ACS dataset (5 epochs co-added into a single image);
previous work by our group (Stanway et al.\ 2003)
was based on single-epoch selections, and the co-added 5-epoch
dataset of the version-1.0 release allows us to push $\sim 1$\,mag
deeper than before, to a level similar to that of Bouwens\,\&\,Illingworth\, 
(2006).  Our $i'$-drop catalog  (Stanway 2004)
was constructed using the Source Extractor program (Bertin \& Arnouts 1996),
training the apertures in the $z'$-band and demanding a colour
of $(i'-z')\,>1.3$\,mag within a small aperture ($0\farcs 3$-diameter) to
minimize contamination effects from nearby sources. 
Candidate $i'$-drops within this list were then scrutinised according to the 
quality and reliability of their detections; we removed detector artifacts,
diffraction spikes from bright stars, and low signal-to-noise ratio candidates
(typically in the edge regions of each tile where fewer epochs overlapped).
We also required that there was no detection at $>3\,\sigma$ in the $B$-band (shortward
of the 912\,\AA\ Lyman limit at $z\sim 6$).
Some of the brighter $i'$-drops are faintly detected in $v$-band
(e.g., 31\_2185/SBM03\#3 with $z_{spec}=5.78$ has $v_{AB}=29$ and $(v-z')_{AB}=4.4$),
consistent with a large Lyman-$\alpha$ forest flux decrement of $D_A>0.95$.
Using the {\em HST}/ACS $z'$-band imaging, bright $i'$-drops with
half-light radii $R_{h} < 0\farcs05$ were identified as probable low-mass
Galactic stars, and these point sources were removed from our catalog;
this included the sources SBM03\#4 \& SBM03\#5\,/\,UDF\,B2104 (Stanway, Bunker \& McMahon 2003;
Bunker et al.\ 2004). At
magnitudes fainter than $z'_{AB}\approx 26.5$ the star--galaxy separation
is less clean, so we keep borderline unresolved fainter sources in our
sample, and use the ACS/near-IR/IRAC SEDs to reject possible stellar interlopers later
(Section~\ref{sec:PHOT}).
 
For consistency
with our other work (Bunker et al.\ 2004; Eyles et al.\ 2005) we have
remeasured the photometry of our $i'$-drops in $0\farcs5$-diameter
apertures, and aperture-corrected them to approximate total magnitudes
(see Section~\ref{sec:PHOT}). We imposed a selection 
criterion based on the $z'$-band magnitudes of our $i'$-drops, selecting only
those with $z'_{AB}<26.9$. These robust detections ($>8\,\sigma$) mean that
the ($i'-z'$) colour is secure, and we should have little contamination from
lower-redshift red galaxies with intrinsic colours just blueward of our colour cut
scattering up into our selection through photometric errors.
This appears to severely affect the $i'$-drop catalogs of other groups who work with
lower significance detections: for example, Dickinson et al.\ (2004) and
Giavaliso et al.\ (2004) analysed the GOODS fields, selecting $i'$-drops down to $S/N=5\,\sigma$
in the $z'$-band, and estimated that 55\% of these would be genuine $z>6$ galaxies on the
basis of simulations. However, Bouwens et al.\ (2005) looked at those sources covered by
the deeper UDF imaging, and found that only 25\% were real $i'$-drops.

Our list of $i'$-drops in GOODS-South has 54 objects with $z'_{AB}<26.9$
(where we treat 3 close pairs seen in {\em HST}/ACS as single objects as they are unresolved
at {\em Spitzer}/IRAC resolution).
We note that the analysis of Yan et al.\ (2006) uses 142 $i'$-drops in GOODS-South
from Giavalisco et al.\ ({\em in prep.}),
but  figure~1 of Yan et al.\ (2006) indicates that their selection extends to $z'_{AB}\approx 27.7$.
This is a factor of two fainter than our limit; the number counts appear to turn over at $z'_{AB}\approx 26.8$, 
so presumably fainter magnitudes suffer large incompleteness.

\begin{figure}
\resizebox{0.48\textwidth}{!}{\includegraphics{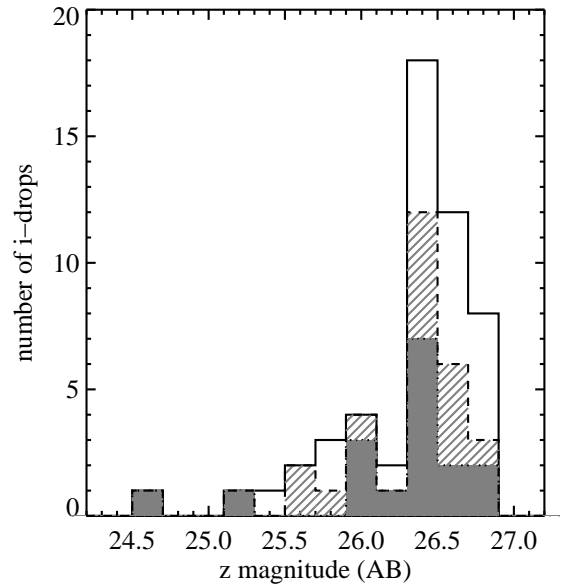}}
\caption{A histogram of the $z'$-band magnitudes of our full catalog of 52
$i'$-drop candidates with $z'_{AB}<26.9$ (total sample,
below continuous line).  From this, 31 objects were found to have either photometric redshifts (from the GOODS-MUSIC catalog) or spectroscopic redshifts (cross-hatched area, within dashed line).  The solid shaded region represents the 17 
objects for which reliable photometry was obtainable and hence were the focus 
of our analysis.  As can be seen in this figure, our sample, although small, 
reasonably represents the total $i'$-drop population in terms of $z'$-band 
magnitude.}
\label{fig:ZMAGHIST}
\end{figure}

We then further refined our list of objects to the subset which had
either spectroscopic or robust photometric redshifts.  
Our group was the first to publish spectroscopic confirmation of the {\em HST}/ACS $i'$-drop selection technique, using Keck/DEIMOS 
(Bunker et al.\ 2003; Stanway et al.\  2004a; see also Dickinson et al.\ 2004), and also Gemini/GMOS spectroscopy 
(GLARE; Stanway et al.\ 2004b).  Three of the four objects with spectroscopic redshifts in our previous 
{\em Spitzer} analysis (Eyles et al.\ 2005)  also appear in our current sample -- $23\_6714$ \& $31\_2185$
(SBM03\#1 \& \#3 respectively), and also $23\_2897$ (GLARE\#3001).  We
note that GLARE\#3011, the fourth (and faintest) object to feature in our 
previous analysis, was omitted from our current $i'$-drop catalog as it has $z'_{AB}=27.15$. 
We  matched our $i'$-drop candidates with the 
GOODS-MUSIC photometric catalog of this field (Grazian et al.\ 2006).  This catalog uses 
13-band SEDs from {\em HST}/ACS and {\em Spitzer}/IRAC photometry along with 
ground-based $U$, $J$ \& $K_s$ to derive photometric redshifts.  
We found 27 of the 54 objects within our 
list of $i'$-drops also featured in the GOODS-MUSIC catalog
with photometric redshifts $z_{phot}>5.6$, of which five have
spectroscopic redshifts from the literature\footnote{The GOODS-MUSIC
survey uses the redshift compilation given in
{\tt http://www.eso.org/science/goods/spectroscopy/CDFS Mastercat/}}. 
A further four of our $i'$-drop catalog had 
confirmed redshifts from VLT/FORS2 
spectroscopy (Vanzella et al.\ 2005; 2006), but were missing from
the GOODS-MUSIC catalog. This gives us a total of 31 objects
with spectroscopic or photometric redshifts at $z>5.6$, which are detailed
in Table~\ref{tab:COORDS}. Of the remaining 23 galaxies in
our $i'$-drop catalog, two 
were likely EROs with $z_{phot}\approx 1.1$, and were removed from our sample (a contamination rate by EROs of $\approx 5$\%). Three GOODS-MUSIC matches
had photometric redshifts of $5.45<z_{phot}<5.6$, below the nominal $i'$-drop selection,
and another five had 
matches in the GOODS-MUSIC catalog, but no photometric
redshift solution had been found ($z_{phot}=-1$).  
The remaining 15
were absent from the GOODS-MUSIC catalog, possibly due to incomplete
photometry in all wavebands in these regions (particularly the $U$, $J$, $H$ \& $K_s$). 
However, the available colours
and the distribution of magnitudes of these galaxies seem
similar to our other $i'$-drops (Figure~\ref{fig:ZMAGHIST}), so we will scale the results from the
spectroscopic$+$photometric sub-sample to our whole GOODS-South $i'$-drop sample 
at $z'<26.9$, excluding bright Galactic stars and lower-redshift EROs (a total of 52 objects).

 By restricting ourselves to the proven $i'$-drop pre-selection, rather
than including {\em all} galaxies with a potential $z\sim 6$
photometric redshift solution, our sample should contain fewer
numbers of spurious high redshift sources which may otherwise scatter
into the selection (e.g., see Dunlop et al.\ 2006).
We note, however, that the
Lyman-break selection we adopt requires at least some ongoing or very recent
star formation to detect the break in the rest-frame UV (a detection
in the $z'$-band) so we are incomplete for galaxies in which star formation
has ceased. Hence our measurements of the global stellar mass
density at $z\sim 6$ will necessarily be firm lower limits for $z\sim 6$ galaxies in the
GOODS-South field (see 
Section~\ref{sec:MASSDENSITY}).  Cosmic variance will of course play
a role, but for the size of the GOODS survey this has been estimated to be
a 20 per cent effect (Somerville et al.\ 2004) if the $i'$-drops have similar
clustering to the Lyman-break galaxies at $z\approx 4$ (the $B$-band drop-outs).

Although, as a consequence of our selection criteria, our sample of
galaxies is smaller than that of Yan et al.\ (2006), in doing so we
preserve the available redshift information, rather than
assigning a common redshift of $z = 6$ to all objects.  
This selection process, based on the quality of {\em HST}/ACS
detections and the availability of robust spectroscopic or
photometric redshifts, resulted in a sub-sample of 31 objects for further analysis.

\begin{figure*}
\resizebox{0.9\textwidth}{!}{\includegraphics{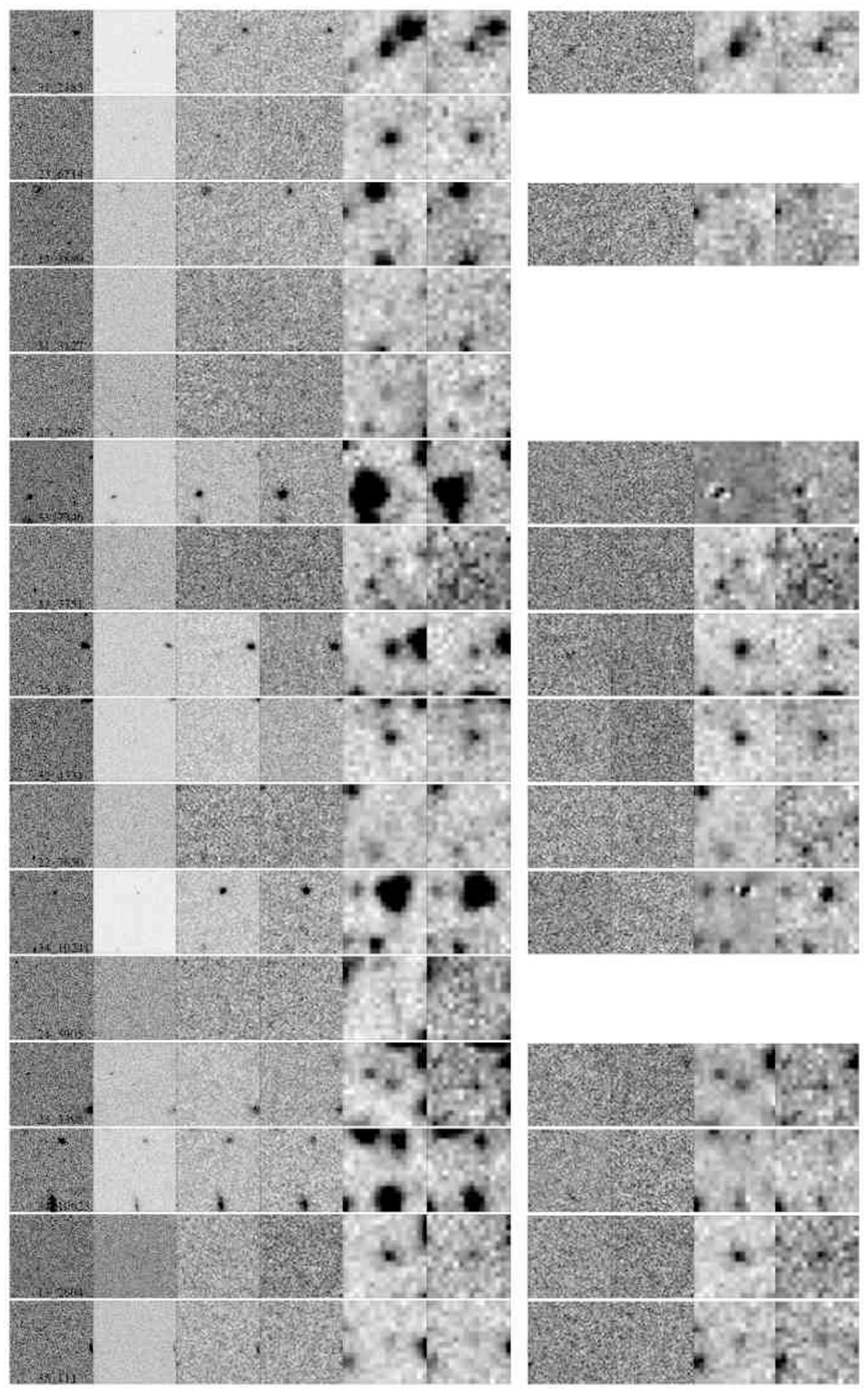}}
\caption{From left to right, postage stamp images of our 16 sources in the $i'$, $z'$, $J$, $K_{s}$, 3.6\,$\mu$m \& 4.5\,$\mu$m filters respectively, {\em before} any confusing source subtraction had been conducted (see Section~\ref{sec:GALFIT}). The stamps are $9''\times 9''$ regions centred on each $i'$-drop. North is up and East is to the left.  On the far right are the postage stamp images of the 12 sources in the $J$, $K_{s}$, 3.6\,$\mu$m \& 4.5\,$\mu$m filters which had neighbouring sources subtracted using GALFIT in the VLT/ISAAC near-infrared and {\em Spitzer}/IRAC imaging. Where no subtraction was necessary (in the case of isolated sources) these panels on the right are left blank.}
\label{fig:ORIGINALSTAMPS}
\end{figure*}

\begin{table*}
\centering
\begin{tabular}{l|c|c|c|c|r|c|l|}
our ID & GOODS-MUSIC ID & RA (J2000) & Dec.\ (J2000) & $z'_{AB}$ (0\farcs5) & ($i'-z'$) & $R_{hl}$ & Redshift\\
\hline\hline
$31\_2185$  & 499 & 03:32:25.605 & --27:55:48.69 & $24.61\pm 0.02$ & $1.59$ & $0\farcs07$ & $5.78^{\diamond}$ \\
$23\_6714$  & 8940 & 03:32:40.012 & --27:48:15.01 & $25.35\pm 0.03$ & $1.43$ & $0\farcs10$ & $5.83^{\diamond}$ \\
$33\_4396^{*}$  & 8079 & 03:32:13.071 & --27:49:00.75 & $25.71\pm 0.05$ & $1.96$ & $0\farcs11$ & $5.77$ \\
$22\_6713^{*}$  & 3912 & 03:32:39.027 & --27:52:23.12 & $25.75\pm 0.05$ & $1.47$ & $0\farcs12$ & $5.83$ \\
$25\_4894^{*}$  & --- & 03:32:33.194 & --27:39:49.11 & $25.84\pm 0.05$ & $2.48$ & $0\farcs14$ & $5.83^{\diamond}$ \\
$13\_3880$  & 10580 & 03:32:49.983 & --27:46:56.20 & $26.10\pm 0.06$ & $1.45$ & $0\farcs11$ & $5.65$ \\
$13\_1487^{*}$  & 7818 & 03:32:54.099 & --27:49:15.88 & $26.13\pm 0.09$ & $2.58$ & $0\farcs10$ & $5.79^{\diamond}$ \\
$31\_3127$  & 959 & 03:32:23.842 & --27:55:11.62 & $26.15\pm 0.07$ & $>2.68$ & $0\farcs08$ & $6.09^{\diamond}$ \\
$23\_2897$  & 7536 & 03:32:46.039 & --27:49:29.71 & $26.16\pm 0.07$ & $1.88$ & $0\farcs14$ & $5.78^{\diamond}$ \\
$33\_7746$  & 9235 & 03:32:14.739 & --27:47:58.75 & $26.36\pm 0.08$ & $1.21$ & $0\farcs14$ & $5.76$ \\
$25\_4498^{*}$  & --- & 03:32:32.460 & --27:40:01.93 & $26.41\pm 0.08$ & $>2.42$ & $0\farcs17$ & $5.97^{\diamond}$ \\
$33\_7751$  & 9234 & 03:32:24.797 & --27:47:58.82 & $26.41\pm 0.09$ & $1.33$ & $0\farcs17$ & $5.87$ \\
$25\_85$    & 16461 & 03:32:37.956 & --27:42:07.53 & $26.44\pm 0.09$ & $>2.39$ & $0\farcs17$ & $6.21$ \\
$32\_4331$  & --- & 03:32:22.282 & --27:52:57.21 & $26.48\pm 0.14$ & $>2.35$ & $0\farcs06$ & $6.20^{\diamond}$ \\
$22\_7650$  & 4264 & 03:32:31.190 & --27:52:06.17 & $26.48\pm 0.09$ & $>2.35$ & $0\farcs17$ & $6.09$ \\
$23\_12643^{*}$ & 10898 & 03:32:36.462 & --27:46:41.40 & $26.56\pm 0.09$ & $>2.28$ & $0\farcs15$ & $5.95^{\diamond}$ \\
$34\_10241$ & 15047 & 03:32:27.888 & --27:43:15.70 & $26.56\pm 0.08$ & $1.84$ & $0\farcs12$ & $5.87$ \\
$24\_3905$ & 13076 & 03:32:35.361 & --27:44:57.18 & $26.57\pm 0.10$ & $1.42$ & $0\farcs17$ & $5.73$ \\
$33\_12465$ & 11002 & 03:32:25.110 & --27:46:35.67 & $26.57\pm 0.10$ & $>2.26$ & $0\farcs06^{*\dagger}$ & $0^{*\dagger}$ \\
$33\_9307^{*}$  & 9858 & 03:32:27.397 & --27:47:28.28 & $26.57\pm 0.10$ & $2.17$ & $0\farcs08$ & $5.95$ \\
$31\_3672^{*}$  & 1366 & 03:32:17.808 & --27:54:41.60 & $26.59\pm 0.13$ & $>2.24$ & $0\farcs09$ & $5.87$ \\
$33\_4724^{*}$  & 8187 & 03:32:18.297 & --27:48:55.64 & $26.59\pm 0.10$ & $>2.24$ & $0\farcs17$ & $6.32$ \\
$13\_3987^{*}$  & --- & 03:32:48.941 & --27:46:51.45 & $26.68\pm 0.11$ & $1.09$ & $0\farcs08$ & $5.79$ \\
$22\_7964^{*}$  & 4361 & 03:32:36.833 & --27:52:01.01 & $26.70\pm 0.11$ & $1.30$ & $0\farcs10$ & $5.83$ \\
$23\_3398$  & 7737 & 03:32:43.348 & --27:49:20.37 & $26.72\pm 0.11$ & $0.850$ & $0\farcs12$ & $5.80$ \\
$24\_9420^{*}$  & 15052 & 03:32:36.342 & --27:43:15.51 & $26.72\pm 0.11$ & $>2.11$ & $0\farcs12$ & $6.12$ \\
$34\_10623$ & 15176 & 03:32:15.257 & --27:43:09.02 & $26.76\pm 0.12$ & $1.75$ & $0\farcs09$ & $6.12$ \\
$33\_11608^{*}$ & 10793 & 03:32:13.414 & --27:46:46.38 & $26.78\pm 0.12$ & $>2.05$ & $0\farcs14$ & $6.09$ \\
$22\_10359^{*}$ & 5207 & 03:32:38.961 & --27:51:16.64 & $26.82\pm 0.12$ & $1.33$ & $0\farcs14$ & $5.64$ \\
$13\_2604$  & 9135 & 03:32:52.214 & --27:48:04.80 & $26.88\pm 0.12$ & $1.53$ & $0\farcs17$ & $5.81$ \\
$35\_111$   & 16458 & 03:32:29.019 & --27:42:07.89 & $26.91\pm 0.13$ & $1.10$ & $0\farcs10$ & $5.94$ \\
\end{tabular}

$^{*}$\,These objects were rejected from further analysis for reasons 
outlined in Section~\ref{sec:GALFIT}.\\
$^{\diamond}$\,These are spectroscopic redshifts.  Those of $23\_6714$ \& 
$31\_2185$ come from Keck/DEIMOS spectroscopy
(Stanway et al.\ 2004; Bunker et al.\ 2003), and for $23\_2897$ from 
Gemini/GMOS spectroscopy (Stanway et al.\ 2004b).  The other 
spectroscopic redshifts are from VLT/FORS2 spectroscopy by
Vanzella et al.\ (2006).\\
$^{\dagger}$\,Object $33\_12465$ has photometry consistent with
being a T-dwarf low-mass star, and is unresolved in the {\em HST}/ACS images.\\
\caption{The coordinates and ($i'-z'$) colours of the sample of $i'$-band drop 
galaxies used in this study. The galaxies are ordered according to $z'$-band
magnitude, with the brightest at the top. $R_{hl}$ is the half-light
radius in arcseconds, measured in the $z'$-band; unresolved
sources have $R_{hl}\approx 0\farcs06$. All redshifts, unless otherwise stated, are 
taken from the GOODS-MUSIC photometric redshift catalog,
which has a typical error of $\Delta z=0.08$ (Grazian et al.\ 2006).
Note that for some objects, their 
($i'-z'$) colours presented here fall outside the colour cut of $> 1.3$ 
used to select $i'$-drop galaxies.  This is due to their initial
selection being based on colours measured in a smaller aperture than the $0\farcs5-$diameter aperture magnitudes presented in this paper.}
\label{tab:COORDS}
\end{table*}

\section{Data Analysis}
\label{sec:REDUCT}

\subsection{Removal of Contaminating Sources}
\label{sec:GALFIT}

Before photometry of individual objects was gathered, it was necessary
to ensure that they did not suffer from contamination due to
neighbouring sources; this is particularly common in
the {\em Spitzer}/IRAC data, due to the low spatial resolution 
(${\rm FWHM}\approx 1\farcs5$) compared with the {\em HST}
(${\rm FWHM}\approx 0\farcs05$).  Each of the $i'$-drops in our
sample was visually inspected in IRAC channels 1 \& 2 (3.6\,$\mu$m \&
4.5\,$\mu$m), and compared to the higher-resolution {\em HST}/ACS
$z'$-image and the VLT/ISAAC $K_{s}$-band in order to identify any
contaminating sources. As in our recent work on the $z\sim 5$ $v$-band
drop-outs in this field (Stark et al.\ 2006), each galaxy was
classified in the {\em Spitzer}/IRAC data as: a) isolated and
detected; b) confused; c) undetected; or d) unobserved
(either not in the field of view,
or in a noisy region close to the edge of the field where
few frames overlap). Four of our
31 objects fell into category d), and so were removed from the sample
and not analysed further.  Of the remaining 27, three sources were
isolated and detected (category a) and one was isolated and
undetected (category c).  Hence 23 of the $i'$-drop galaxies in our sample were
classified as confused, and for these objects attempts were made to
subtract out the contribution from neighbouring sources, using the
``GALFIT'' software package (Peng et al.\ 2002) to model the surface
brightness profiles of nearby sources in the IRAC field. We
emphasize that we use GALFIT to fit the neighbouring sources to
subtract out, not the $i'$-drops themselves. 

In order to perform the subtraction of contaminating sources with GALFIT, it
was necessary to obtain the point spread function (PSF) for each
waveband.  For the {\em Spitzer}/IRAC data, we used in-flight PSFs
which were oversampled by a factor of two relative to the `drizzled'
data\footnote{see \newline {\tt
    http://ssc.spitzer.caltech.edu/irac/psf.html}}.
These PSFs were then rotated to match the orientation of the
drizzled IRAC GOODS-South images; epoch 2 differed from epoch 1 by a roll angle of
$180^{\circ}$.  The PSFs of the $J$ \& $K_{s}$-band data were created
by trimming $12''\times 12''$ regions around bright yet unsaturated
stars in the field of view; due to variation in seeing between
different pointings in the mosaic, a separate PSF for each individual
image tile was created.  The stars used for these PSFs were then later
employed in calculating aperture corrections for the
photometry (see Section~\ref{sec:PHOT}).

GALFIT constructs a two-dimensional model of each confusing galaxy for various
surface brightness profiles (e.g.,\ exponential disk, de Vaucouleurs,
S\'ersic) according to several initial input parameters specified by
the user (e.g.,\ magnitude, centre, axial ratio, position angle),
which can be held fixed or allowed to vary. The model is then
convolved with the instrument PSF (we used two-times oversampled PSFs
for IRAC).  Perturbing the model parameters from the initial guess, an
iterative $\chi^{2}$ minimisation process converges on a best-fit
solution.  The best-fit parameters are returned, along with images of the
2-D fitted model and of the residual flux when the model is subtracted from
the original galaxy image.

For the 23 confused sources in our sample, $12''\times 12''$ regions
surrounding each individual object in the $J$, $K_{s}$, 3.6\,$\mu$m \&
4.5\,$\mu$m wavebands were used to begin the following process.
Initial GALFIT input parameters for neighbouring sources were estimated 
from the $K_{s}$-band
images, as this waveband typically had the highest spatial resolution of the
infrared images. We used the IRAF task {\tt imexamine} to obtain
magnitude and coordinate estimates of the neighbouring sources.
GALFIT was then run on the $K_s$-band image with all input parameters
allowed to vary and a generic S\'ersic profile adopted, where
$\log({\rm Surface\ Brightness)}\propto r^{1/n}$ (where $n$ is the S\'ersic index). 
Using the output
parameters from the $K_{s}$-band fits as input estimates, GALFIT was
then applied to the other infrared filters ($J$, 3.6\,$\mu$m \& 4.5\,$\mu$m).
In these wavebands, three versions of the fitting process were carried
out.  For one version, all input parameters as determined from the 
$K_s$-band were held fixed, except the magnitudes (this would be
appropriate if the morphology was independent of wavelength, i.e.\ no
colour gradients).  In the second version, we repeated the GALFIT procedure, 
but updated the input
centres and magnitudes of the surrounding sources with those measured
directly from the $J$, 3.6\,$\mu$m \& 4.5\,$\mu$m images, and again
fixed the galaxy shape parameters (axial ratio, PA, S\'ersic index,
scale length) from the $K_s$-band fitting.  Finally, we also re-ran
GALFIT on the $J$, 3.6\,$\mu$m \& 4.5\,$\mu$m images allowing {\em
  all} of the input parameters to vary.  Each resulting subtraction
was then visually inspected, and the best fit for each waveband was
selected.  For those confused objects which appear in both epochs of
the {\em Spitzer}/IRAC data, the subtraction of neighbouring objects was attempted on each epoch
independently, as the PSF differed greatly. Some galaxies
appeared in the central overlap region between the two epochs;
we used the deeper of the two, and where the depth was the
same we used
the epoch which yielded the better GALFIT subtraction of confusing
sources.  In several cases (particularly when used on the {\em
  Spitzer}/IRAC data), GALFIT failed to satisfactorily remove the
confusing source(s); this resulted in the removal of a further 10
sources from our sample, leaving a total of 17 sources for which
reliable photometry was obtainable.  From this list, object $33\_12465$ was 
later rejected as its photometry is more consistent with being a low-mass
T-dwarf star than a $z\sim 6$ galaxy (see Section~\ref{sec:PHOT}),
and it is unresolved in the {\em HST}/ACS images.  Figure~\ref{fig:ORIGINALSTAMPS}
shows the multi-waveband images of the
remaining 16 objects of our sub-sample, before subtraction of
confusing sources was conducted (left-hand side), and also the results of 
GALFIT for those confused objects for which subtraction of neighbouring
sources was successful (right-hand side).
In Section~\ref{sec:MASSDENSITY}, we discuss the correction
to the total stellar mass density of those objects for which attempts at subtracting the effect of neighbouring confusing sources were
unsuccessful. In Figure~\ref{fig:ZMAGHIST}
we show the distribution of
$z'$-band magnitudes for our 17 $i'$-drops with clean photometry, compared
to the full sub-sample of 31 $i'$-drop galaxies with spectroscopic or photometric redshifts,
and with our complete catalog of 52 $i'$-drops with $z'_{AB}<26.9$ (see
Section~\ref{sec:IDROPS}).  
The magnitude distribution of the sub-sample analysed does not show any obvious bias; a Kolmogorov-Smirnov test shows
that there is a 99 per cent probability of the 31 galaxies with
redshift information having the same $z'$-band magnitude
distribution as the total $i'$-drop sample
of 52; and that there is a 94 per cent chance that the 17 $i'$-drops
analysed have the same magnitude distribution as all
the $i'$-drops with redshift information.

\begin{figure}
\resizebox{0.48\textwidth}{!}{\includegraphics{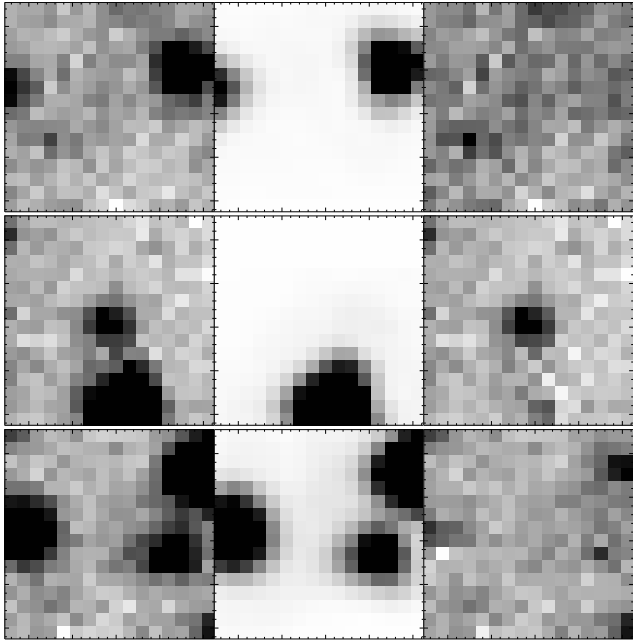}}
\caption{Examples of subtraction of neighbouring confusing
sources using GALFIT in IRAC channel 1. From top to bottom, objects $13\_3880$, $25\_85$ \& $34\_10623$, with the original images, the models fitted and the residual images from left to right respectively. Each box is $9''$ across. North is up and East is to the left.}
\label{fig:GALFITEXAMPLES}
\end{figure}

\subsection{Photometry}
\label{sec:PHOT}

As shown in our previous work, the typical half-light radius of a $z
\sim 6$ $i'$-drop galaxy is $r_{hl}<0\farcs2$ (Bunker et al.\ 2004;
see also Ferguson et al.\ 2004, Bouwens et al.\ 2004a), and these
objects are often barely resolved in the {\em HST}/ACS images.  Hence
when viewed at the poorer resolution of the VLT/ISAAC data, and in
particular the {\em Spitzer}/IRAC images, they are essentially
unresolved, and so we treat them as point sources for the purposes of 
photometry.  {\em HST}/ACS photometry of our full catalog of 52
$i'$-drop candidates (see Section~\ref{sec:IDROPS}) had been
previously gathered by our group (Stanway 2004; Bunker et al.\ 2004),
using $0\farcs5$ diameter apertures, and had been
both aperture \& extinction-corrected for Galactic dust; we used these 
magnitudes for this analysis.  Magnitudes for
each of our selected $i'$-drops in the near-IR and IRAC wavebands were 
obtained using the IRAF {\tt digiphot.phot} package to perform aperture 
photometry, measuring the enclosed flux at the coordinates determined from 
the astrometry of the GOODS version-1.0 $z'$-band data.  Aperture
corrections were then applied to convert the measured aperture
magnitudes to approximate total magnitudes, appropriate for point
sources.  Photometry
in the ground-based VLT/ISAAC $J$ \& $K_{s}$ images was performed
using $1''$-diameter apertures, removing the residual sky background
determined from an annulus between $12''$ and $24''$ radius.  The
seeing was generally good (${\rm FWHM}=0\farcs4-0\farcs5$) but varied
from tile to tile. Hence a unique aperture correction for each
individual tile was calculated, using the PSF stars used
in the subtraction of confused sources with GALFIT
(Section~\ref{sec:GALFIT}), measuring the flux out to a diameter
of $6''$.  The aperture corrections for our fixed
$1''$-diameter apertures were found to be in the range $\approx
0.3-0.5$\,mag.
Noise estimates were also made on a
tile-to-tile basis to account for the differing seeing conditions and
exposure times, producing $3\sigma$ limiting total magnitudes of
$J(AB)\approx 26.0$ \& $K_{s}(AB)\approx 25.6$.

For the {\em Spitzer}/IRAC photometry, we used an aperture of diameter
$\approx 1.5\times {\rm FWHM}$, which maximises the signal-to-noise
ratio ($S/N$) for unresolved objects.  For both IRAC channels 1 \& 2,
the FWHM (as determined from unsaturated point sources in the GOODS-South
IRAC images) was 2.5
drizzled pixels or $1\farcs5$.  Hence we set the photometric
aperture diameter to be 4 drizzled pixels ($2\farcs4$).  The local background
was determined from an annulus between 20 and 40 pixels radius ($12''$
and $24''$).  We note that for object $35\_111$, a smaller background
annulus was employed (between 7 and 10 pixels).  This was due to the
fact that this source resides in a particularly crowded region of the
GOODS-South field, where the larger sky annulus was significantly contaminated
by several bright objects.  As determined in
our previous study, the corrections for the IRAC 3.6\,$\mu$m \&
4.5\,$\mu$m channels to convert from our aperture magnitudes to
approximate total magnitudes were found to be $\approx 0.7$\,mag,
measuring out to $18''$ diameter.  These are consistent with those derived for the
First Look Survey (Lacy et al.\ 2005).  The noise estimates adopted in
this study mirror those used in Eyles et al.\ (2005) and
Stark et al.\ (2006) -- $3\sigma$ limiting magnitudes of $26.5$ \&
$26.1$ for channels 1 \& 2 respectively, in $2\farcs4$ diameter
apertures.

Photometry was gathered on the 17 sources in our sub-sample, with 13 of
them having had neighbouring objects subtracted with GALFIT (see
Section~\ref{sec:GALFIT}).  Of these 17 objects, 10 were detected 
at $>3\,\sigma$ in IRAC channel 1 and 9 in channel 2 (with source marginally 
detected in channel 2).  Table~\ref{tab:MAGS} lists the measured AB 
magnitudes in the $i'$, $z'$, $J$, $K_{s}$, $3.6\,\mu$m \& $4.5\,\mu$m 
filters of the 17
objects for which reliable photometry was gathered.  These values have
had the appropriate aperture corrections applied, and so represent
estimated total magnitudes, which were then used to conduct spectral energy 
distribution modelling (Section~\ref{sec:SEDS}).

The photometry
of object $33\_12465$ appears to be anomalous for a $z\sim 6$ galaxy,
with very peculiar colours -- twice as bright in $J$ and IRAC channel 2
than in $K_s$ and IRAC channel 1, quite unlike the SEDs of our spectroscopically-confirmed
$i'$-drops. The 
source appears to be unresolved in the {\em HST}/ACS imaging
suggesting that it may be a Galactic point source (although the GOODS-MUSIC catalog does
allocate a photometric redshift of $z=6.93$). It was not
removed from the initial $i'$-drop list because its faint magnitude ($z'_{AB}=26.6$)
is where star-galaxy separation just begins to become unreliable.
However, in Figure~\ref{fig:TDWARF} we overplot the SED on a spectrum of
the T7 Dwarf Gl 229B\footnote{Available from {\tt
    http://www.jach.hawaii.edu/$\sim$skl/spectra/ \newline
    T7$\_$Gl229B.txt}} (Leggett et al.\ 2002).  The photometry of
$33\_12465$ closely resembles that of a T-dwarf, with
H$_2$0 and CH$_4$ spectral features significantly contributing to the detected $J$-band
and $4.5\,\mu$m fluxes.  Comparing our measured photometry
to the colours of T-dwarfs in Patten et al.\ (2006), and converting
our AB magnitudes to the Vega system used, the $K_s$, 
3.6\,$\mu$m and 4.5\,$\mu$m colours appear to be consistent 
with a T5$-$7 dwarf. The only other $i'$-drop in our subsample
to have a half-light radius as small as $R_{hl}=0\farcs 06$ 
(as for the T-dwarf $33\_12465$) is $32\_4331$, which has a spectroscopic redshift
of $z=6.2$ and colours unlike the T-dwarf (i.e., probably not a star) .

Hence object $33\_12465$ was also eliminated from
inclusion in any further analysis as a probable T-dwarf star,
resulting in a final selection of
16 $i'$-drop galaxies on which SED fitting was subsequently conducted.
For objects $23\_6714$, $31\_2185$ \& $23\_2897$ (SBM03\#1, SBM03\#3
\& GLARE\#3001 respectively), comparisons were made with the
photometry collected in our previous study (Eyles et al.\ 2005); that
of $23\_6714$ \& $31\_2185$ appears to be consistent with our original
data.  It should be noted that in our earlier study, deeper {\em
  HST}/NICMOS F110W (`$J$-band') \& F160W (`$H$-band') data were used in place
of the VLT/ISAAC $J$-band for $23\_6714$ (SBM03\#1); this is the only
object in our selection which falls within the HUDF field.  Also, at
the time, no IRAC $4.5\,\mu$m data were available for $31\_2185$ -- it
was not in the channel 2 field of view in the epoch 1 data release.
The $4.5\,\mu$m detection of object $23\_2897$ in the deeper epoch 2
data is slightly discrepant with respect to our earlier analysis, in
which it was undetected in epoch 1. The stellar ages and mass estimates
from stellar population fits presented in Eyles et al.\ (2005)
are not significantly altered with this new photometry (Section~\ref{sec:ANALYSIS}).

We have also compared our photometry with that in the GOODS-MUSIC catalog,
which used PSF-matching of the space- and ground-based data with
the `ConvPhot' routine. For our $i'$-drop sample, we find in general good
agreement for the magnitudes;
where discrepancies were noted, visual inspection showed that this
was most likely through confusing sources which we have tried
hard to eliminate by GALFIT modelling and subtraction.
We have chosen to  use our aperture magnitudes in the current analysis, corrected to approximate
total magnitudes through aperture corrections, as these are most appropriate
to barely-resolved sources (such as our $i'$-drops in the infrared) and are
simple and reproducible. We note however that agreement with the independently-determined
GOODS-MUSIC catalog magnitudes is good, and adopting these magnitudes
for most sources would not qualitatively affect
our stellar age and mass determinations from population synthesis model fits.

\begin{table*}
\centering
\begin{tabular}{l|c|c|c|c|c|c|c|c|}
ID & $i'$ & $z'$ & $J$ & $K_{s}$ & 3.6\,$\mu$m & 4.5\,$\mu$m\\
\hline\hline
$31\_2185^{\dagger}$  & $26.21\pm 0.05$ & $24.61\pm 0.02$ & $24.53\pm 0.15$ & $25.66\pm 0.37$ & $23.93\pm 0.08$ & $24.35\pm 0.14^{\diamond}$ \\
$23\_6714$  & $26.77\pm 0.09$ & $25.35\pm 0.03$ & $25.34\pm 0.20$ & $24.99\pm 0.22$ & $24.26\pm 0.09$ & $24.34\pm 0.15$ \\
$13\_3880^{\dagger}$  & $27.55\pm 0.18$ & $26.10\pm 0.06$ & $>26.1$ ($3\,\sigma$) & $25.09\pm 0.27$ & $>26.5$ ($3\,\sigma$) & $>26.1$ ($3\,\sigma$) \\
$31\_3127$  & $>28.8$ ($3\,\sigma$) & $26.15\pm 0.07$ & $>25.8$ ($3\,\sigma$) & $>25.4$ ($3\,\sigma$) & $>26.5$ ($3\,\sigma$) & $>26.1$ ($3\,\sigma$) \\
$23\_2897$  & $28.04\pm 0.26$ & $26.16\pm 0.07$ & $26.09\pm 0.38$ & $25.40\pm 0.35$ & $26.01\pm 0.51$ & $25.17\pm 0.31$ \\
$33\_7746^{\dagger}$  & $27.57\pm 0.18$ & $26.36\pm 0.08$ & $>25.8$ ($3\,\sigma$) & $>25.1$ ($3\,\sigma$) & $>26.5$ ($3\,\sigma$) & $>26.1$ ($3\,\sigma$) \\
$33\_7751^{\dagger}$  & $27.75\pm 0.21$ & $26.41\pm 0.09$ & $>25.7$ ($3\,\sigma$) & $>25.6$ ($3\,\sigma$) & $>26.5$ ($3\,\sigma$) & $>26.1$ ($3\,\sigma$) \\
$25\_85^{\dagger}$    & $>28.8$ ($3\,\sigma$) & $26.44\pm 0.09$ & $25.22\pm 0.18$ & $>25.0$ ($3\,\sigma$) & $24.18\pm 0.09$ & $24.67\pm 0.20^{\diamond}$ \\
$32\_4331^{\dagger}$ & $>28.8$ ($3\,\sigma$) & $26.48\pm 0.14$ & $25.55\pm 0.29$ & $25.68\pm 0.29$ & $24.60\pm 0.14$ & $24.61\pm 0.19$ \\
$22\_7650^{\dagger}$  & $>28.8$ ($3\,\sigma$) & $26.48\pm 0.09$ & $26.48\pm 0.50$ & $25.85\pm 0.50$ & $>26.5$ ($3\,\sigma$) & $>26.1$ ($3\,\sigma$) \\
$34\_10241^{\dagger}$ & $28.40\pm 0.31$ & $26.56\pm 0.08$ & $>26.0$ ($3\,\sigma$) & $25.78\pm 0.73$ & $>26.5$ ($3\,\sigma$) & $>26.1$ ($3\,\sigma$) \\
$24\_3905$ & $27.99\pm 0.25$ & $26.57\pm 0.10$ & $>26.1$ ($3\,\sigma$)  & $>25.8$ ($3\,\sigma$) & $25.94\pm 0.38$ & $26.13\pm 0.65$ \\
$33\_12465^{*}$ & $>28.8$ ($3\,\sigma$) & $26.57\pm 0.10$ & $24.13\pm 0.08$ & $24.74\pm 0.18$ & $24.68\pm 0.17$ & $24.07\pm 0.11$ \\
$23\_3398^{\dagger}$  & $27.57\pm 0.17$ & $26.72\pm 0.11$ & $26.47\pm 0.52$ & $>25.6$ ($3\,\sigma$) & $24.91\pm 0.16$ & $25.26\pm 0.42$ \\
$34\_10623^{\dagger}$ & $28.51\pm 0.41$ & $26.76\pm 0.12$ & $>25.9$ ($3\,\sigma$) & $>25.3$ ($3\,\sigma$) & $>26.5$ ($3\,\sigma$) & $>26.1$ ($3\,\sigma$) \\
$13\_2604^{\dagger}$  & $28.41\pm 0.36$ & $26.88\pm 0.12$ & $26.37\pm 0.54$ & $>25.6$ ($3\,\sigma$) & $24.54\pm 0.15$ & $25.18\pm 0.43^{\diamond}$ \\
$35\_111^{\dagger}$   & $28.01\pm 0.26$ & $26.91\pm 0.13$ & $26.13\pm 0.44$ & $>25.2$ ($3\,\sigma$) & $25.92\pm 0.35$ & $>26.1^{\diamond}$ ($3\,\sigma$) \\
\end{tabular}

$^{\dagger}$\,These objects are those which were noted to be confused, and 
subsequently had the neighbouring sources successfully subtracted by GALFIT. \\
$^{\diamond}$\,IRAC channel 2 (4.5\,$\mu$m) is anomalously
faint compared to IRAC channel 1 (3.6\,$\mu$m).\\
$^{*}$\,This object is most likely a low-mass T-dwarf object (see Section~\ref{sec:PHOT}).  
\caption{Estimated total magnitudes (AB system) of our sample of 17 $i'$-band drop candidates 
for which reliable photometry was obtained (including a probable T-dwarf).  Any non-detection is represented 
by the corresponding $3\,\sigma$ limiting magnitude.} 
\label{tab:MAGS}
\end{table*}

\subsection{Spectral Energy Distribution Fitting}
\label{sec:SEDS}

Once magnitudes in each of the different wavebands had been obtained,
the photometric data were then used to construct SEDs for each of our
selected sources.  As in our previous work, we made use of the latest
Bruzual \& Charlot (2003, hereafter B\&C) isochrone synthesis code,
utilising the Padova-1994 evolutionary tracks (preferred by B\&C).  
The models span a range of 221 age steps approximately
logarithmically spaced, from $10^5$\,yr to $2\times 10^{10}$\,yr,
although here we discount solutions older than $\sim 10^9$\,yr
(the age of the Universe at $z\approx 6$).
The B\&C models have 6900 wavelength steps, with high resolution (FWHM 3\,\AA ) and 1\,\AA\
pixels over the wavelength range 3300\,\AA\ to 9500\,\AA\, and
unevenly spaced outside this range.  We opted to primarily explore
models with the Salpeter (1955) initial mass function (IMF), and of
solar metallicity. We considered the effects of sub-solar metallicity
and adopting the Chabrier (2003) IMF in Eyles et al.\ (2005,
see also Section~\ref{sec:DUST}). From the
range of possible star formation histories (SFH) available, we
considered a single stellar population (SSP; an instantaneous burst),
a constant star formation rate (SFR), and several
exponentially-decaying star formation rate ``$\tau$-models". These had ${\rm
  SFR = SFR_0} e^{-t/\tau}$, where ${\rm SFR_0}$ is the star formation
rate at time $t=0$ (the formation of the galaxy). We considered
exponential star formation histories with decay constants in the range
$\tau=10-1000$\,Myr, specifically $\tau=10, 30, 70, 100, 300, 500\ {\rm \&} 1000$\,Myr.

For an SSP (instantaneous burst) model, the B\&C synthetic spectra are
normalised to an initial total mass of $1\,M_{\odot}$. For the
constant SFR model, the B\&C template normalization is an SFR of
$1\,M_{\odot}\,{\rm yr}^{-1}$.  Additionally, we investigated a model
composed of two distinct stellar populations: an ongoing starburst 
with constant SFR at
the time of observation, and an underlying older population that
formed via an instantaneous burst sometime previously.  We also
consider the possibility that the red optical--infrared colours of
objects within our sample could be due to intrinsic dust reddening,
rather than an age-sensitive spectral break (see
Section~\ref{sec:ANALYSIS}).  As in our previous work, we adopted the
empirical reddening model of Calzetti (1997), suitable for starburst
galaxies.

For each $i'$-drop in our sample, the filters were corrected to their
rest-frame wavelengths by the appropriate redshift factor.  The
measured flux in each waveband was folded through the corresponding
filter transmission profile, and the best-fit age model was computed
via minimisation of the reduced $\chi^{2}$, using the errors measured
on the magnitudes. The flux of the models
below Lyman-$\alpha$ ($\lambda_{\rm rest}=1216$\,\AA ) was 
reduced to correct for blanketting by intervening Lyman-$\alpha$ 
forest absorption ($D_A=0.95$ was assumed at $z\approx 6$).
Some of our data points, particularly those
from the {\em HST}/ACS imaging, have $S/N>10$.  However, as done
previously, the minimum magnitude error is set to $\Delta({\rm mag}) =
0.1$, to account for calibration uncertainties.  Non-detections
in the VLT/ISAAC $J$ \& $K_{s}$-bands were
treated in two different ways. First, the fitting routine was run
with all non-detections set to the corresponding $1\,\sigma$ detection
limits, and a magnitude error of $\Delta({\rm mag}) = 1.0$ was
imposed, giving these filters a very low statistical weighting during
the fitting process. 
Secondly, the
fitting was re-run, omitting any filters with non-detections.  Checks
were carried out on the returned best fit SEDs to ensure that the flux
density in the wavelength region of the omitted filter was not in
conflict with the corresponding $3\,\sigma$ upper limit: this was
found to be the case, so we adopted the fits excluding each near-IR 
waveband in which an object was undetected. In seven of the 16 galaxies in 
our sub-sample, there was no IRAC detection at $>3\,\sigma$ in either channel 1
or 2 (Figure~\ref{fig:MULTIGALPLOTSundetect}). We present a stacking 
analysis of these sources in Section~\ref{sec:NONDETECTIONS}.

\begin{figure*}
\resizebox{0.26\textwidth}{!}{\includegraphics{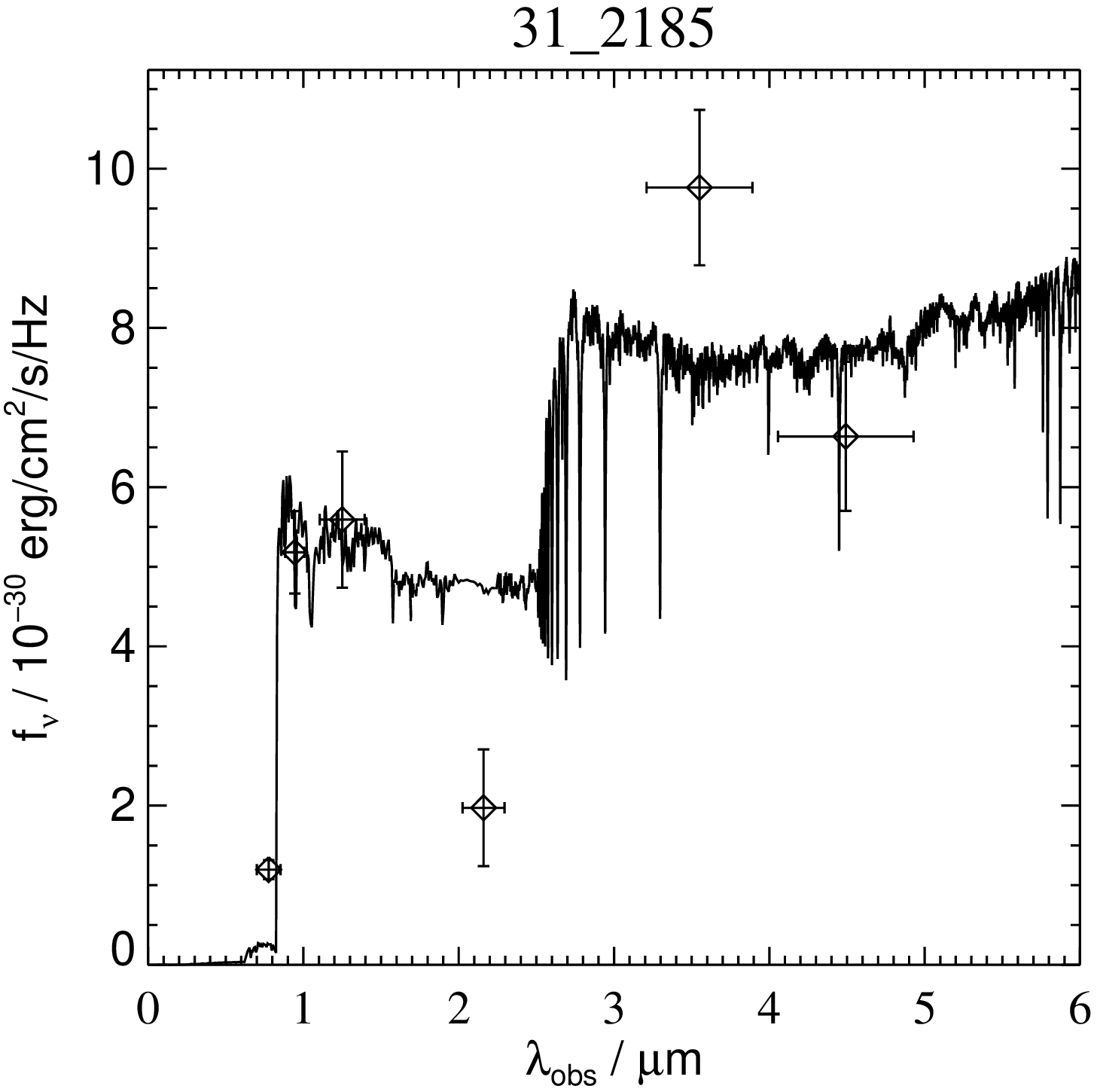}}
\resizebox{0.26\textwidth}{!}{\includegraphics{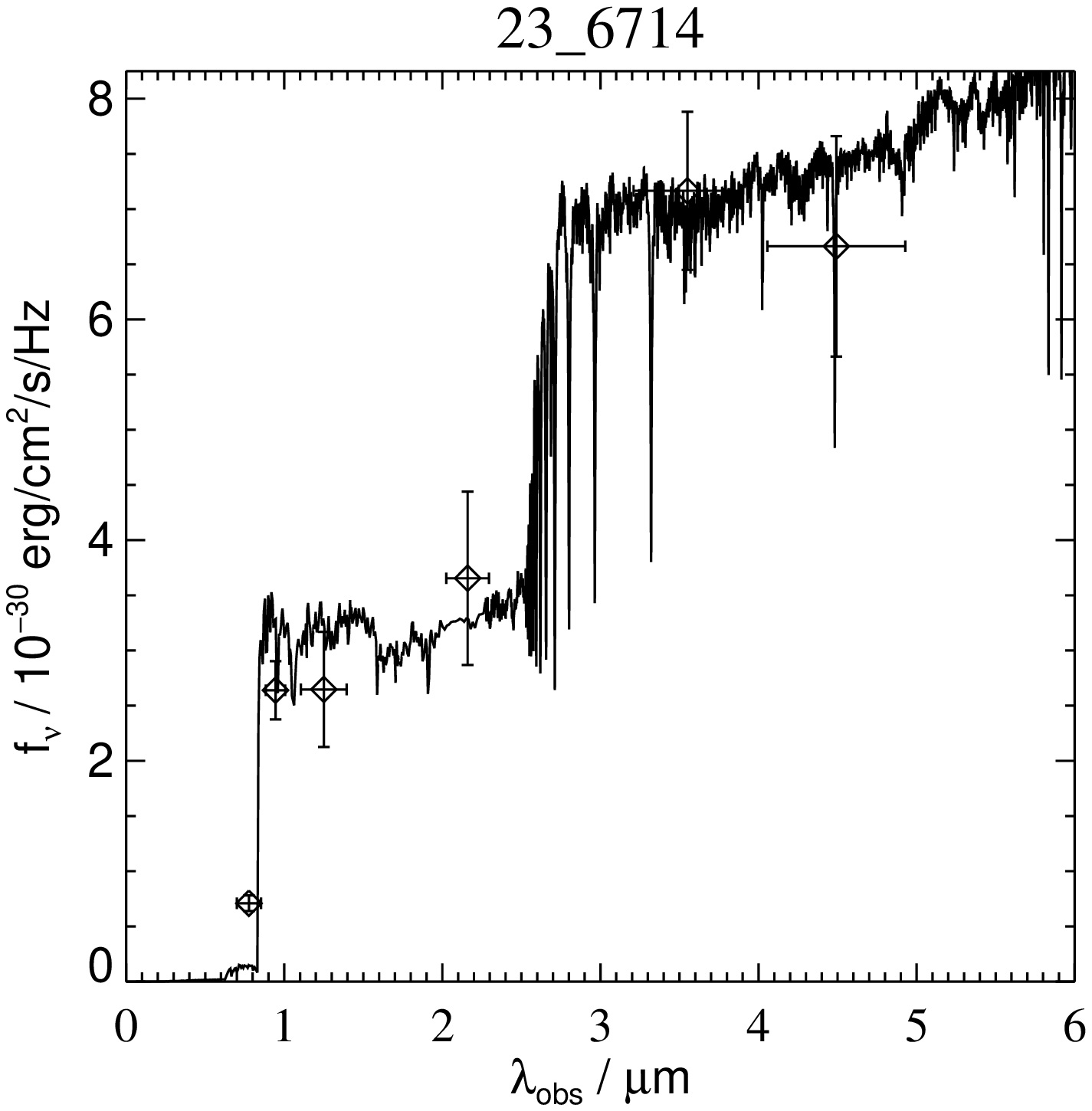}}
\resizebox{0.26\textwidth}{!}{\includegraphics{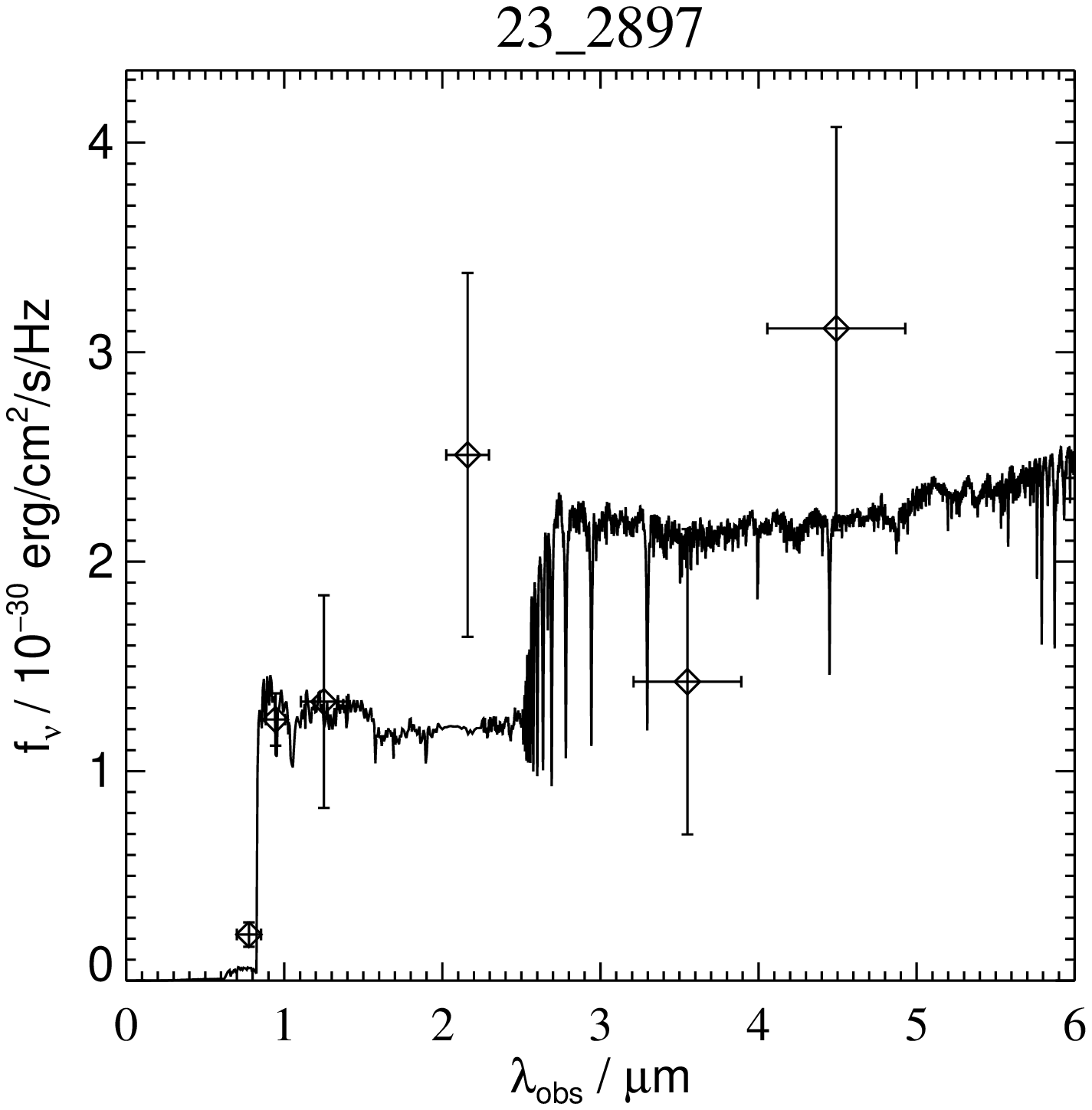}}
\resizebox{0.26\textwidth}{!}{\includegraphics{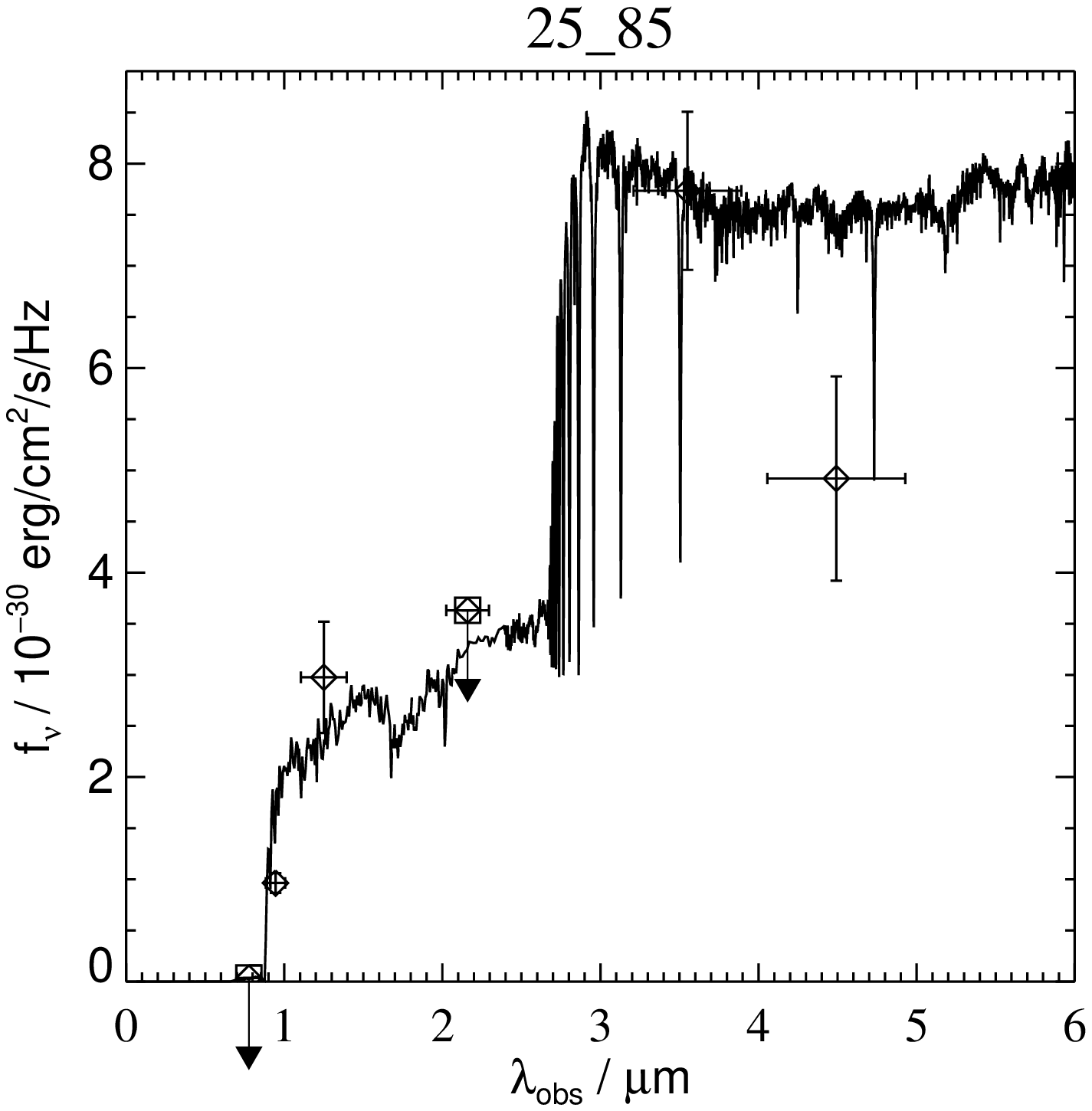}}
\resizebox{0.26\textwidth}{!}{\includegraphics{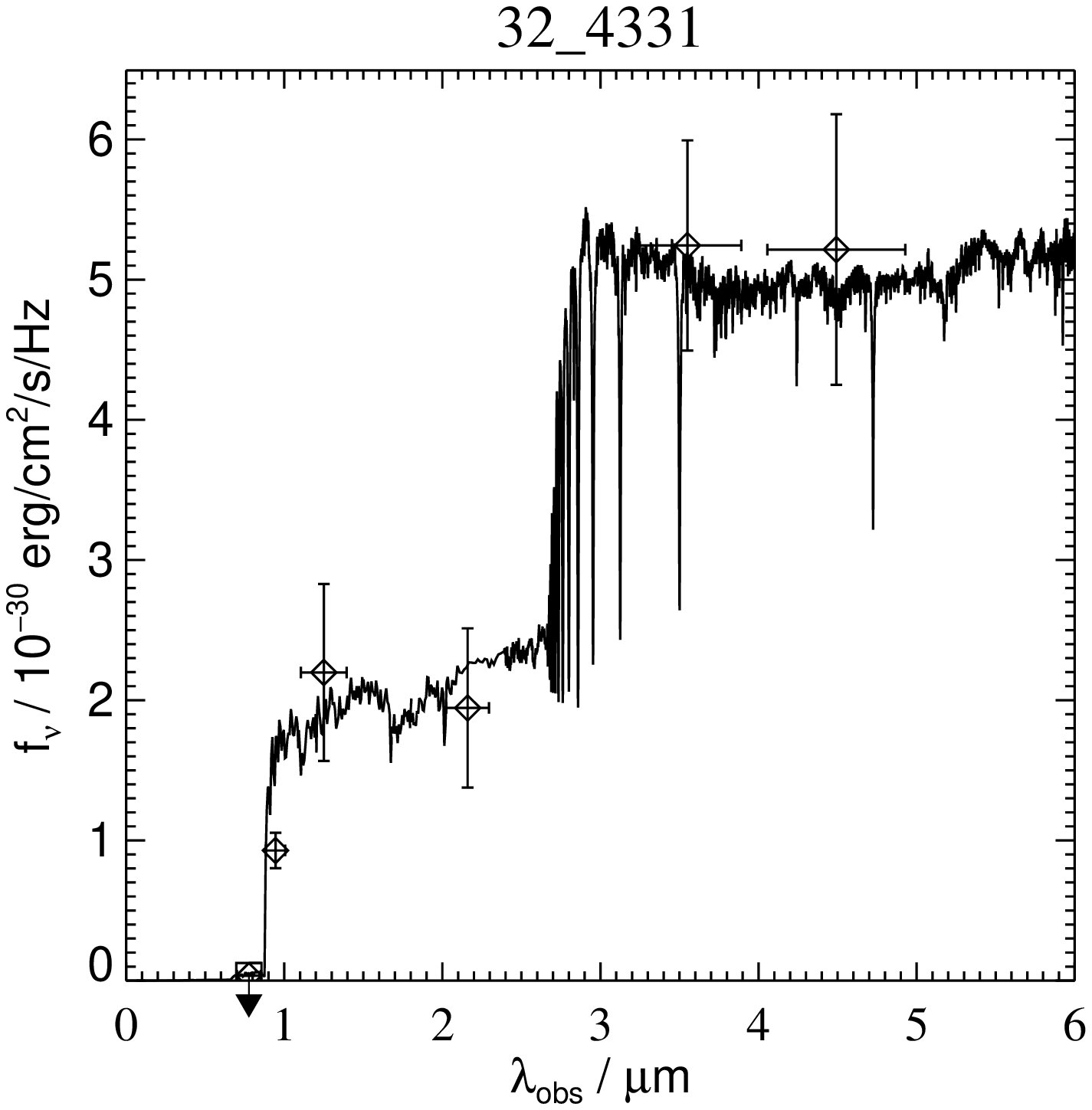}}
\resizebox{0.26\textwidth}{!}{\includegraphics{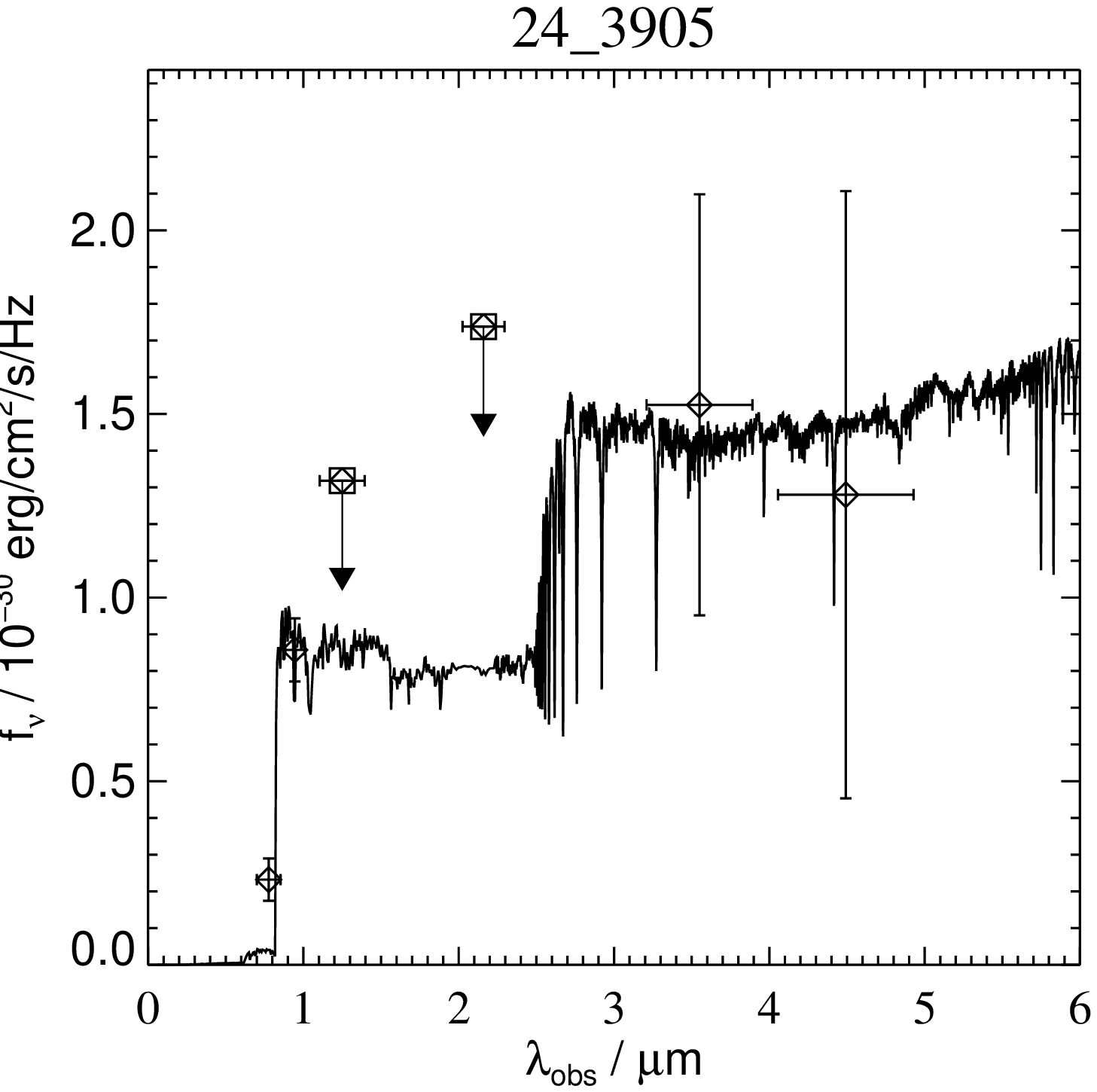}}
\resizebox{0.26\textwidth}{!}{\includegraphics{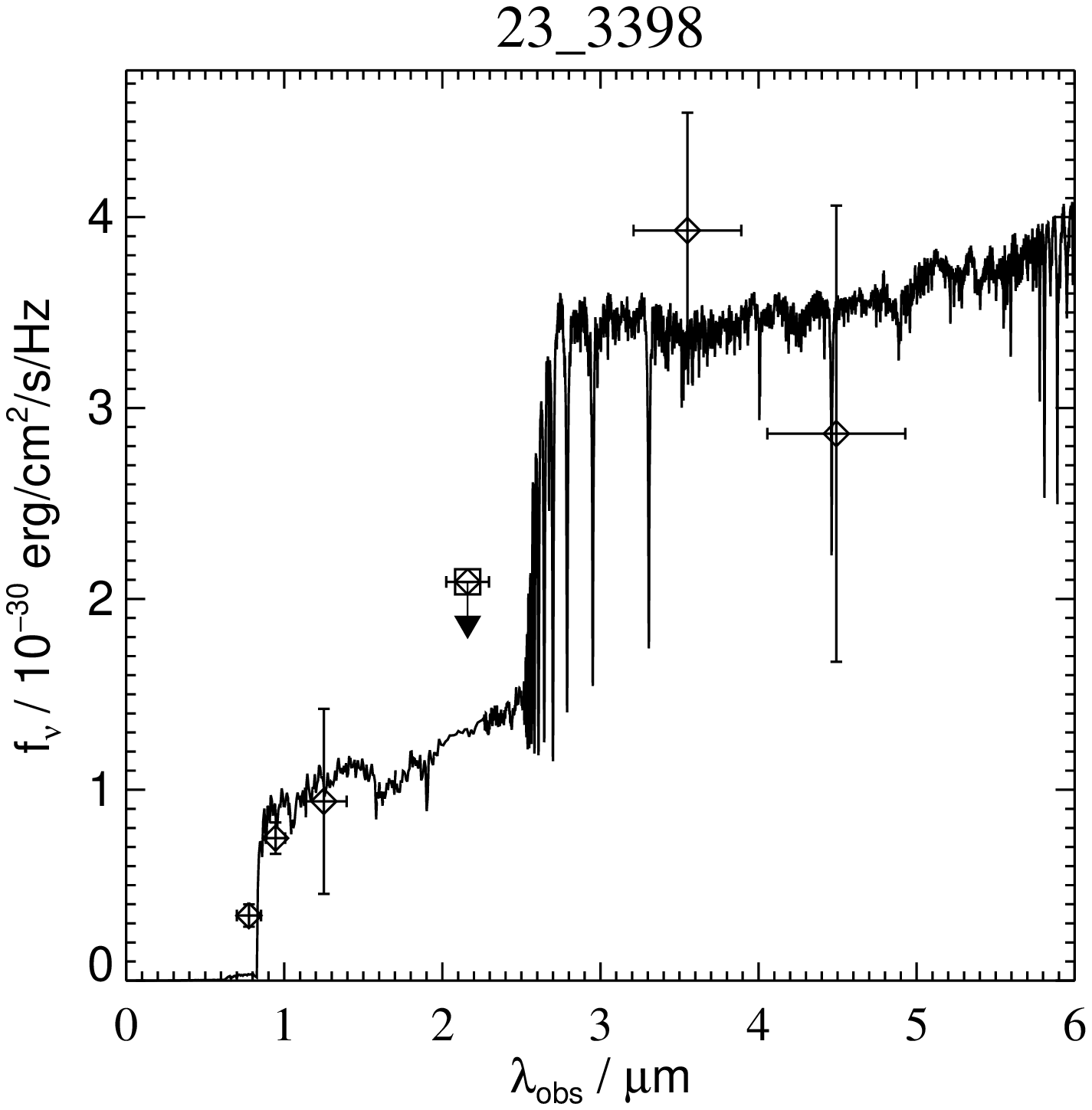}}
\resizebox{0.26\textwidth}{!}{\includegraphics{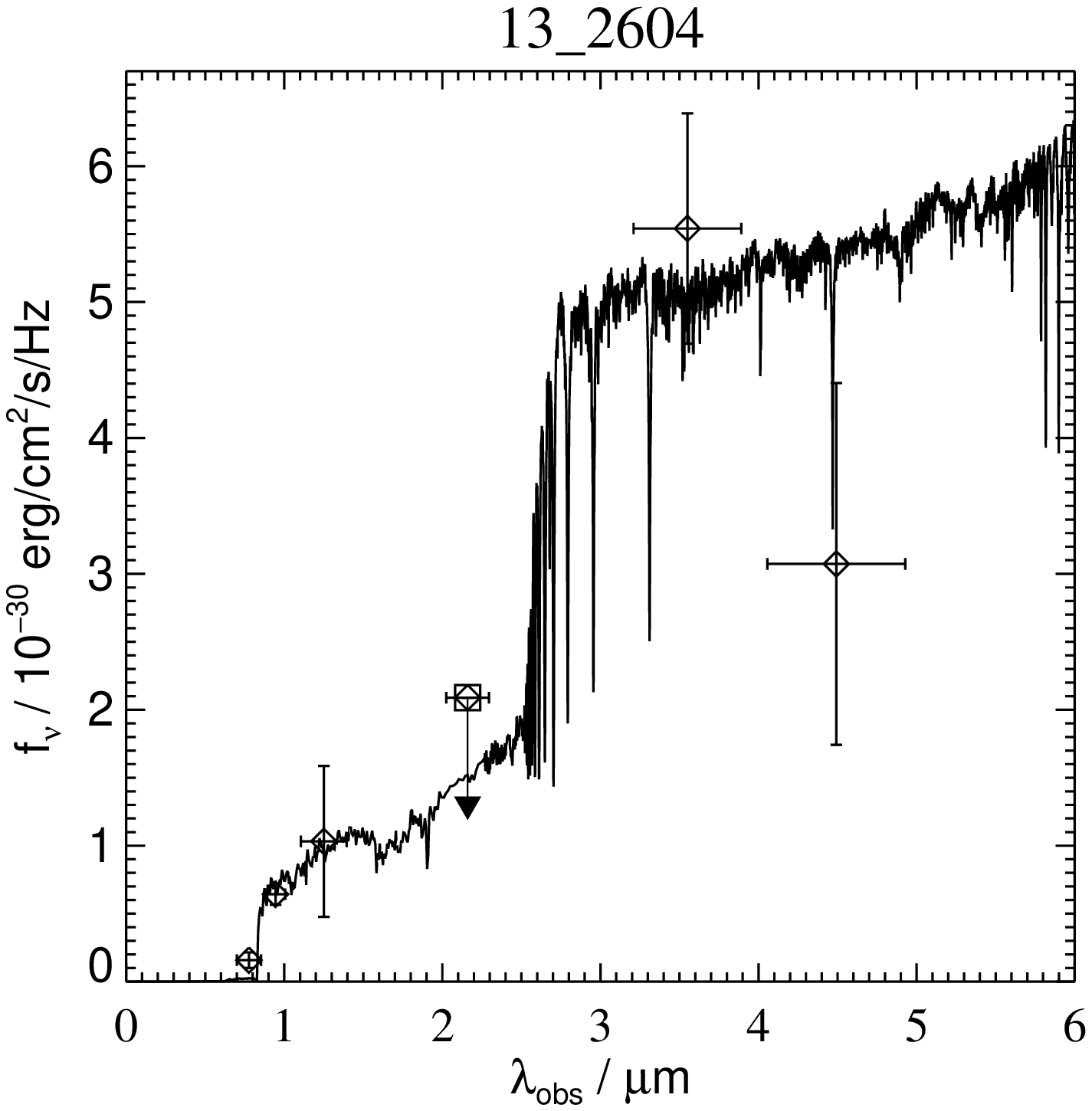}}
\resizebox{0.26\textwidth}{!}{\includegraphics{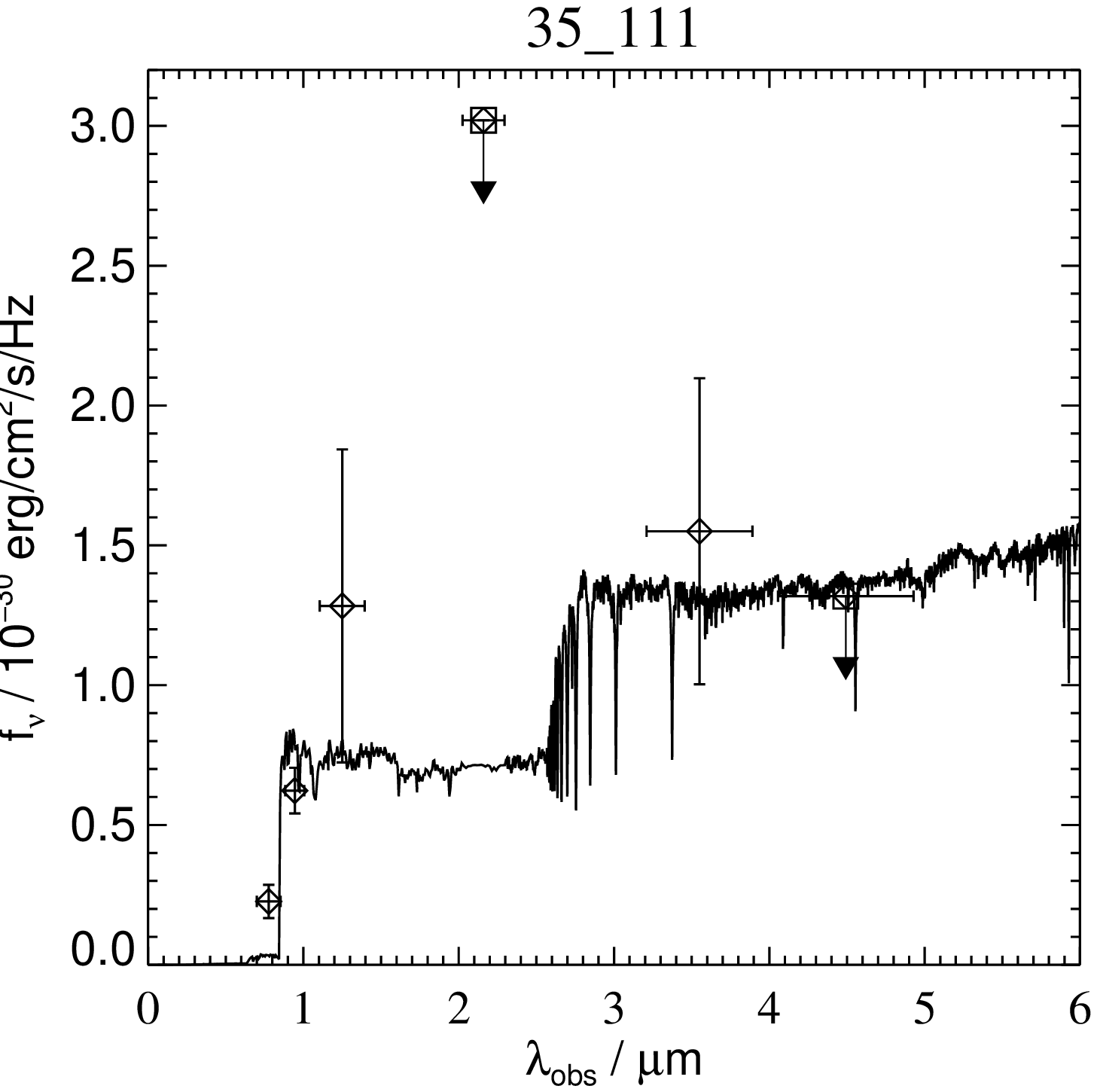}}
\caption{The best-fit SED models (excluding dust reddening) to the photometric datapoints for the nine $z\sim 6$ sources
with IRAC detections. Non-detections are represented by their $3\,\sigma$ upper limits. In the case of $25\_85$, the
anomalously faint IRAC channel 2 (4.5\,$\mu$m) magnitude
has been excluded from the fit.}
\label{fig:MULTIGALPLOTS}
\end{figure*}

\begin{figure*}
\resizebox{0.26\textwidth}{!}{\includegraphics{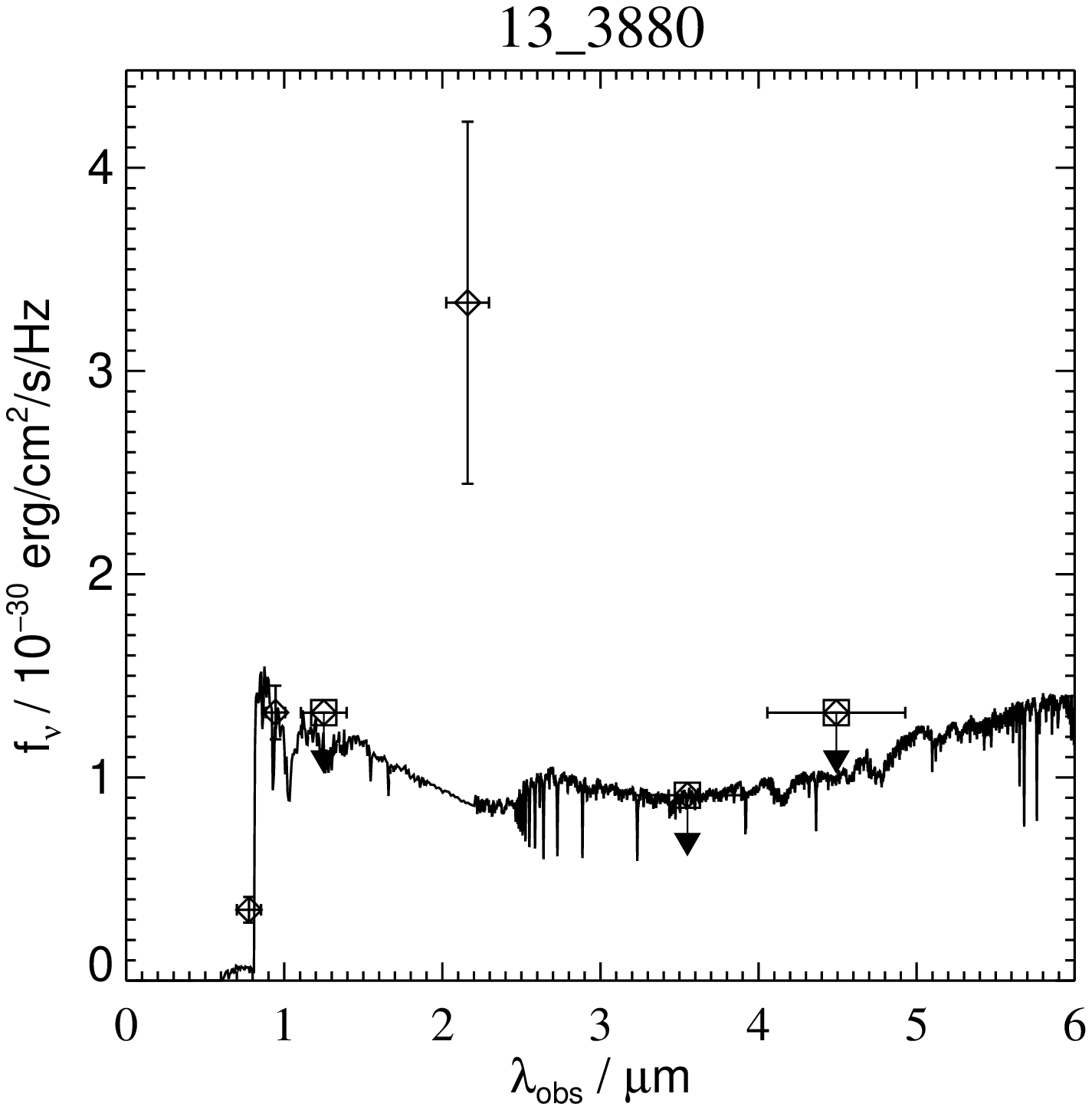}}
\resizebox{0.26\textwidth}{!}{\includegraphics{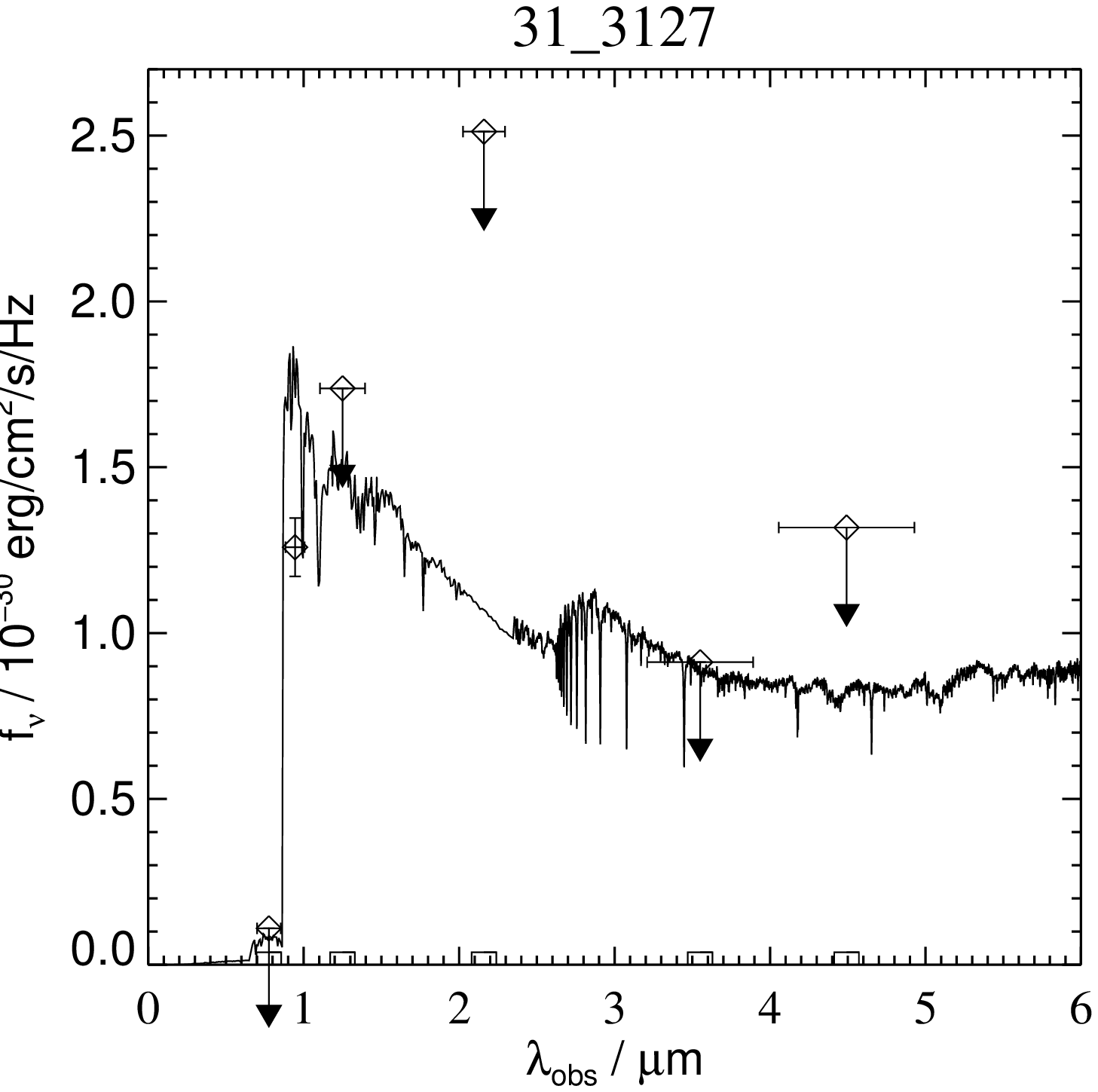}}
\resizebox{0.26\textwidth}{!}{\includegraphics{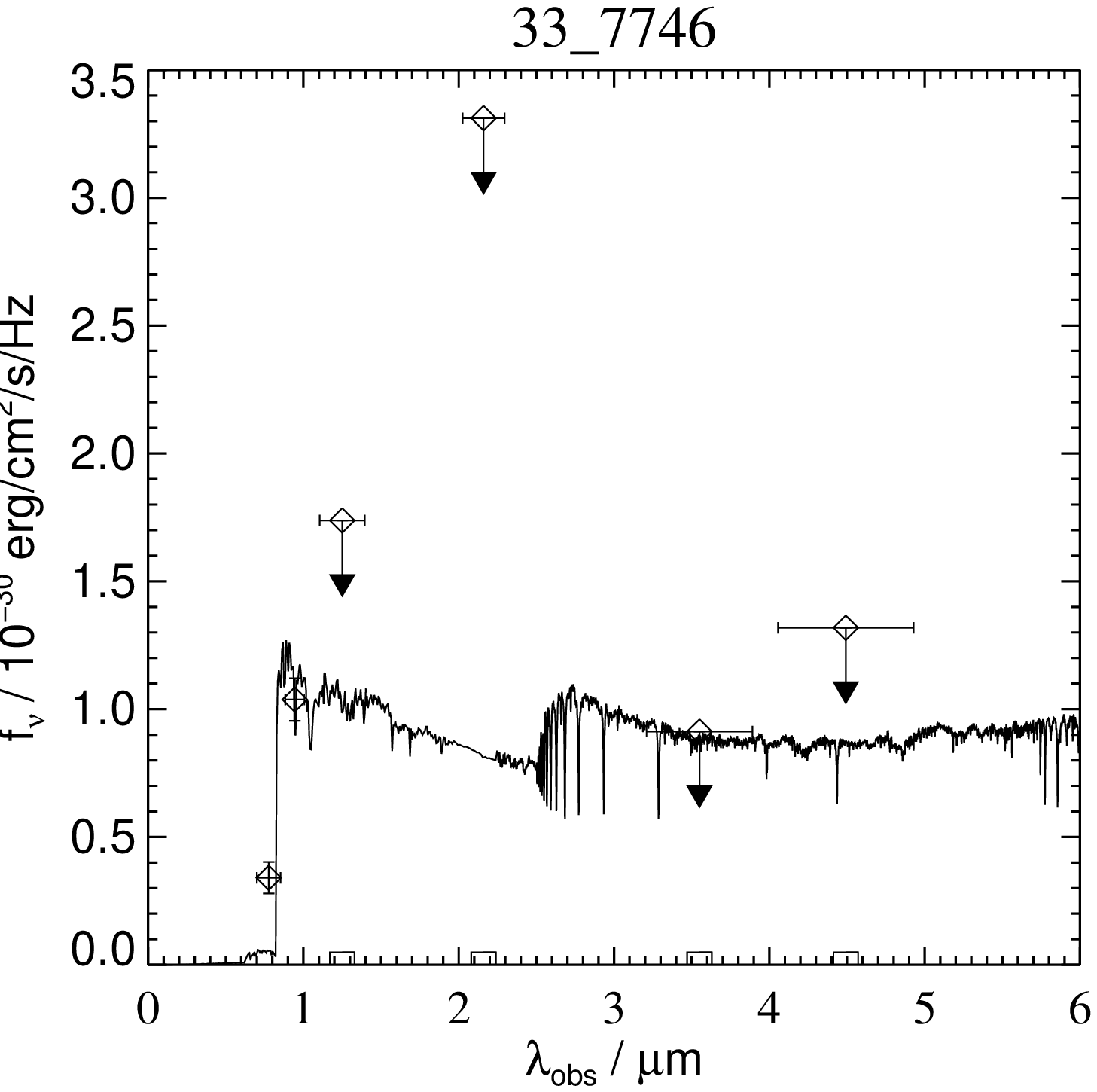}}
\resizebox{0.26\textwidth}{!}{\includegraphics{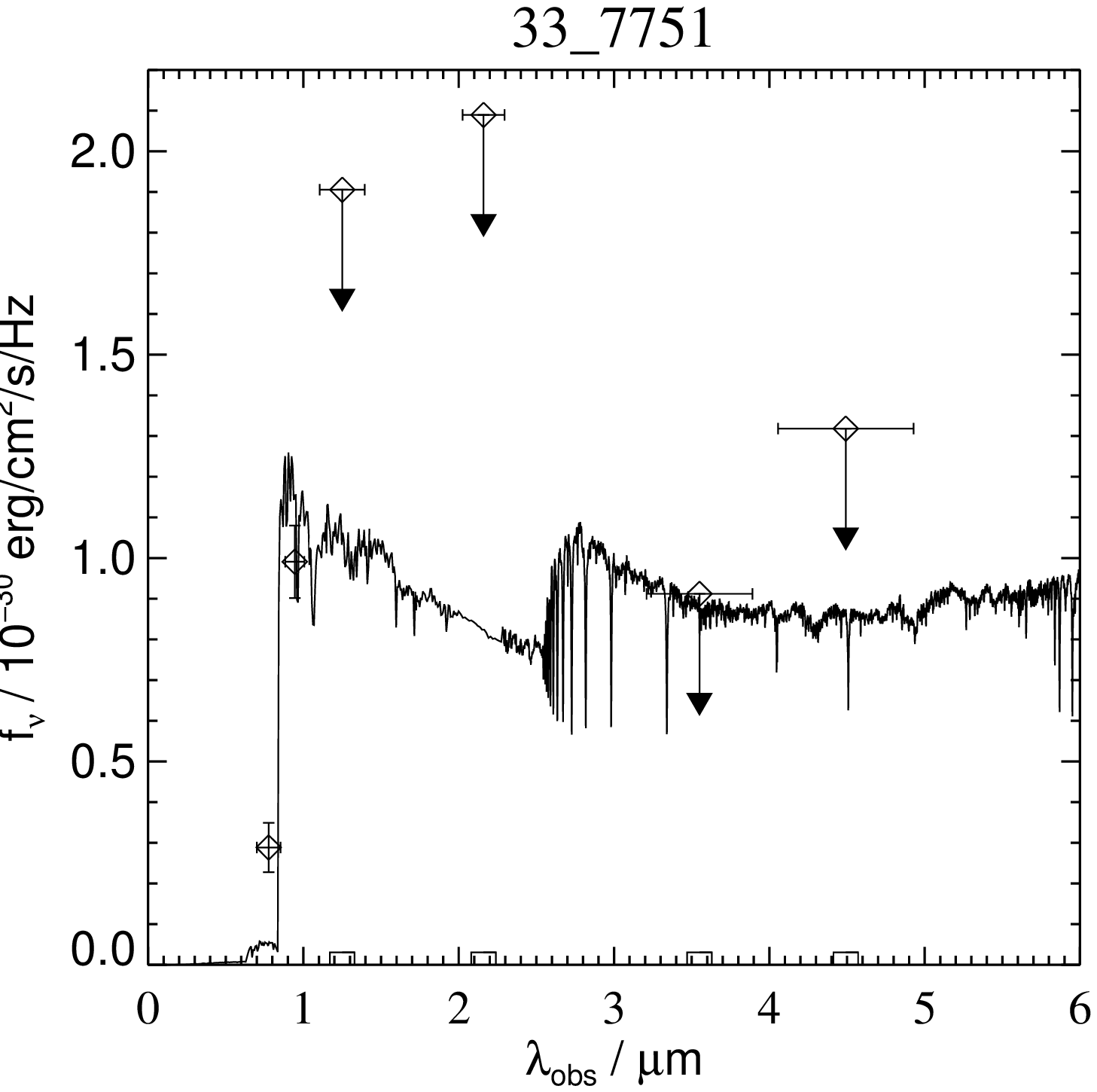}}
\resizebox{0.26\textwidth}{!}{\includegraphics{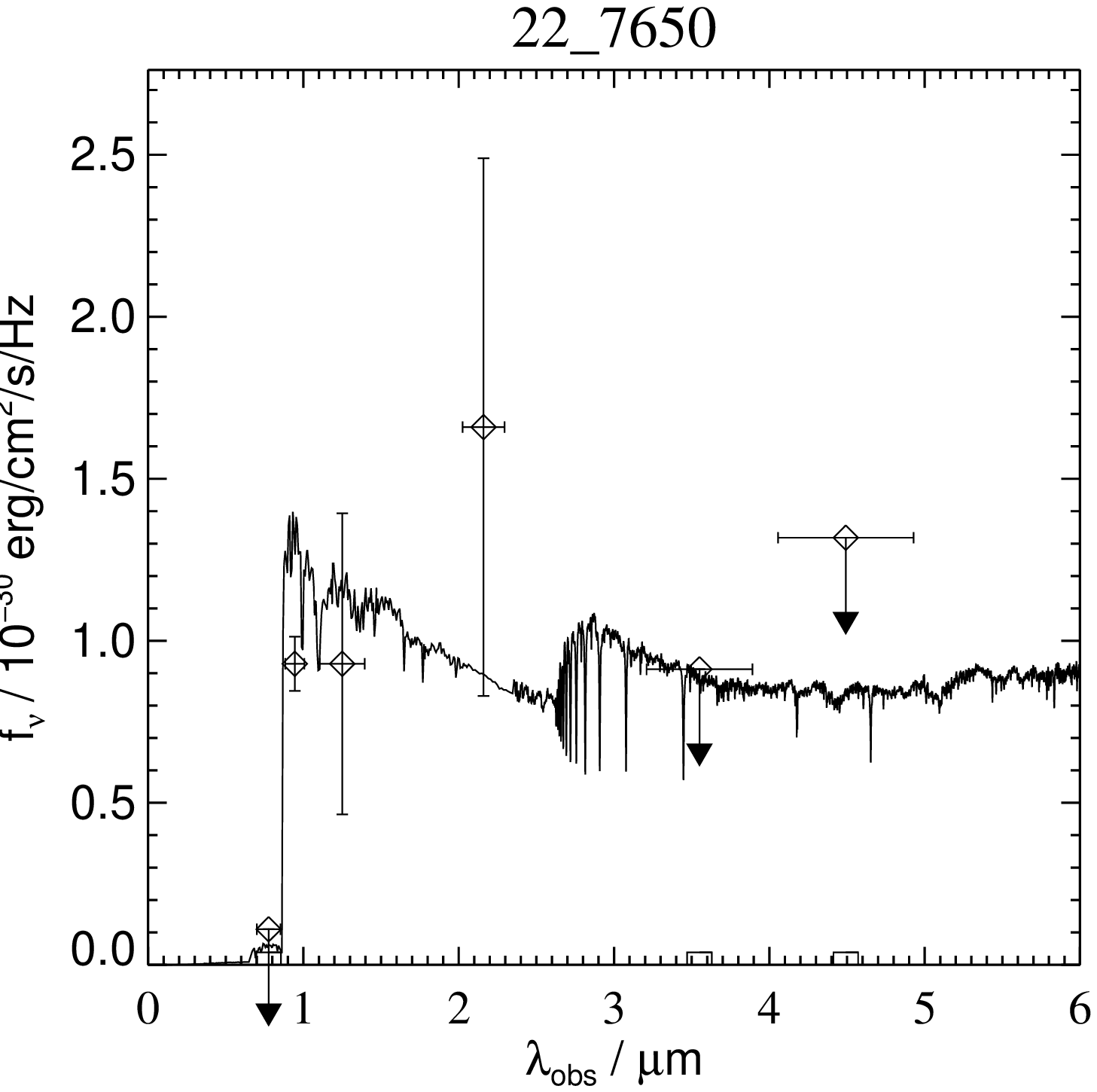}}
\resizebox{0.26\textwidth}{!}{\includegraphics{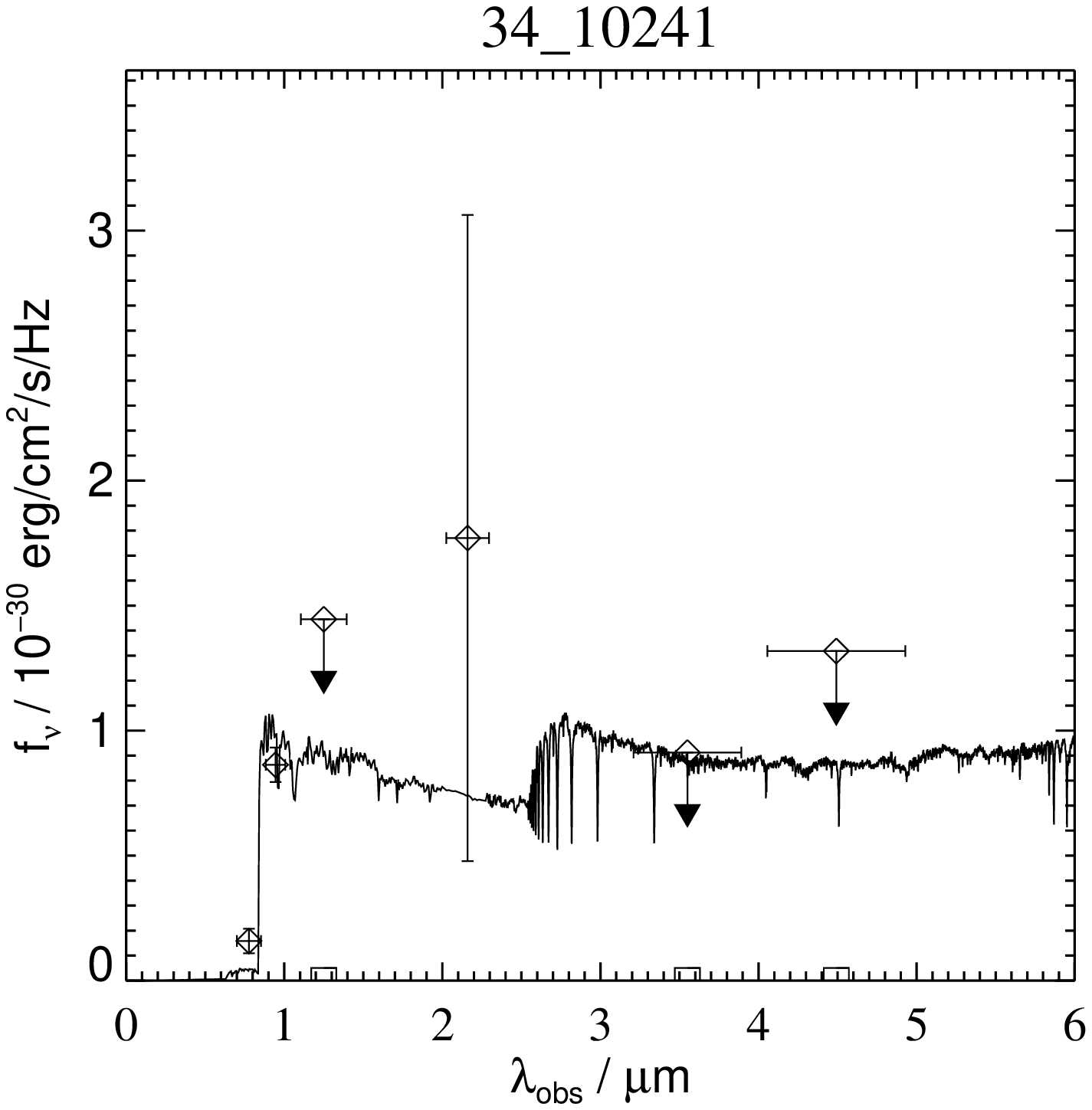}}
\resizebox{0.26\textwidth}{!}{\includegraphics{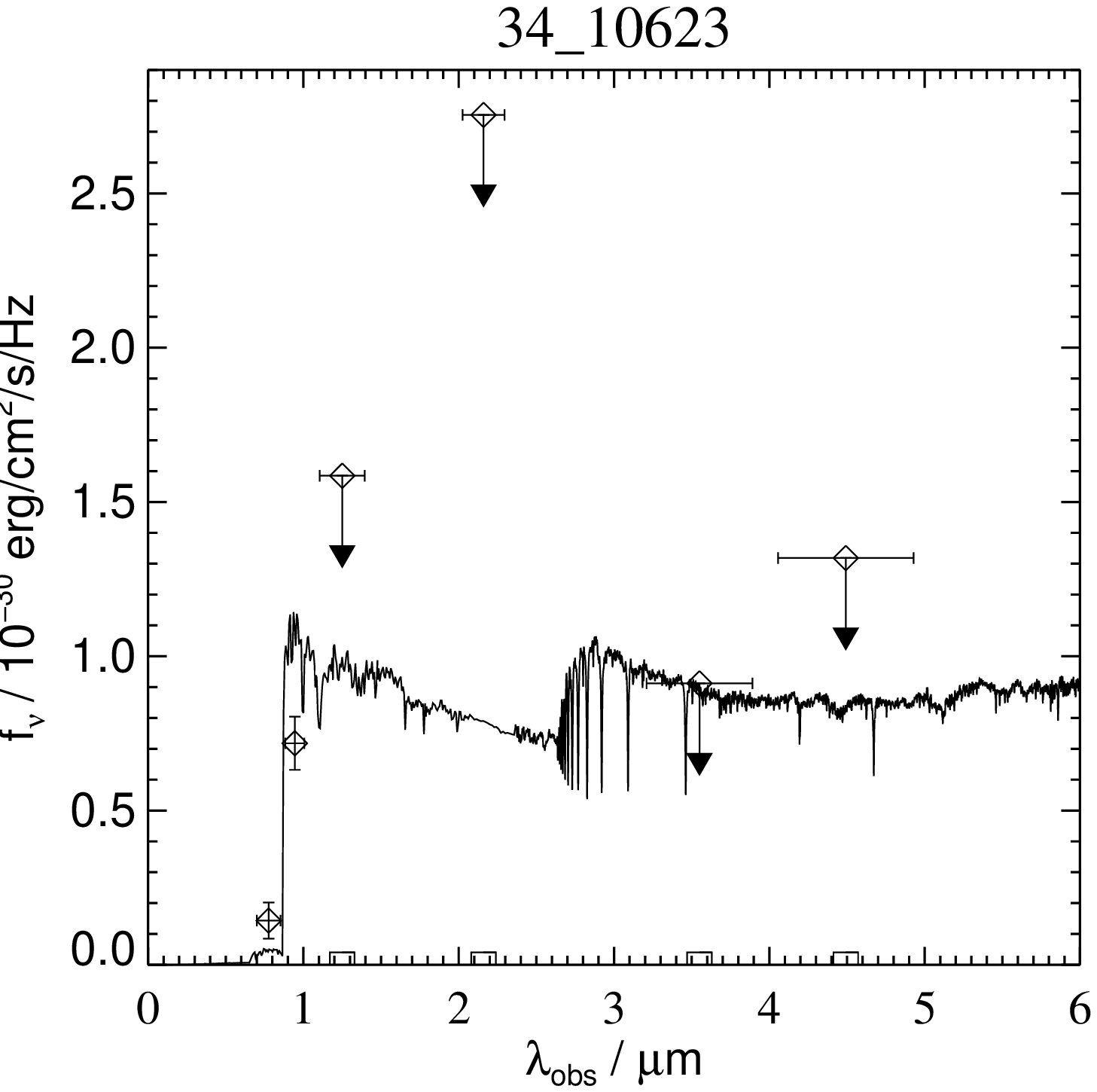}}
\caption{SED fits for the IRAC-undetected $z\sim 6$ sources; the $3.6\,\mu$m upper
limit is the best constraint on the maximum stellar mass, so we force the fit
to go through this. Non-detections are represented by their $3\,\sigma$ upper limits.}
\label{fig:MULTIGALPLOTSundetect}
\end{figure*}

\begin{figure*}
\resizebox{0.26\textwidth}{!}{\includegraphics{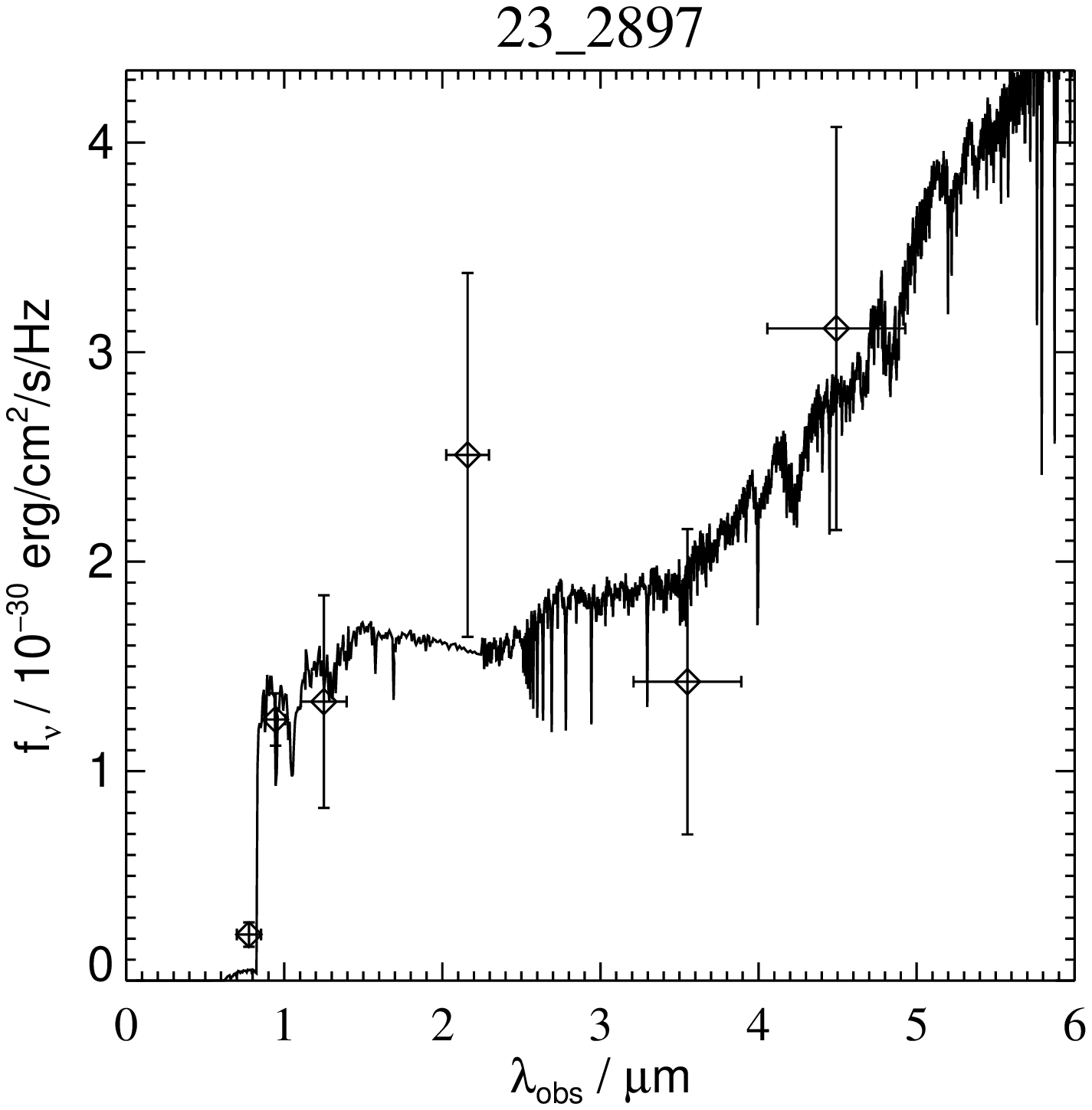}}
\resizebox{0.26\textwidth}{!}{\includegraphics{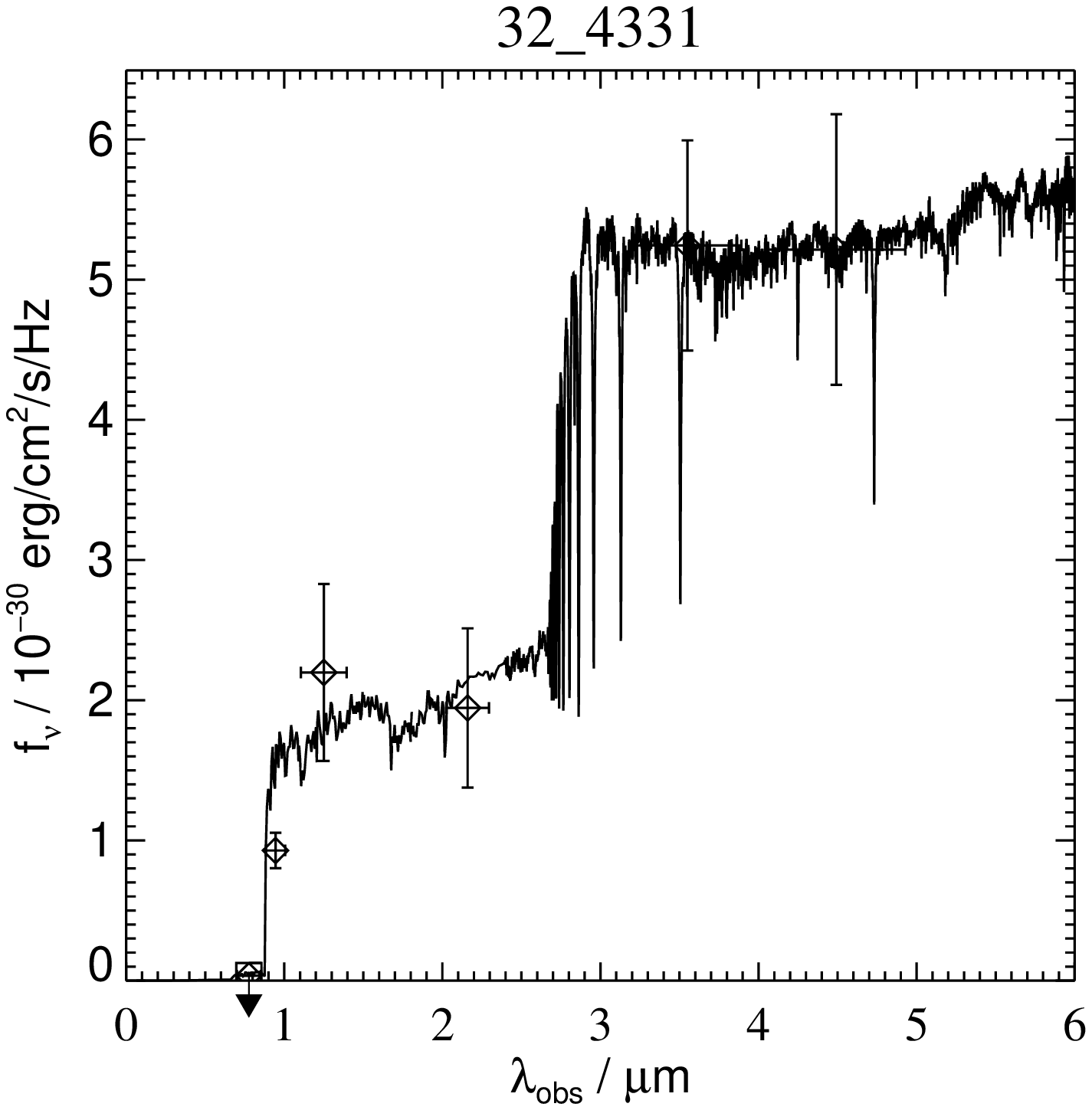}}
\resizebox{0.26\textwidth}{!}{\includegraphics{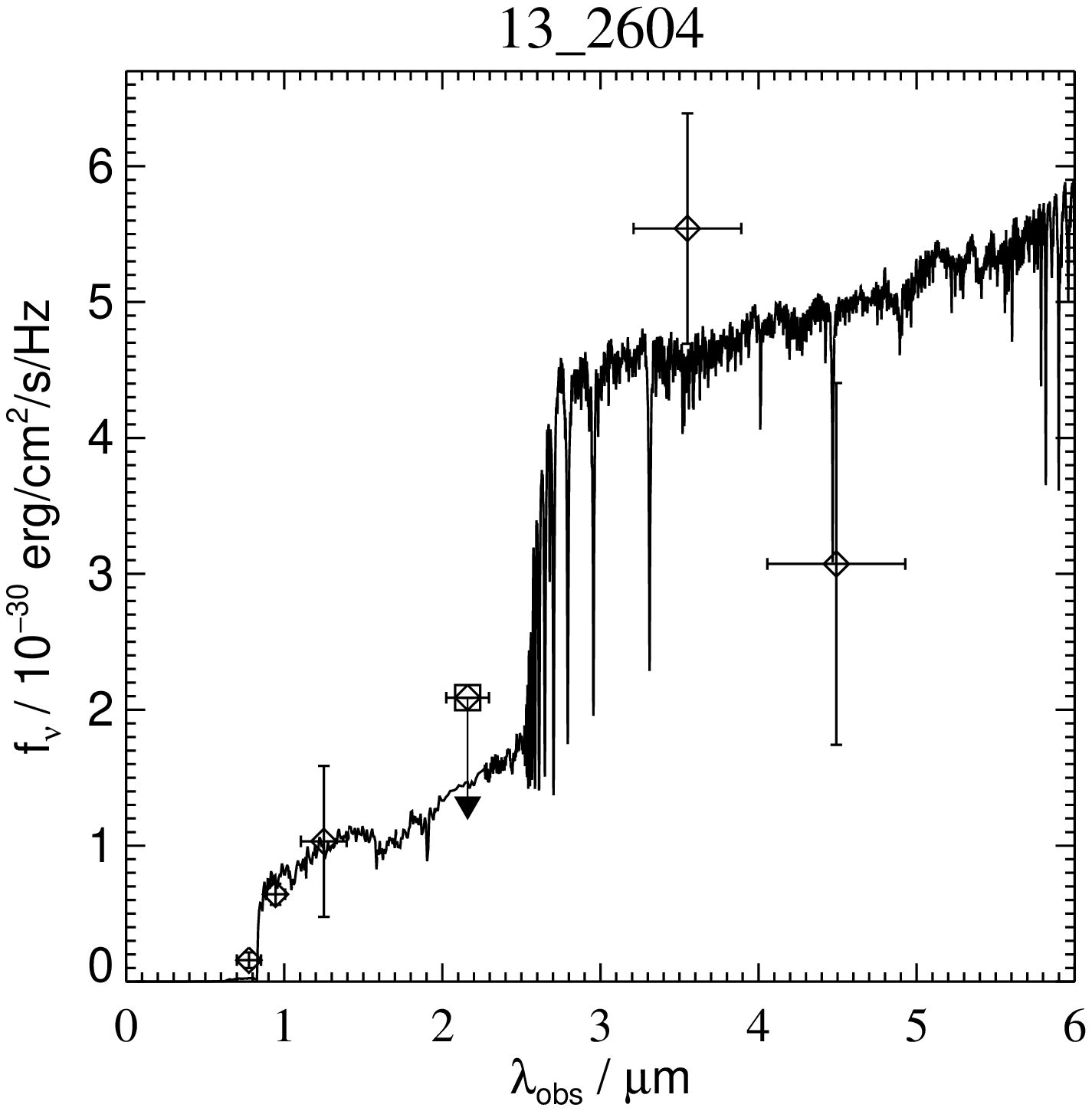}}
\caption{The best-fit SED models, including the effects of dust reddening, to the photometric datapoints for our $z\sim 6$ 
sources. Only those with a non-zero best-fit extinction are shown --
the others are identical to the plots in Figure~\ref{fig:MULTIGALPLOTS}. 
Non-detections are represented by their $3\,\sigma$ upper limits.}
\label{fig:MULTIDUSTPLOTS}
\end{figure*}

\begin{figure*}
\resizebox{0.26\textwidth}{!}{\includegraphics{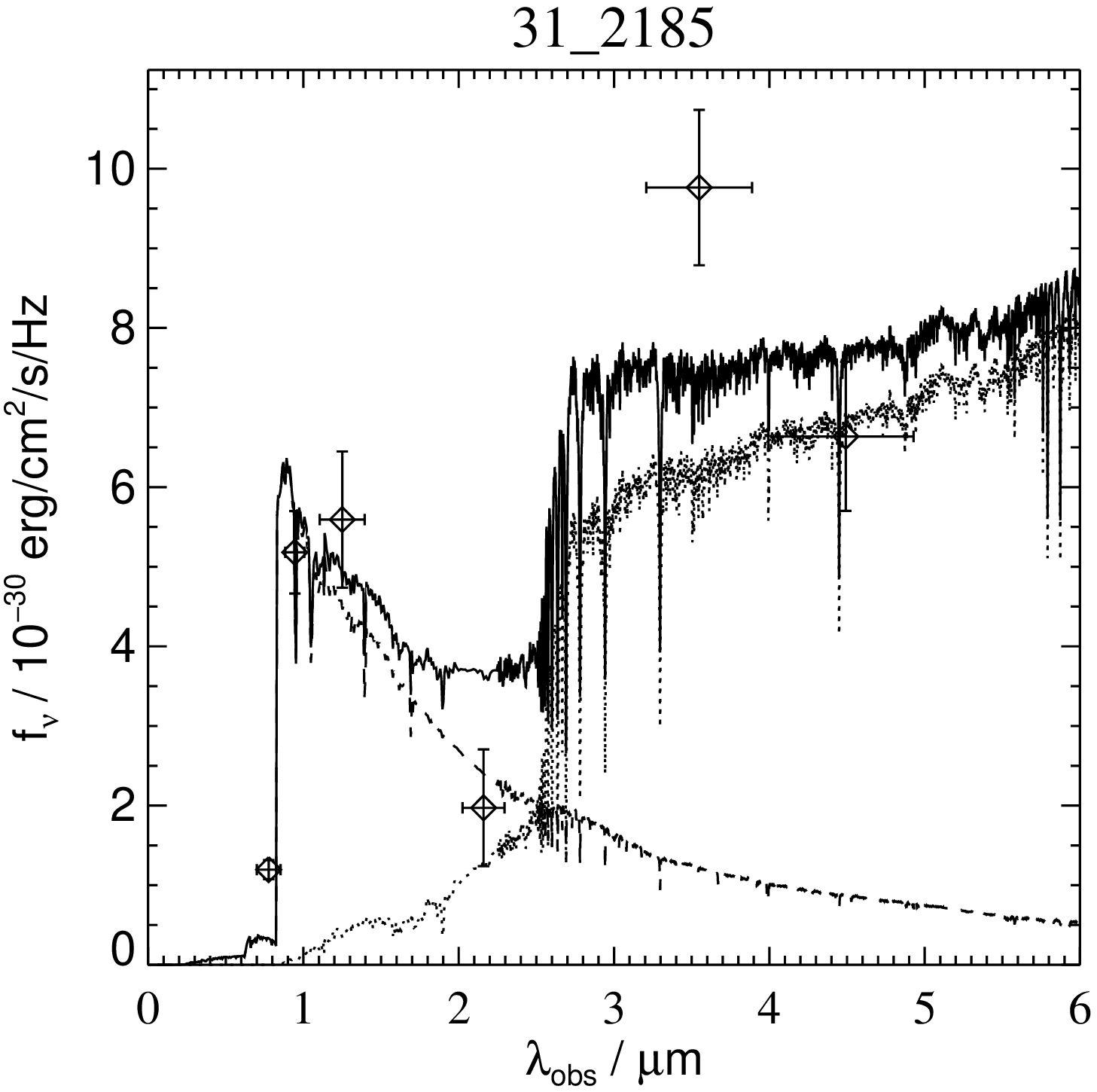}}
\resizebox{0.26\textwidth}{!}{\includegraphics{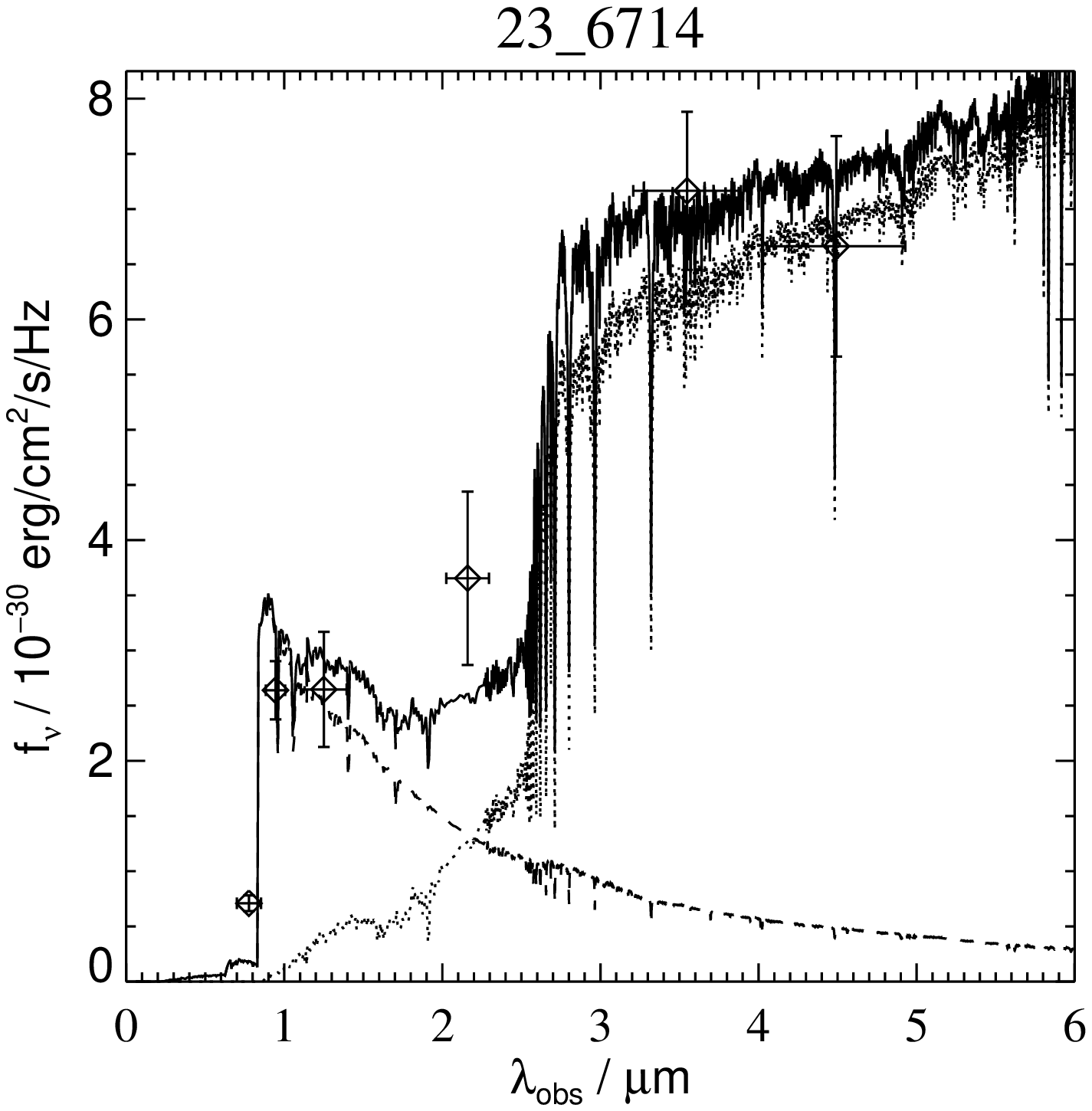}}
\resizebox{0.26\textwidth}{!}{\includegraphics{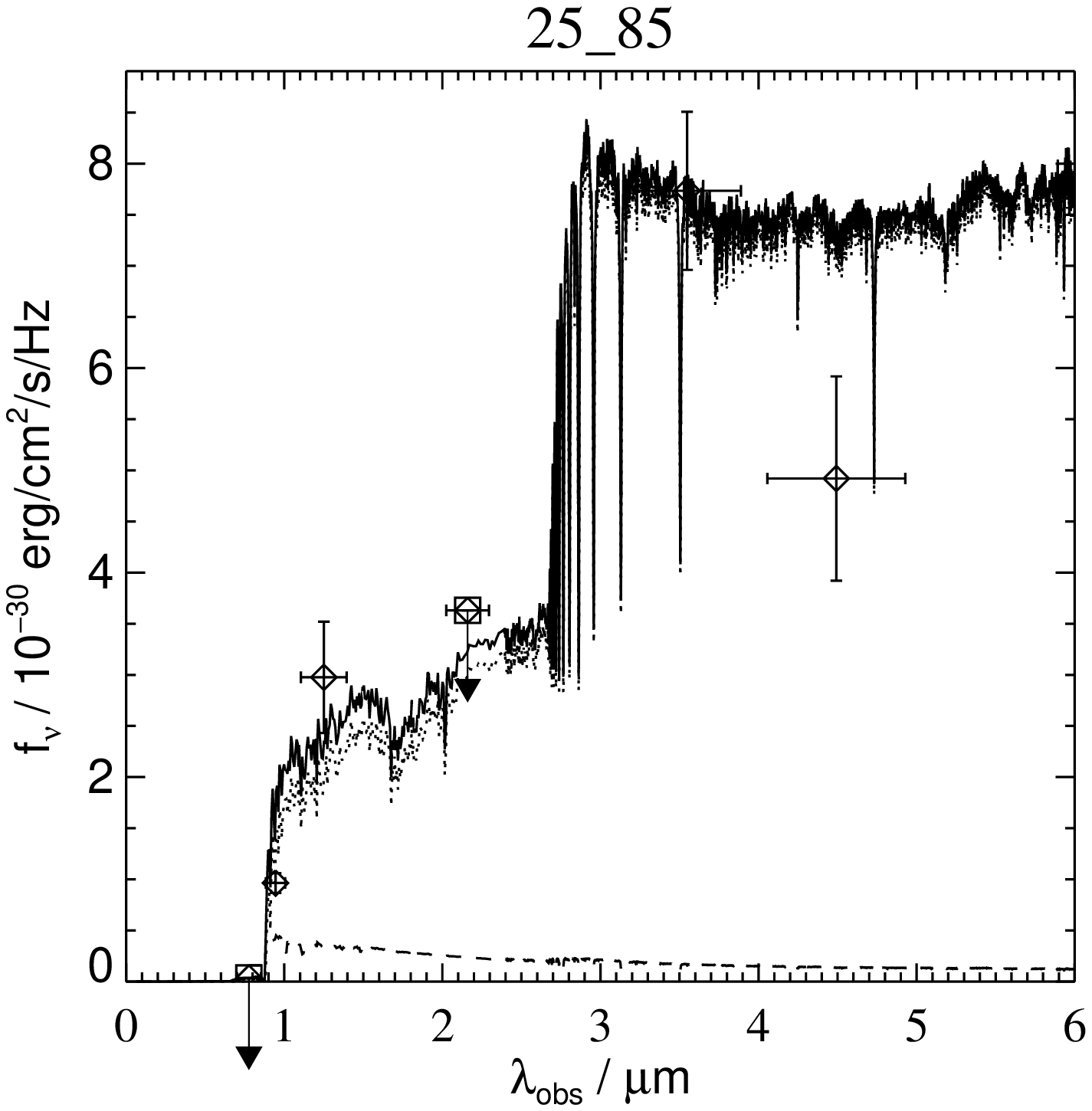}}
\resizebox{0.26\textwidth}{!}{\includegraphics{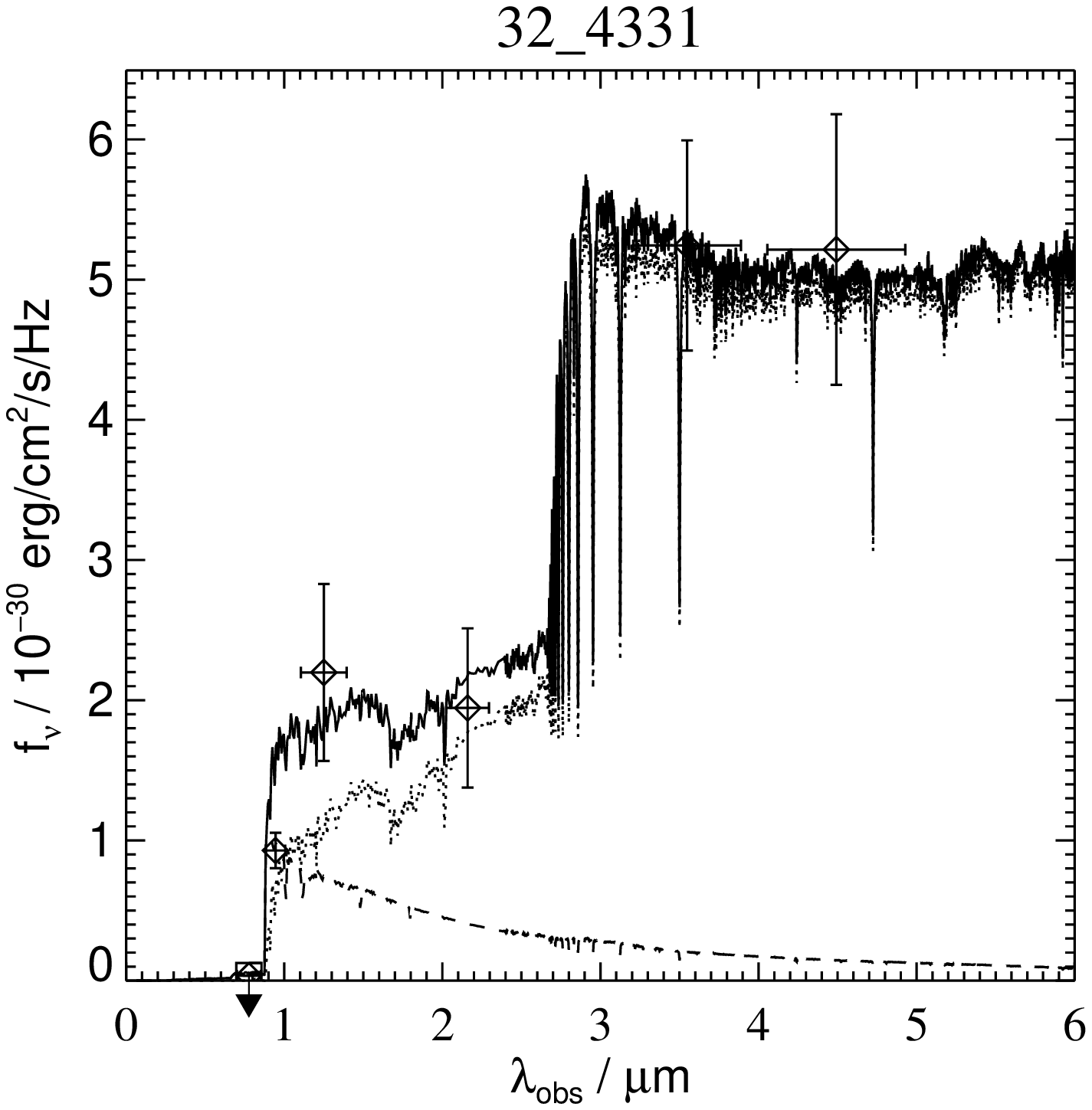}}
\resizebox{0.26\textwidth}{!}{\includegraphics{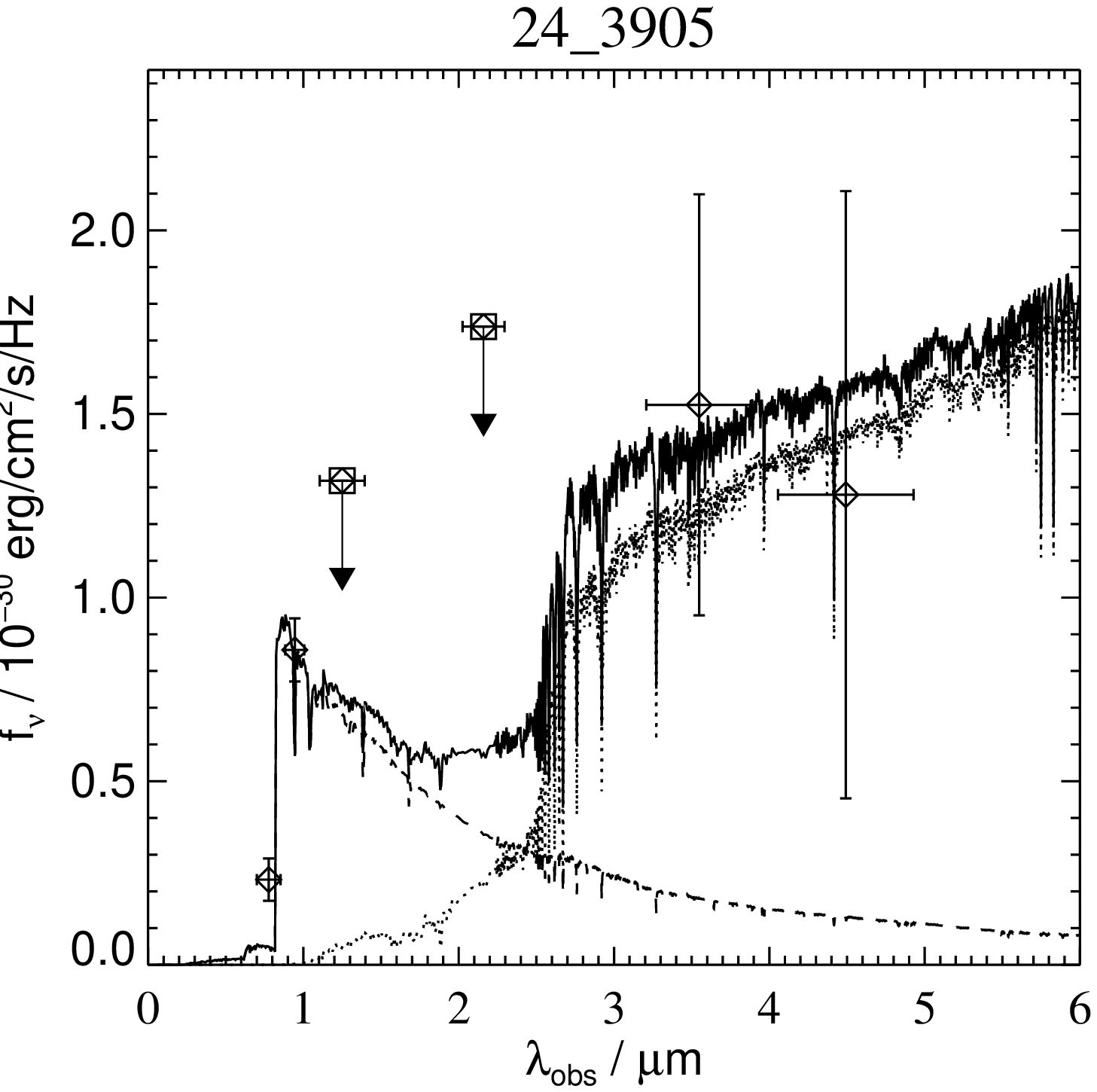}}
\resizebox{0.26\textwidth}{!}{\includegraphics{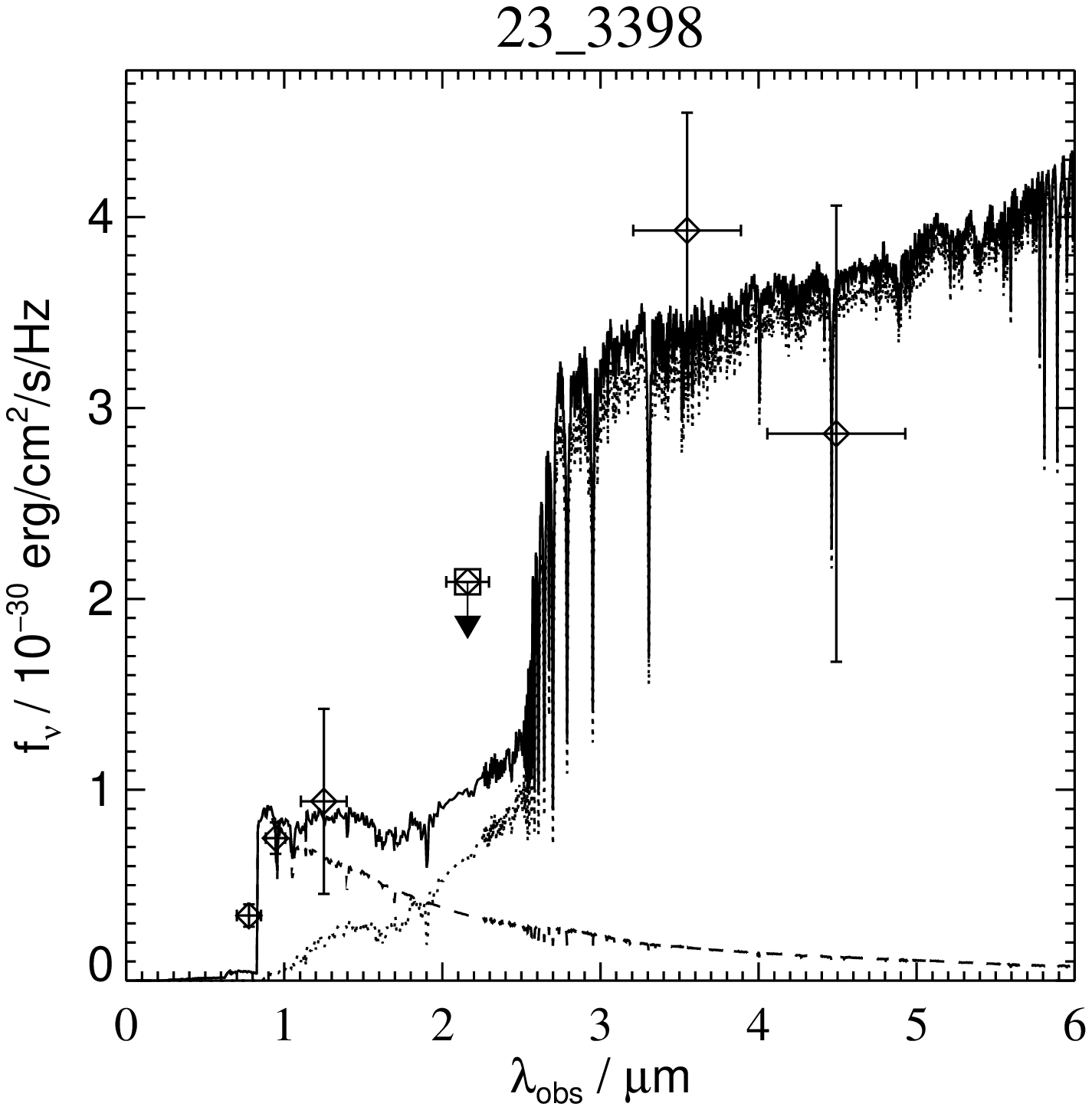}}
\resizebox{0.26\textwidth}{!}{\includegraphics{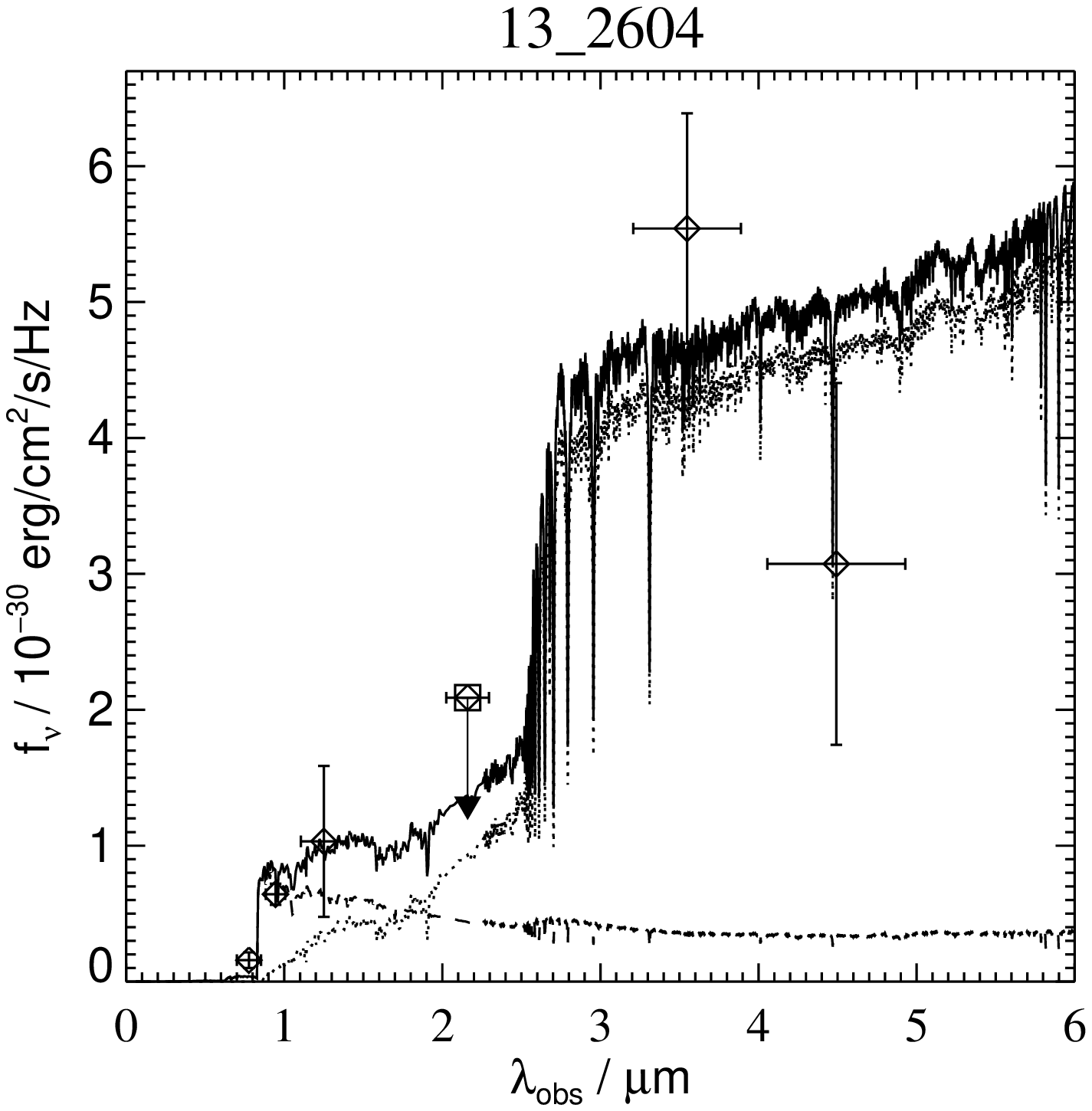}}
\resizebox{0.26\textwidth}{!}{\includegraphics{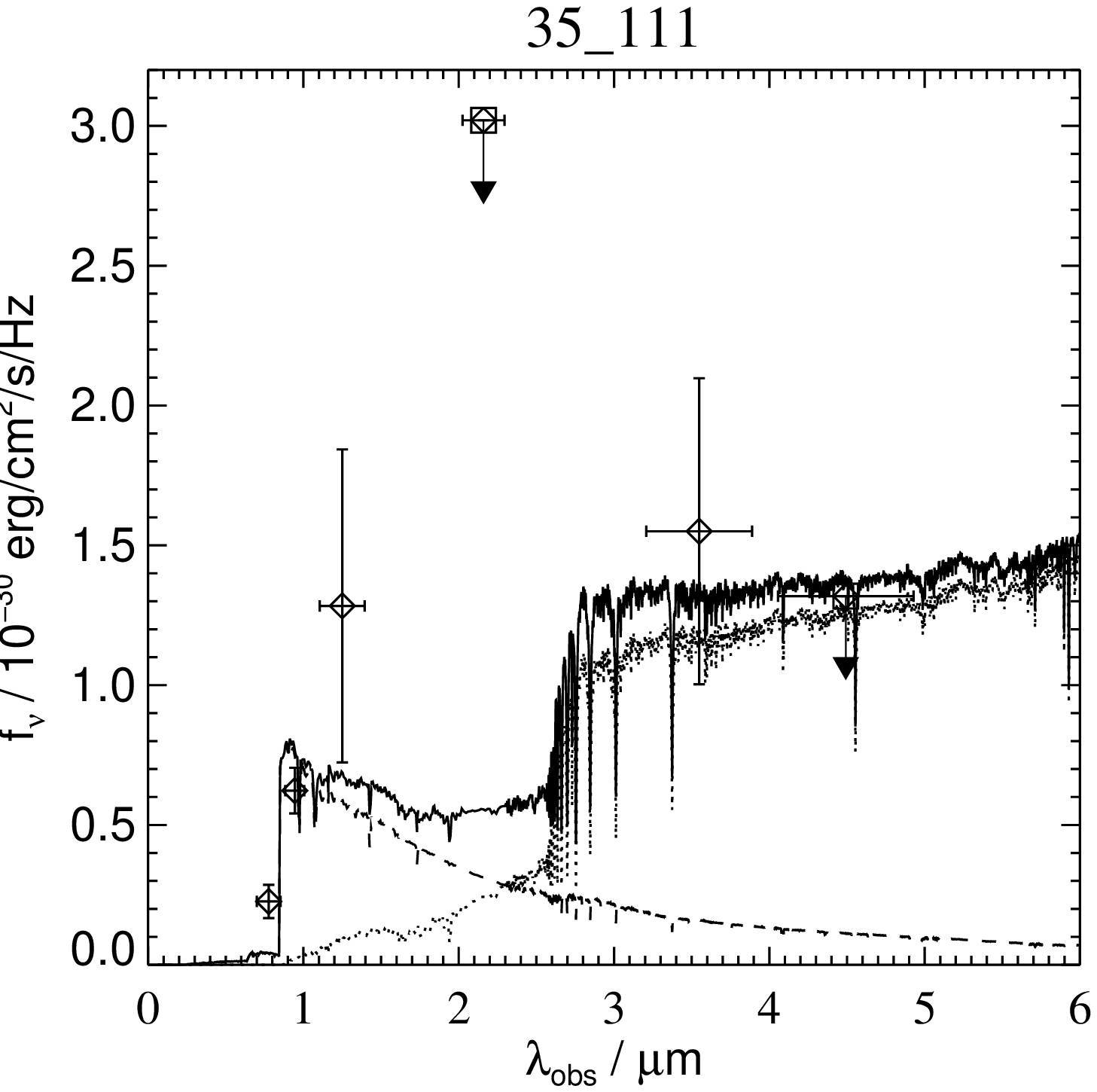}}
\caption{The best-fit two population composite models to the photometric datapoints for our $i'$-drops, showing the young
ongoing starburst (dashed line), an older component (dotted line)
and the resulting total spectrum (solid line).
Only those with non-zero current starburst mass fractions are shown -- that 
of $23\_2897$ is identical to the plot in Figure~\ref{fig:MULTIGALPLOTS}. 
Non-detections are represented by their $3\,\sigma$ upper limits.
In the case of $25\_85$, the
anomalously faint IRAC channel 2 (4.5\,$\mu$m) magnitude
has been excluded from the fit.}
\label{fig:MULTITWOPOPPLOTS}
\end{figure*}

\begin{figure}
\resizebox{0.48\textwidth}{!}{\includegraphics{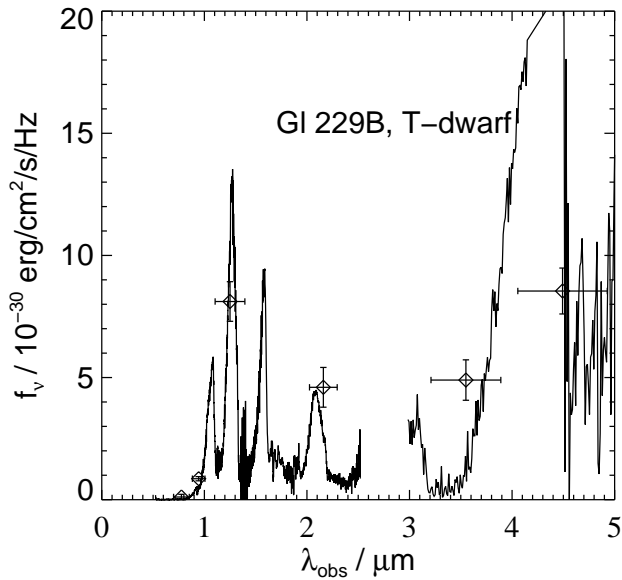}}
\caption{The photometry of object $33\_12465$ overplotted on the spectrum of 
T7 Dwarf Gl 229B.  The peculiar magnitudes for this source, measured in the 
$J$, $K_{s}$, 3.6\,$\mu$m \& 4.5\,$\mu$m filters, may well be explained by the 
presence of several substantial spectral features, if it is 
indeed a T-dwarf object.}
\label{fig:TDWARF}
\end{figure}

The normalisations (scalings) for the models that produced the best
fits to the broadband photometry were returned by the fitting code,
and these were then used to calculate the corresponding best-fit total
masses (see Section~\ref{sec:DETECTIONS}), using the luminosity distance
for the redshift of each $i'$-drop.  For each fit, the number of
degrees of freedom was taken to be the number of independent data
points (i.e., magnitudes in different filters).  When considering
models other than a SSP (instantaneous burst), it was necessary to
correct the total `mass' values output by the fitting routine.  For a
constant SFR model, each of these masses needed to be multiplied by
the corresponding best-fit age, since the B\&C template normalization
has the mass grow by $1\,M_{\odot}\,{\rm yr}^{-1}$.  For the $\tau$
models, the returned `mass' values were corrected by a factor
$(1-e^{-t/\tau})$, accounting for the decay timescale and the
normalization of the B\&C models (where $M\rightarrow 1\,M_{\odot}$
as $t\rightarrow \infty$).  The fits to the B\&C models returned
`total masses' which were the sum of the mass currently in stars, in
stellar remnants, and in gas returned to the interstellar medium (ISM)
by evolved stars.  For each best-fit model, we also calculated the
mass currently in stars for every galaxy, again using information from
the B\&C population synthesis code.
The results of the SED fitting for the nine galaxies with IRAC detections are presented in
Tables\,\ref{tab:MULTIGALTABLE},\,\ref{tab:MULTIDUSTTABLE}\,\&\,\ref{tab:MULTITWOPOPTABLE},
where we list both the `stellar masses' and `total masses' (stars,
recycled gas and stellar remnants) for each galaxy.  The best-fit stellar populations
are shown in Figure~\ref{fig:MULTIGALPLOTS} (for simple star formation histories), Figure~\ref{fig:MULTIDUSTPLOTS} (incorporating dust reddening) and
Figure~\ref{fig:MULTITWOPOPPLOTS}  (for composite two-component models).

\begin{table*}
\centering
\begin{tabular}{l|c|c|c|c|c|c|c|c|c|}
ID & Model & $\chi^{2}$ & Age & Total mass & Stellar mass & Current SFR & SSFR\\
& & & /Myr & /$M_{\odot}$ & /$M_{\odot}$ & /$M_{\odot}\,{\rm yr}^{-1}$ & /$10^{-10}\,{\rm yr}^{-1}$\\

\hline\hline
$31\_2185$ & const.\ SFR & $5.05$ & $640$ & $2.2\times 10^{10}$ & $1.8\times 10^{10}$ & $34$ & $19$ \\
$23\_6714$ & $\tau=500$\,Myr & $4.29$ & $720$ & $3.0\times 10^{10}$ & $2.4\times 10^{10}$ & $19$ & $8.0$ \\
$23\_2897$ & $\tau=1000$\,Myr & $0.73$ & $640$ & $7.1\times 10^{9}$ & $5.7\times 10^{9}$ & $7.9$ & $14$ \\
$25\_85^{\diamond}$ & $\tau=30$\,Myr & $0.82$ & $180$ & $2.4\times 10^{10}$ & $2.0\times 10^{10}$ & $1.9$ & $0.99$ \\
$32\_4331$ & $\tau=70$\,Myr & $0.15$ & $260$ & $1.6\times 10^{10}$ & $1.3\times 10^{10}$ & $6.3$ & $4.7$ \\
$24\_3905$ & $\tau=1000$\,Myr & $0.51$ & $640$ & $4.7\times 10^{9}$ & $3.8\times 10^{9}$ & $5.3$ & $14$ \\
$23\_3398$ & $\tau=100$\,Myr & $3.30$ & $360$ & $1.3\times 10^{10}$ & $1.0\times 10^{10}$ & $3.5$ & $3.5$ \\
$13\_2604$ & $\tau=100$\,Myr & $1.18$ & $450$ & $2.4\times 10^{10}$ & $1.9\times 10^{10}$ & $2.6$ & $1.4$ \\
$35\_111$ & $\tau=1000$\,Myr & $2.00$ & $720$ & $5.0\times 10^{9}$ & $4.0\times 10^{9}$ & $4.8$ & $12$ \\
\hline
$13\_3880$ & inst.\ burst & $0.03$ & $11$ & $4.2\times 10^{8}$ & $3.9\times 10^{8}$ & $0.0$ & $0.0$ \\
weak sources & const.\ SFR & $0.5$ & 57 & $5.4\times 10^8$ & $4.9\times 10^{8}$ & $9.5$ & $193$ \\
\end{tabular}

$^{\diamond}$\,The anomalously-faint IRAC channel 2 (4.5\,$\mu$m) was excluded from the fit.\\

\caption{Tabulated are the best-fit results from the SED fitting, for each of our IRAC-detected $i'$-drops, without consideration of the effects of dust reddening. The ``weak sources'' are the objects which
are individually undetected by IRAC; the derived average properties come from the stacking analysis described
in Section~\ref{sec:NONDETECTIONS}. The specific star formation
rate (SSFR) is a measure of the fraction of the total stellar mass currently being born as stars -- see Section~\ref{sec:SFH}.} 
\label{tab:MULTIGALTABLE}
\end{table*}

\begin{table*}
\centering
\begin{tabular}{l|c|c|c|c|c|c|c|c|c|c|}
ID & Model & $\chi^{2}$ & Age & Total mass & Stellar mass & $E(B-V)$ & Current SFR & SSFR\\
& & & /Myr & /$M_{\odot}$ & /$M_{\odot}$ & /mag & /$M_{\odot}\,{\rm yr}^{-1}$ & /$10^{-10}\,{\rm yr}^{-1}$\\
\hline\hline
$31\_2185$ & const.\ SFR & $5.05$ & $640$ & $2.2\times 10^{10}$ & $1.8\times 10^{10}$ & $0.00$ & $34$ & $19$ \\
$23\_6714$ & $\tau=500$\,Myr & $4.29$ & $720$ & $3.0\times 10^{10}$ & $2.4\times 10^{10}$ & $0.00$ & $19$ & $8.0$ \\
$23\_2897$ & inst.\ burst & $0.40$ & $8.7$ & $1.8\times 10^{9}$ & $1.7\times 10^{9}$ & $0.16$ & $0.0$ & $0.0$ \\
$25\_85^{\diamond}$ & $\tau=30$\,Myr & $0.82$ & $180$ & $2.4\times 10^{10}$ & $2.0\times 10^{10}$ & $0.00$ & $1.9$ & $0.99$ \\
$32\_4331$ & $\tau=100$\,Myr & $0.12$ & $320$ & $2.0\times 10^{10}$ & $1.6\times 10^{10}$ & $0.01$ & $8.6$ & $5.2$ \\
$24\_3905$ & $\tau=1000$\,Myr & $0.51$ & $640$ & $4.7\times 10^{9}$ & $3.8\times 10^{9}$ & $0.00$ & $5.3$ & $14$ \\
$23\_3398$ & $\tau=100$\,Myr & $3.30$ & $360$ & $1.3\times 10^{10}$ & $1.0\times 10^{10}$ & $0.00$ & $3.5$ & $3.5$ \\
$13\_2604$ & $\tau=100$\,Myr & $1.12$ & $400$ & $2.3\times 10^{10}$ & $1.8\times 10^{10}$ & $0.03$ & $4.0$ & $2.2$ \\
$35\_111$ & $\tau=1000$\,Myr & $2.00$ & $720$ & $5.0\times 10^{9}$ & $4.0\times 10^{9}$ & $0.00$ & $4.8$ & $12$ \\
\end{tabular}

$^{\diamond}$\,The anomalously-faint IRAC channel 2 (4.5\,$\mu$m) was excluded from the fit.\\

\caption{Presented in this table are the best-fit results from the SED modelling for each IRAC-detected $i'$-drop, including the effects of dust reddening.  The specific star formation
rate (SSFR) is a measure of the fraction of the total stellar mass currently being born as stars -- see Section~\ref{sec:SFH}.} 
\label{tab:MULTIDUSTTABLE}
\end{table*}

\begin{table*}
\centering
\begin{tabular}{l|c|c|c|c|c|c|c|c|c|c|}
ID & Model & $\chi^{2}$ & Age & Total mass & Stellar mass & burst fraction & Current SFR & SSFR\\
& & & /Myr & /$M_{\odot}$ & /$M_{\odot}$ &  & /$M_{\odot}\,{\rm yr}^{-1}$ & /$10^{-10}\,{\rm yr}^{-1}$\\
\hline\hline
$31\_2185$ & 3\,Myr burst & $2.73$ & $450$ & $3.8\times 10^{10}$ & $3.0\times 10^{10}$ & $0.014$ & $140$ & $47$ \\
$23\_6714$ & 3\,Myr burst & $2.97$ & $450$ & $3.9\times 10^{10}$ & $3.1\times 10^{10}$ & $0.007$ & $78$ & $25$ \\
$23\_2897^{\dagger}$ & 3\,Myr burst & $0.73$ & $40$ & $2.5\times 10^{9}$ & $2.2\times 10^{9}$ & $0.000$ & $0.0$ & $0.0$ \\
$25\_85^{\diamond}$ & 10\,Myr burst & $0.80$ & $140$ & $2.3\times 10^{10}$ & $1.9\times 10^{10}$ & $0.002$ & $4.6$ & $2.4$ \\
$32\_4331$ & 3\,Myr burst & $0.10$ & $180$ & $1.8\times 10^{10}$ & $1.5\times 10^{10}$ & $0.005$ & $24$ & $16$ \\
$24\_3905$ & 3\,Myr burst & $0.31$ & $570$ & $8.9\times 10^{9}$ & $6.9\times 10^{9}$ & $0.009$ & $21$ & $30$ \\
$23\_3398$ & 3\,Myr burst & $2.45$ & $450$ & $2.0\times 10^{10}$ & $1.6\times 10^{10}$ & $0.004$ & $20$ & $13$ \\
$13\_2604$ & 30\,Myr burst & $0.95$ & $400$ & $2.4\times 10^{10}$ & $1.9\times 10^{10}$ & $0.007$ & $4.9$ & $2.5$ \\
$35\_111$ & 3\,Myr burst & $1.81$ & $400$ & $6.9\times 10^{9}$ & $5.4\times 10^{9}$ & $0.010$ & $18$ & $34$ \\
\end{tabular}

$^{\diamond}$\,The anomalously-faint IRAC channel 2 (4.5\,$\mu$m) was excluded from the fit.\\

$^{\dagger}$\,This object was found to produce degenerate results, as each best fit model infers none of its stellar mass is involved in an ongoing (current) starburst. Hence the two population SED model is non-applicable, and so for each burst duration, the same results are produced.
\caption{The results corresponding to the best-fit two population composite models of each IRAC-detected $i'$-drop.  The burst fraction denotes the fraction of the stellar mass that is involved in the current starburst occurring within each galaxy. The specific star formation
rate (SSFR) is a measure of the fraction of the total stellar mass currently being born as stars -- see Section~\ref{sec:SFH}.} 
\label{tab:MULTITWOPOPTABLE}
\end{table*}

Upon comparison of the results for objects $23\_6714$ \& $31\_2185$
(SBM03\#1 \& \#3 respectively) with those of our previous analysis (Eyles et al.\ 2005), 
we find that the derived stellar ages and masses are broadly consistent; variations
are primarily due to differences in the photometry used, as discussed
in Section~\ref{sec:PHOT}.  

\section{Analysis}
\label{sec:ANALYSIS}

\subsection{Properties of IRAC-detected $i'$-drop Galaxies -- 
Balmer/4000\,\AA\ Breaks}
\label{sec:DETECTIONS}

The Balmer/4000\,\AA\ spectral break is sensitive to
the luminosity-weighted stellar age\footnote{That is,
the average age of the stars in the galaxy with each
star's contribution weighted by its luminosity.} of a galaxy;
the presence of a substantial break
indicates that the observed system is being viewed some time after an
epoch of substantial star formation.  The Balmer break, at a rest
wavelength of 3648\,\AA , is most pronounced when the main-sequence
turn-off has reached the A-star population (after a few hundred Myr),
whilst the 4000\,\AA\ break is due to metal line blanketting in
late-type stars.  Our SED fitting process is sensitive to the
Balmer/4000\,\AA\  break,
which falls between the $K_s$ and IRAC-3.6\,$\mu$m filters at $z\sim
6$. For our nine galaxies with significant detections in the
{\em Spitzer}/IRAC imaging, we present the results of our population
synthesis fitting to the SEDs for a range of star formation histories
in
Tables\,\ref{tab:MULTIGALTABLE},\,\ref{tab:MULTIDUSTTABLE}\,\&\,\ref{tab:MULTITWOPOPTABLE}.
We ignore those models which returned best-fit ages $\ge 1$\,Gyr, as
these exceed the age of the Universe at $z\sim 6$ and hence are
unphysical.

We find evidence for the presence of substantial Balmer/4000\,\AA\ breaks in 
the SEDs of six of our galaxies (31$\_$2185, 23$\_$6714, 25$\_$85, 32$\_$4331, 23$\_$3398 \& 13$\_$2604), each brightening, on average, by a factor of $\sim 2 - 3$ in
flux density ($f_{\nu}$) from the near-infrared ($\approx 0.9-2.2\,\mu$m) to 3.6\,$\mu$m.
These break amplitudes are comparable to those observed at $z\approx
0$ in the Sloan Digital Sky Survey (SDSS; e.g., Kauffmann et al.\
2003), and by Le Borgne et al.\ (2006) in $z\sim 1$ massive
post-starburst galaxies in the Gemini Deep Deep Survey (GDDS).  For
our $i'$-drops it is known that there has been at least some recent or
ongoing star formation, due to the fact that the $i'$-drop Lyman break
technique relies on a magnitude-limited selection in the $z'$-band,
sampling the rest-frame UV (see Section~\ref{sec:MASSDENSITY}) which
is dominated by young, hot, massive stars.  Additionally, for the two
bright sources SBM03\#1\,\&\,\#3 which featured in our previous paper (and which are
re-analysed here), Keck/Gemini spectra (Bunker
et al.\ 2003; Stanway et al.\ 2004a) display Lyman-$\alpha$ emission,
caused by the photoionisation of hydrogen by populations of
short-lived OB stars.  However, in order to produce the observed
Balmer/4000\,\AA\ break amplitudes in the SEDs of these six
$i'$-drops, it is likely that the bulk of their stellar mass formed
well before the current period of star formation. If even a modest
fraction of the mass is involved in current star formation then it
will tend to dilute the break amplitude.

From the fitting of the B\&C spectral synthesis models to our
photometric data, the inferred luminosity-weighted ages of these six
sources all lie in the range $180 - 720$\,Myr, suggesting formation
redshifts in the range $7\le z_{f} \le 18$.  Our SED modelling also
shows that these $i'$-drops with Balmer/4000\,\AA\ breaks have considerable
stellar masses, in the range $1 - 3\times 10^{10}\,M_{\odot}$.  We
can draw comparison of these inferred stellar masses to that of an
$L^{*}$ galaxy today, by taking $L^{*}_{r} = -21.21$ from analysis of
the SDSS by Blanton et al.\ (2003).  Using $M/L_{V}\approx
5\,M_{\odot}/L_{\odot}$ appropriate for a $\approx 10$\,Gyr old
population from the B\&C models (Salpeter IMF) we calculate the stellar mass
of an $L^{*}$ galaxy today to be $M^{*}=1.2\times 10^{11}\,M_{\odot}$,
comparable to the estimate of Cole et al.\ (2001).  Hence we find that
our six IRAC-detected sources with Balmer/4000\,\AA\ breaks have
best-fit stellar masses $\sim 10 - 30$\% that of an $L^{*}$ galaxy
observed today.

Exploring the results from the two population composite modelling
(Table~\ref{tab:MULTITWOPOPTABLE}), we find that for these objects,
only a small fraction (in the range 0.1\% - 1.5\%) of their stellar mass
is involved in the current star formation episode, indicating that in
these cases underlying older stellar populations dominate the mass.
This may require previous epochs of extremely intense star formation
(see Section~\ref{sec:SFH}).  Assuming the sources for which SED
fitting was conducted are representative of the entire $i'$-drop
population, our results suggest that a significant fraction ($\sim
40$\%) of the $i'$-drop galaxies with $z'_{AB}<27$ contain old,
established stellar populations that formed at $z>6$.  Our results
also suggest that in the first Gyr after the Big Bang the number
density of massive galaxies containing evolved stellar populations is
in fact rather high, perhaps contrary to the expectations of some
hierarchical models of galaxy formation. For example,
the stellar mass function presented in Bower et al.\ (2005)
from the GALFORM model
indicates a space density of $1.7\times 10^{-5}\,{\rm Mpc}^{-3}$
for galaxies at $z=6$ with stellar masses $>1.6\times 10^{10}\,M_{\odot}$ (interpolating for $z=6$ between the space density
values at $z=5.3$ of $9.9\times 10^{-5}\,h^{3}\,{\rm Mpc}^{-3}$
and $2.22\times 10^{-6}\,h^{3}\,{\rm Mpc}^{-3}$ at $z=7.88$,
with $h=0.7$ in our adopted cosmology). This would suggest that in the volume of the GOODS-South field there should theoretically be 3 galaxies with
stellar masses $>1.6\times 10^{10}\,M_{\odot}$,
using an effective volume of $1.8\times 10^5\,{\rm Mpc}^3$ for
$i'$-drops at $z\approx 6$ (from Stanway, Bunker \& McMahon 2003, see also Section~\ref{sec:MASSDENSITY}). In our sub-sample of SED fits alone, we have 4 galaxies of this stellar mass; scaling for the incompleteness due to confusion and galaxies outside
the GOODS-MUSIC photometric redshift catalog, we infer
that there should be $\approx 13$ such massive galaxies,
a factor of 4 higher than the GALFORM model.
The recent Yan et al.\ (2006) paper compares the number density of massive
$i'$-drops to other $\Lambda$CDM simulations by Night et al.\ (2006). These simulations
predict a space density for galaxies with stellar masses $>1.6\times 10^{10}\,M_{\odot}$ of
$4-10\times 10^{-5}\,{\rm Mpc}^{-3}$ at $z\approx 6$, a factor of $2-6$ higher space density
than the models in Bower et al.\ (2005), and broadly consistent with our measurements.

We have nine robust detections with IRAC of $i'$-drops, and we have
discussed the six that show probable Balmer breaks.  
The remaining three sources with IRAC detections (23$\_$2897, 24$\_$3905 and
35$\_$111) are fainter and have inconclusive stellar population fits
to the photometry, which are compatible with a wide range of SEDs from
flat spectra in $f_{\nu}$ to a brightening across a spectral break by
up to a factor of $2-3$ in $f_{\nu}$.

In order to assess the uncertainties on our mass values, we use the
method employed in Eyles et al.\ (2005) and also adopted by Stark et
al.\ (2006) for $z\sim 5$ galaxies.  For a particular SED model, we
took the best-fit stellar mass, and allowed it to vary over the range
$(0.1 - 3)\,\times$ M$_{stellar}$, recalculating a reduced $\chi^{2}$
for each variation in order to map the confidence intervals for the
masses and ages returned by our fitting code.  We find typical
uncertainties in these properties of each $i'$-drop to be $\sim 30-50$
percent, and so it is reasonable to assume a 50 percent uncertainty in
these values, and subsequently in the stellar mass density (see
Section~\ref{sec:MASSDENSITY}).

\begin{figure}
\resizebox{0.48\textwidth}{!}{\includegraphics{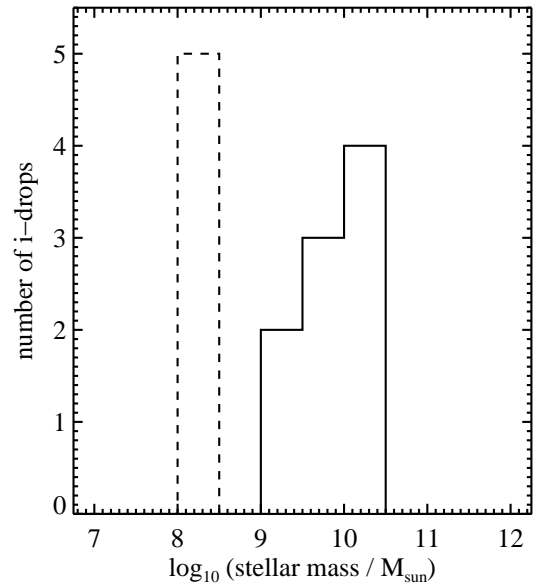}}
\caption{The distribution of stellar masses for our nine IRAC-detected 
$i'$-drops, for the best-fit SEDs without inclusion of dust reddening (solid line).  The masses of our four stacked undetected objects and also $13\_3880$ are represented by the dashed line.}
\label{fig:STELMASSHIST}
\end{figure}

\begin{figure}
\resizebox{0.48\textwidth}{!}{\includegraphics{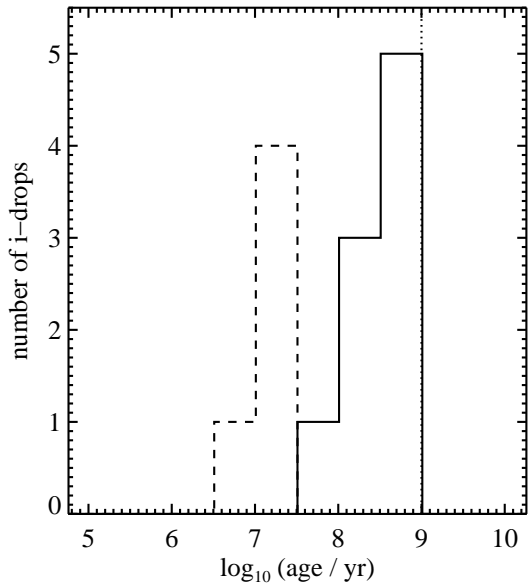}}
\caption{A histogram depicting the distribution of galaxy ages for our nine 
IRAC-detected $i'$-drops (solid line), when considering SEDs without consideration of 
reddening due to intrinsic dust. The ages of our four stacked undetected objects and $13\_3880$ are represented by the dashed line; this should be regarded as an upper limit on the stacked objects ages, as we have taken a constant star formation rate (instantaneous burst and declining star formation rate models yield younger ages).  The vertical dotted line marks the age of the Universe at $z\sim 6$.}
\label{fig:AGESHIST}
\end{figure}

Histograms of the best-fit stellar masses and ages of these nine
IRAC-detected objects are given in
Figures\,\ref{fig:STELMASSHIST}\,\&\,\ref{fig:AGESHIST}, respectively.
In summary, we find that these IRAC-detected
$i'$-drops have ages between $\approx 9 -
720$\,Myr; several galaxies have ages a significant fraction of the
age of the Universe at the $z\sim 6$ epoch (1 Gyr).  Contained within
these $i'$-drops is a significant amount of stellar mass -- adding the
best-fit masses we obtain a total of $\approx 1.4\times 10^{11}M_{\odot}$
present in these nine objects.  The majority of this stellar mass is
contained within older underlying stellar populations which,
considering the inferred ages of six of these $i'$-drops, must have formed 
during earlier vigorous star formation episodes ($z>6$).

\begin{figure}
\resizebox{0.48\textwidth}{!}{\includegraphics{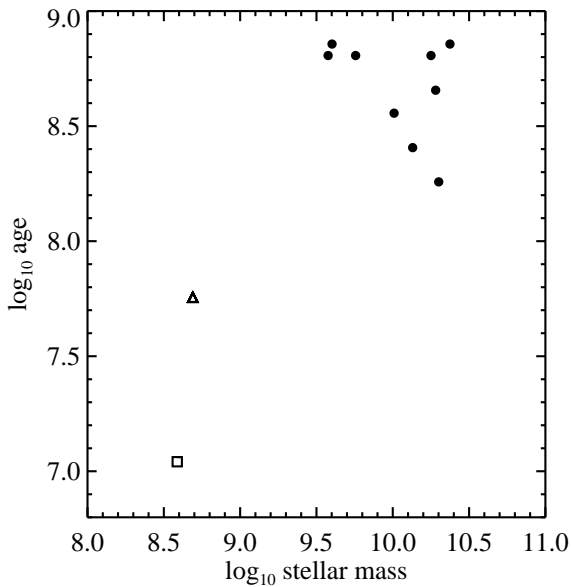}}
\caption{The distribution of stellar masses with ages, as inferred by the SED 
fitting for our nine IRAC-detected sources (circles), with $13\_3880$ (square) and our four stacked undetected galaxies (triangle).  The 
diagonal trend shown here is likely to be a selection effect rather than a 
genuine correlation (see Section~\ref{sec:MASSDENSITY}).}
\label{fig:AGEVMASS}
\end{figure}

\subsection{Properties of IRAC-undetected $i'$-drops}
\label{sec:NONDETECTIONS}

The fitting of photometry to the B\&C models returned unconstrained 
SEDs the seven $i'$-drops which are not detected at IRAC wavelengths 
(although some of these can be seen in the IRAC images, they are
fainter than the $3\,\sigma$ limiting magnitudes).  Hence we are unable
to constrain their properties to the level that is obtainable for
IRAC-detected objects.  However, we can still draw some useful
information from their analysis.  By considering the upper limits on
their ($z' - 3.6\,\mu$m) colours, it is possible to estimate the
maximum possible Balmer/4000\,\AA\ break amplitudes that may be
present in their spectral energy distributions. In five cases, we find
that the 3.6\,$\mu$m 3\,$\sigma$ limiting magnitude is inconsistent with
the presence of a Balmer/4000\,\AA\ spectral break, as the
corresponding 3.6\,$\mu$m limiting flux is actually less than that
detected in the $z'$-band. For the remaining two IRAC-undetected objects, their
3.6\,$\mu$m non-detections are consistent with flat spectral energy
distributions, and the maximum break amplitudes are constrained to be $<
0.6$\,magnitudes ($AB$).

The ($z' - 3.6\,\mu$m) colours exhibited by these IRAC-undetected
objects in our sample are noticeably bluer than most of the IRAC-detected
sample, as depicted in Figure~\ref{fig:COLMAG}, and this is not simply
an effect of limiting magnitude; many of the IRAC-detected sample are
fainter in $z'$-band than the IRAC-undetected sources.  None of our
galaxies fall into the `IRAC-selected Extremely Red Objects' (IERO)
colour cut of ($z' - 3.6\,\mu$m) $> 3.25$\,mag (Yan et al.\ 2004).

\begin{figure}
\resizebox{0.23\textwidth}{!}{\includegraphics{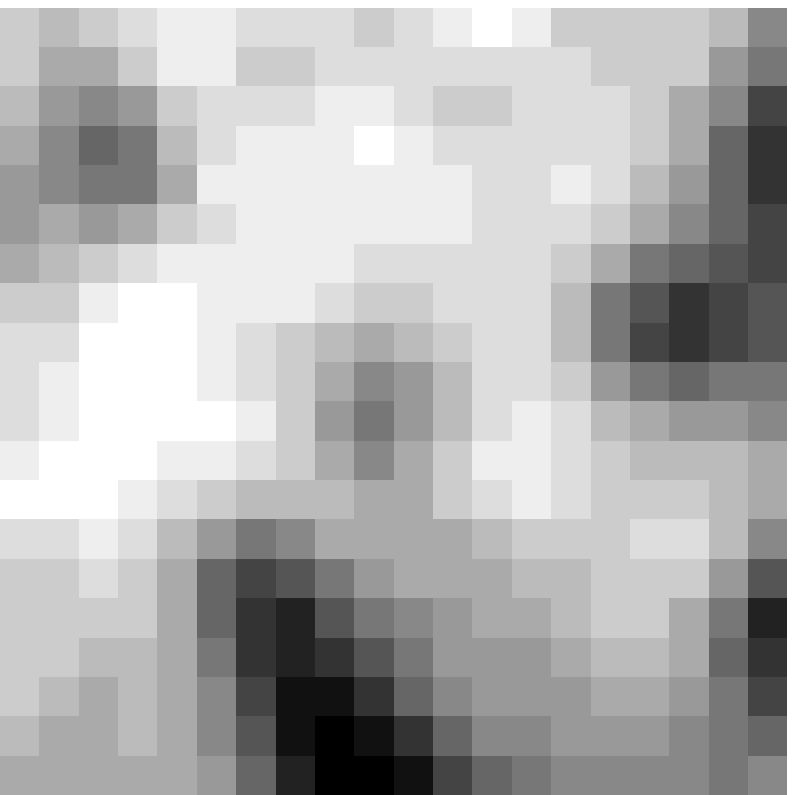}}
\resizebox{0.23\textwidth}{!}{\includegraphics{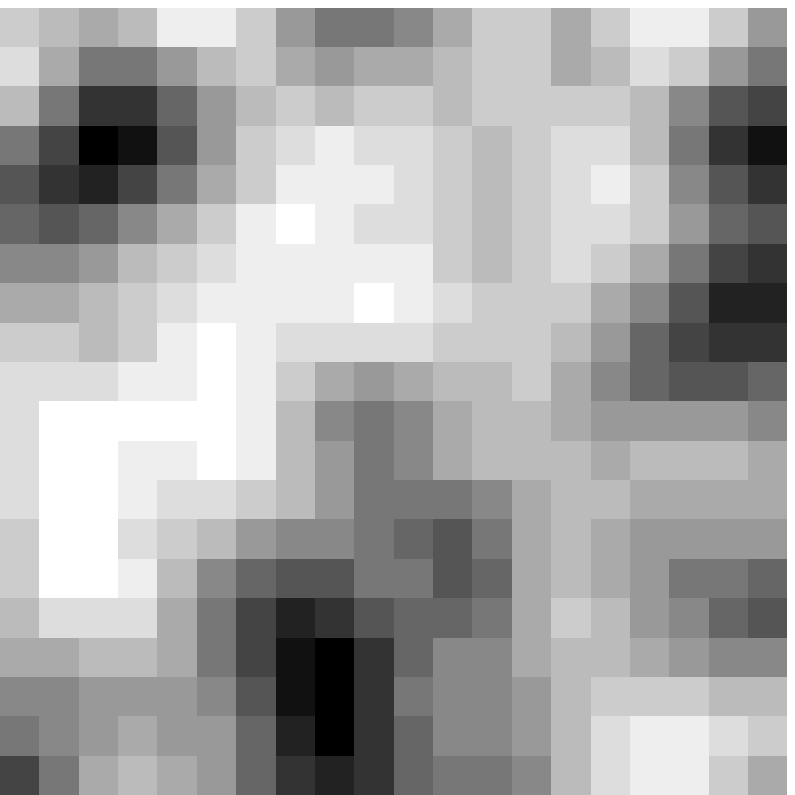}}
\caption{Stacking analysis of the four IRAC-undetected $i'$-drops with
the lowest contamination from neighbouring sources. A composite
detection is seen in channel 1 (left), and a more marginal signal
in channel 2 (right) at the $2.7\,\sigma$ level. The images have
been smoothed through convolution with a Gaussian of $\sigma=1$\,pixel.}
\label{fig:NONDETECTIMS}
\end{figure}

\begin{figure}
\resizebox{0.48\textwidth}{!}{\includegraphics{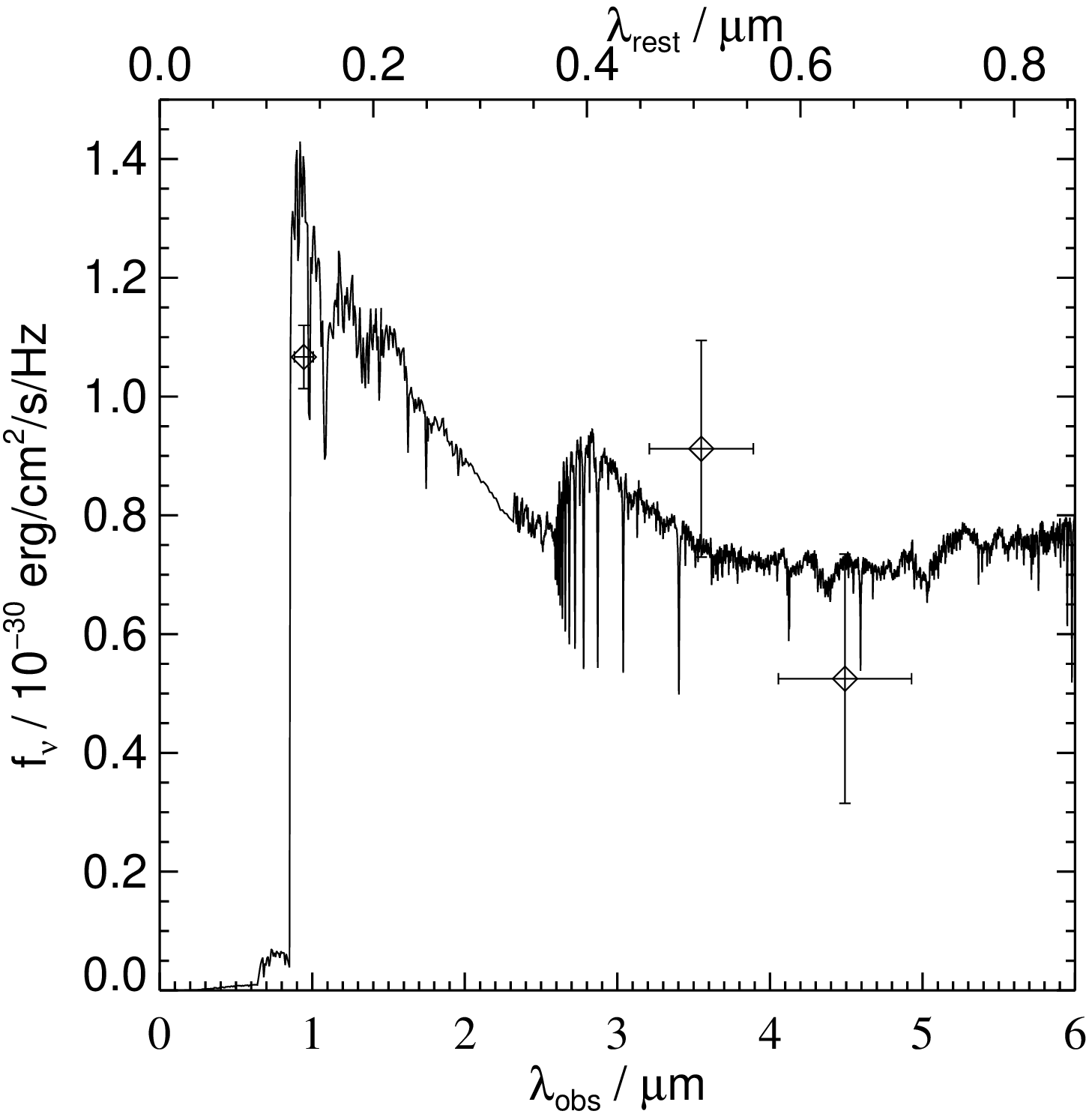}}
\resizebox{0.48\textwidth}{!}{\includegraphics{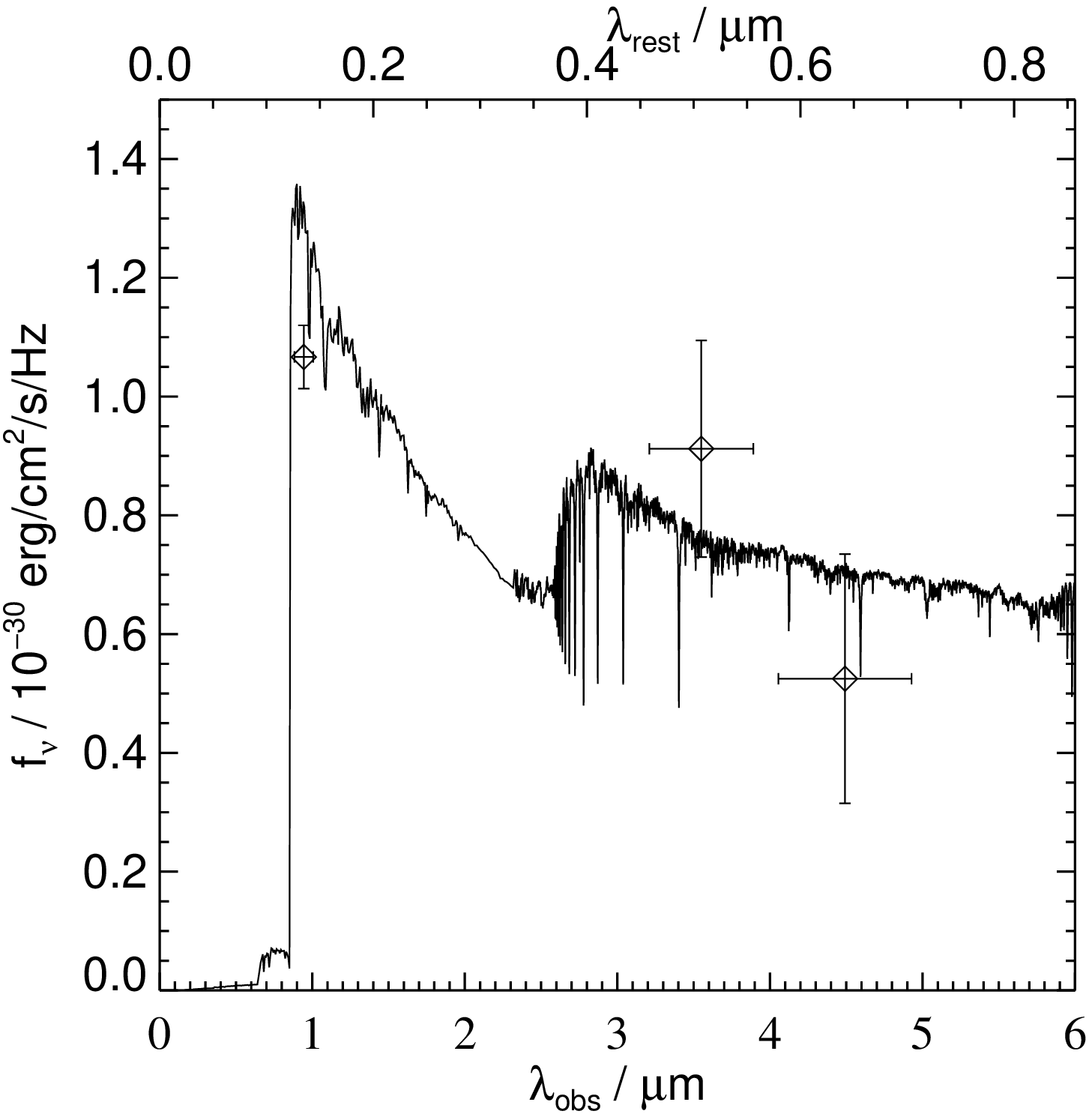}}
\caption{The best-fitting continuous star formation rate models from B\&C
for the SEDs of the stack of $i'$-drops individually undetected in IRAC.
Top panel is solar metallicity ($Z_{\odot}$), and the bottom panel is $0.2\,Z_{\odot}$.
Averages ages are 60\,Myr, and the stellar masses are $5\times 10^8\,M_{\odot}$.}
\label{fig:NONDETECTSED}
\end{figure}

\begin{figure}
\resizebox{0.48\textwidth}{!}{\includegraphics{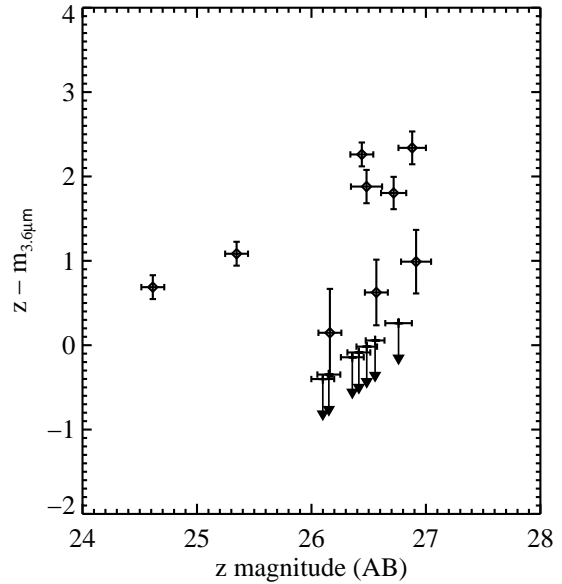}}
\caption{A colour-magnitude plot of the $z' - m_{\rm 3.6\,\mu m}$ colours of our 
16 selected $i'$-drop galaxies versus their $z'$-band magnitudes.  The colours 
of the IRAC-visible subset of these galaxies appear significantly redder than 
those of the IRAC-undetected sources (plotted here as upper limits, 
represented by downward-pointing arrows).  
Some objects are significantly redder, older and more 
massive (IRAC-detected) than those that are undetected at IRAC wavelengths
($3\,\sigma$ limiting magnitudes are shown as arrows).}
\label{fig:COLMAG}
\end{figure}

We have also stacked four non-detections that are not badly 
confused (31$\_$3127, 22$\_$7650, 34$\_$10623 \& 33$\_$7751 [two epochs]) to
improve the signal-to-noise (Figure~\ref{fig:NONDETECTIMS})
with outlier
rejection to reduce the effect of contamination by
neighbouring sources. 
This resulted in a detection
at 3.6\,$\mu$m at 26.5$\pm 0.3$ mag, and a marginal $\approx 2.7\sigma$ 
detection at 4.5\,$\mu$m of
27.1 mag. The average $z'$ magnitude of these objects was 26.33. This
confirms the very blue ($z' - 3.6\,\mu$m) color of the IRAC-weak population.
Figure~\ref{fig:NONDETECTSED} shows the fit to this composite SED. For solar metallicity, the 
best fit is a 57\,Myr old continuous star formation model with a total mass of
$5.4\times
10^{8}\,M_{\odot}$ and a stellar mass of $4.9\times10^{8}\,M_{\odot}$ (reduced $\chi^2 = 0.5$). For a metallicity of 0.2 solar
the best fit is a 67\,Myr old continuous star formation model, also with a total mass of
$5.4\times 10^{8}\,M_{\odot}$ and a stellar mass of $4.9\times10^{8}\,M_{\odot}$ (reduced $\chi^2 = 0.5$). 
Hence at $z\sim 6$, we suggest that whilst a
large proportion of $i'$-drop sources contain mature stellar
populations of considerable stellar mass, there is also a significant
fraction of young star-bursting galaxies which have colours
consistent with being `protogalaxies' experiencing their first throes
of star formation.

\subsection{Effects of Dust, Metallicity and IMF}
\label{sec:DUST}

To investigate the effects of differing metallicities on our results,
in addition to using solar ($Z_{\odot}$) metallicity models, we also
considered a sub-solar model (0.2$Z_{\odot}$).  We find that for both
these metallicities, the ages and masses of our $i'$-drops output by
our fitting code are similar, with the sub-solar models returning
slightly smaller reduced $\chi^{2}_{min}$ values.

We wish to address whether or not these sources suffer from
significant dust reddening; our earlier two case studies showed little
or no evidence for the presence of any intrinsic dust.  However, the
studies of a spectroscopically confirmed lensed galaxy at $z=6.56$
(Chary et al.\ 2005; Schaerer \& Pello 2005), and also of a $z=6.295$ GRB host galaxy (Berger
et al. 2006) suggest that these objects suffer from measurable dust
extinction ($A_{V}\approx 1$ mag).  If $z\sim 6$ objects are generally
found to be dusty, then their star formation rates, and hence the
global star formation density, will have been underestimated.  To
investigate this, we re-ran our SED fitting routine whilst
incorporating the Calzetti (1997) dust reddening law, appropriate for
starburst galaxies.  For each of the B\&C spectral template age
steps, we varied the reddening over the range $E(B-V) = 0.00 - 1.00$
mag, in steps of $0.01$ mag, and computed the reduced $\chi^{2}$ at
each step.  Inspecting the results presented in
Table~\ref{tab:MULTIDUSTTABLE}, we find little or no evidence for
intrinsic dust reddening in our selection of $i'$-drop galaxies, over
the wavelength range studied (out to $\approx 5\,\mu$m), with object
$23\_2897$ displaying the highest E(B-V) value of $0.16$, and several
other sources having formal best fits with no reddening.  This
supports a spectral break interpretation for those objects which
significantly brighten in $f_{\nu}$ flux between $\approx
0.9-2.2\,\mu$m to 3.6\,$\mu$m, rather than the smoother increase in
the continuum that would be produced by dust reddening, indicating the
presence of older stellar populations in these galaxies (see
Section~\ref{sec:DETECTIONS}).

\begin{figure}
\resizebox{0.48\textwidth}{!}{\includegraphics{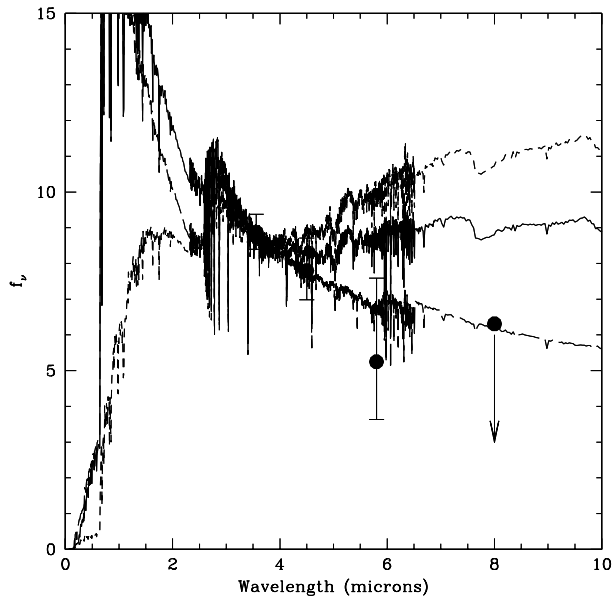}}
\caption{The composite IRAC SED points with a 30\,Myr constant star 
formation rate model overplotted as a solid line. The short-dashed line
is the same model reddened by $E(B-V)=0.1$, and 
the long-dashed line is the model but with a 0.2 solar metallicity.}
\label{fig:CH34LIMITS}
\end{figure}

We were also able to use data at 5.8\,$\mu$m and 8.0\,$\mu$m to constrain the 
average reddening in our sample. Cutouts of six brightest 3.6$\mu$m 
galaxies (all objects with $m_{3.6\,\mu\,m}(AB)<25$\,mag, excluding the T-dwarf) were taken from 
the IRAC images in all four channels. These images in each channel were scaled
by the ratio of each galaxy's 3.6$\mu$m flux to that of the brightest 
object in the sample and combined, with weighting by 
inverse-variance of the scaled images.  
To remove confusion, the highest and lowest pixels in the 
pixel stack were rejected. Aperture photometry was then 
carried out on these combined images to obtain a composite SED for the 
brighter IRAC detections. Larger apertures of $3\farcs0$ and $3\farcs 6$ pixel diameter
were used at 5.8\,$\mu$m and 8.0\,$\mu$m to allow for the larger PSF (see Eyles et 
al.\ 2005 for details).
At 5.8$\mu$m a detection is present at 
$\approx 3\sigma$, whilst at 8.0$\mu$m we fail to detect the galaxies but are
able to obtain a useful limit. 
This composite SED allows us to rule out reddening of 
$E(B-V)\stackrel{>}{_{\sim}} 0.1$, assuming a solar metallicity and an 
SED typical of our IRAC detections (see Figure~\ref{fig:CH34LIMITS}). In 
fact, our composite is somewhat 
bluer than the solar metallicity SEDs at rest-frame wavelengths of 0.5-1.0
$\mu$m, and favours a low 
metallicity ($\sim 0.2$ solar), though solar metallicity cannot be ruled out
on the basis of these data.

All of the results quoted in this work were produced via the fitting
of B\&C spectral synthesis models to our photometry using a Salpeter (1955) power law IMF.  To test the effects of differing IMF
models on the derived properties of our $i'$-drops, for the limiting
cases of an instantaneous (SSP) burst model and that of a continuous
SFR we also employed a Chabrier (2003) IMF.  We find that for a
Chabrier IMF, the returned ages of our objects were very similar, but
that the stellar masses were $\approx 30\%$ less than those produced
using a Salpeter IMF.  We note, as in Eyles et al.  (2005), that the
variations in the derived stellar masses primarily stem from a mass
re-scaling, arising from differing mass fractions contained in
low-mass long-lived stars. The inferred ages of our sources do not
significantly vary as this re-scaling is independent of the assumed
SFH.

\subsection{Line Contamination}
\label{sec:CONTAMINATION}

It is possible that some of our photometric datapoints may be
contaminated due to the presence of strong spectral lines.  We cannot
definitively remove any contribution by such lines from the measured
fluxes without spectroscopy of these emission lines; the most
prominent of these would lie at $\lambda>3\,\mu$m, where sufficiently
deep spectroscopy is not currently viable, but could be done with
NIRSpec on the {\em James Webb Space Telescope (JWST)}. For the
moment, we use our estimates of the current star formation rates in
each individual $z\sim 6$ galaxy to assess the possible impact of line
contamination on our measurement of the spectral energy distributions,
and on the fitting of the stellar ages and masses.

We consider first the effect of Lyman-$\alpha$ on the $i'$-band and
$z'$-band magnitudes. For three of the galaxies considered here, we
have detected Lyman-$\alpha$ emission in Keck/DEIMOS and Gemini/GMOS
spectroscopy. In the galaxies 23$\_$6714, 31$\_$2185 \& 23$\_$2897 at
$z\approx 5.8$ we measure rest-frame equivalent widths of 30, 20 \&
30\,\AA\ (Stanway et al.\ 2004a; Bunker et al.\ 2003; Stanway et al.\
2004b). This line emission would cause the $i'$- and $z'$-band
photometry to appear $\sim 0.1-0.3$\,mag brighter than for a
pure-continuum source. We note that at $z\approx 3$, only 25\% of
Lyman-break galaxies have Lyman-$\alpha$
rest-frame equivalent widths of $>20$\,\AA\ (Shapley et al.\ 2006).

We now consider the effect of H$\alpha$, which is redshifted into IRAC
channel 2 ($\lambda_{\rm cent}=4.5\,\mu$m) at $z=5-6.5$, most of the
redshift range selected through the $i'$-drop technique. We use the
rest-frame UV flux density to derive the unobscured star formation
rate, and then convert this to an H$\alpha$ line luminosity using the
relations in Kennicutt (1998). We then remove this line contribution
to the IRAC channel 2 filter and recompute the magnitude.  In all
cases, the effect was a dimming of $0.03-0.1$\,mag (at most a 10\%
effect).  This is consistent with the range in H$\alpha$ rest-frame equivalent
widths 
of $\sim 100-1000$\,\AA\ at $z\sim 2$ reported by Erb et al.\ (2006).
We also consider the effect of dust causing the star
formation rate to be underestimated from the rest-UV by a factor of
$\sim 3$ relative to H$\alpha$ (see Erb et al.\ 2005), which would
cause at most an overestimate of $0.3$\,mag in the channel 2
photometry (comparable to the typical error on the detection). We have
re-run the stellar population fitting to the `line-free' SEDs, and get
similar results (within the $1\,\sigma$ confidence intervals) for the
stellar ages and stellar masses; this is largely because our
constraint on the break amplitude comes mainly from the IRAC channel
1, which is much less affected by line contamination that channel 2.
For the higher-redshift portion of our sample ($z>6$), channel 1
may be affected by [OIII]\,5007/4959\,\AA\ and H$\beta$\,4861\,\AA ,
although these are typically weaker than H$\alpha$ in star-forming
galaxies. We note that for most of our $i'$-drop sample, the
IRAC channel 1 flux density is slightly brighter than channel 2 (in $f_{\nu}$, see Table~\ref{tab:MAGS}),
which again argues against significant line contamination, as the
brighter line (H$\alpha$) is centrally located in channel 2.

\subsection{Stellar Mass Density}
\label{sec:MASSDENSITY}

We have made measurements of the stellar masses and
luminosity-weighted ages for a number of probable $z\sim 6$ galaxies
in the GOODS-South field. In order to progress from our previous
individual case studies to a measurement of the global stellar mass
density at this epoch, consideration of several selection effects is
required, most notably: i) incompleteness due to foreground confusion;
ii) our apparent magnitude limit; and iii) our pre-selection of
star-forming galaxies in the rest-frame UV through the $i'$-drop Lyman
break technique. Due to these, our sample of $i'$-drop galaxies will
necessarily provide a lower limit on this stellar mass density at
$z\approx 6$.  Where possible, we attempt to correct for these
effects, as described below.

Our sample of 17 objects with clean {\em Spitzer}/IRAC photometry was
taken from a list of 31 sources for which either spectroscopic or reliable
photometric redshifts were available, out of our full sample of 52 $i'$-drops
(excluding two probable EROs).  We were able to infer the
properties of nine of these $z\sim 6$ galaxies (IRAC-detected), whilst
one was found to exhibit peculiar colours (see
Figure~\ref{fig:TDWARF}) and was ignored as a likely low-mass T-dwarf object.  
The remaining seven IRAC-undetected objects are consistent with
having young ages and low stellar masses.  As discussed in
Section~\ref{sec:GALFIT}, the remaining 14 of the 31 objects were
discarded due to the severity of confusion in the IRAC imaging;
attempts to remove the neighbouring sources using GALFIT proved unsuccessful,
and hence reliable photometry was unobtainable.  However, inspection
of the $z'$-band magnitudes and also the ($i' - z'$) colours
(Table~\ref{tab:COORDS}) using small apertures for
the {\em HST}/ACS suggests that 
the {\em Spitzer}-confused galaxies, and the galaxies
without GOODS-MUSIC photometric redshifts, have
the same distribution of $z'$-magnitudes and colours as those 16 galaxies which were isolated or had neighbouring sources
successfully subtracted.  Hence we are able to compensate for the loss
of these objects from our sample, by making the reasonable assumption
that the stellar mass distribution of the $i'$-drop population is
independent of IRAC contamination (i.e., we assume that the occurrence
of a foreground confusing source in the IRAC image is random and not
related to the intrinsic properties of the $z\sim 6$
$i'$-drop). Hence we can now correct for the stellar mass present in
those IRAC-confused sources and those without GOODS-MUSIC
photometric redshifts.  Taking
the sum of the best-fit masses for our 16 $i'$-drops with clean photometry, we
find a total stellar mass of $1.4\times 10^{11}M_{\odot}$ (most
of this coming from the nine IRAC-detected objects).  Scaling
this to account for all 51 of our $i'$-drops in GOODS-South with $z'_{AB}<26.9$
(after removing two lower-redshift EROS and the T-dwarf star),
we find that the  minimum total stellar
mass contained in these sources is approximately $4.5\times
10^{11}M_{\odot}$.

Another selection effect involved in our study is due to the nature of
the Lyman break technique.  The $i'$-drop method is reliant upon the
detection of a source at rest-frame UV wavelengths; for $z\sim 6$
galaxies, in the {\em HST}/ACS $z'$-band.  As the majority of the flux
in this wavelength region is due to hot, short-lived OB stars, all
galaxies selected via the $i'$-drop technique must have at least some
ongoing or very recent star formation (within $\sim 10$\,Myr).  Hence
dormant objects (pure post-starburst), experiencing no current or
recent star formation episodes will not be included amongst the
$i'$-drop population, and the stellar mass contained within them will
not be accounted for.  The abundance of such objects at $z\sim 6$ is
essentially unknown, and highly uncertain even for redshifts $0 \le z
\le 1$; cluster surveys suggest that post-starburst (`E+A') galaxies
may comprise as much as 20\% of the total cluster population (e.g.,
Dressler et al.\ 1999; Tran et al.\ 2003), although other estimates
suggest they are present in far fewer numbers, at $\sim 2$\% of the
total population at low redshift (Balogh et al.\ 1999).  At slightly
higher redshifts, Doherty et al.\ (2005) find that a significant
fraction ($>30$\%) of the `Extremely Red Object' population at $z\sim
1$ exhibit post-starburst spectra. Hence it is impossible to
accurately account for these post-starburst galaxies and their
contribution to the global stellar mass density and SFR. As a
consequence our derived values in this work are necessarily lower
limits.  This distribution of stellar ages and masses is shown in Figure~\ref{fig:AGEVMASS}.
Two selection effects combine to restrict the objects in Figure~\ref{fig:AGEVMASS} to
near the diagonal of the plot. Objects in the upper left, with low
stellar masses and high ages will be below our IRAC detection limits.
Objects in the lower right of the plot, with high stellar masses and
high star formation rates would be very bright in $z'$-band, and our
survey volume is not large enough to find such luminous objects.

A fraction of $i'$-drop galaxies are likely to be below our selection
threshold ($z'_{AB}<26.9$\,mag).  We now attempt to estimate the stellar
mass contribution of those objects which lie at the faint end of the
UV luminosity function, below our limit.  The faintest $z'$-band
magnitude of an object in our selection of 16 $i'$-drop galaxies is
$z'_{AB}=26.9$. This would correspond to a luminosity limit of $\sim 0.3\,L^{*}$,
if $L^*$ is the same for the Lyman-break galaxies at $z\sim 6$ as for
the well-studied populations at $z\sim 3-4$, where
$L^{*}_{\rm 1500\AA }=-21.1$ mag (AB), and $SFR^{*}=15\,M_{\odot}\,{\rm
  yr}^{-1}$ (Steidel et al.\ 1999).  If in fact $L^*$ at $z\approx 6$
is a factor of $\sim 2$ fainter than at $z\sim 3$ (as has been
suggested by Bouwens et al.\ 2005) then our limit corresponds to $\sim
0.6\,L^*$. Assuming a faint end slope with a steep $\alpha = -1.8$
(Bunker et al.\ 2004), we integrate the Schechter function, truncating
it at $0.1\,L^{*}$ (as in Steidel et al.\ 1999), and consequently obtain a factor by which we need
to scale our measurement of the total stellar mass in order to account
for objects which lie below our luminosity limit.  We note that
although our selection of objects reside at different redshifts, for
the sake of simplicity we assume a fixed redshift ($z = 6$) so that
only a single luminosity cut is used (using an ``effective volume''
accounts for the effect of different redshifts and hence luminosities
for a fixed limiting magnitude).  From this, we estimate the
scaling factor to be $\approx 1.8-3.2$ (for the Steidel and Bouwens
values of $L^*_{UV}$), resulting in a possible total (corrected)
stellar mass of $8-14\times 10^{11}M_{\odot}$.  
The range of correction factors change only slightly if the
faint end slope appropriate for Lyman break galaxies at
$z\sim 3-4$ is adopted ($\alpha=-1.6$, Steidel et al.\ 1999)
rather than $\alpha =-1.8$ (the corrections are $1.6-2.6$
compared to $1.8-3.2$, a $10-20$ per cent effect).

However, performing
this kind of correction is fraught with uncertainty.  The rest-frame
UV, as discussed previously, reflects the current star formation
distribution.  Yet from our results, we find a range of star formation
histories for our 16 selected $i'$-drop sources, with some galaxies
showing evidence for significant older populations, and others
containing very young stellar populations.  Therefore, correcting the
observed mass function of these objects for those which are below our
selection limit is likely to be highly inaccurate, as the relation
between the mass function and the rest-UV luminosity function is
non-trivial.

We now move to calculate the comoving stellar mass density at $z\sim
6$.  We reiterate our value will be a lower limit, as we can only
consider those objects which satisfy our selection criteria and
detection capabilities.  Taking a redshift interval of $5.7\le z\le
7.0$, the effective comoving volume for the GOODS-south field is
$1.8\times 10^5\,{\rm Mpc}^{3}$, as in Stanway, Bunker \& McMahon
(2003), which takes into account the luminosity bias against higher-redshift
sources in our redshift selection window due to the apparent magnitude
cut in $z'$-band.  Using our confusion-corrected total stellar mass of
$4.5\times 10^{11}\,M_{\odot}$, we find the lower limit to the $z\sim
6$ stellar mass density to be $2.5\times 10^{6}\,M_{\odot}\,{\rm
  Mpc}^{-3}$.  This value may be around $5-8\times
10^{6}\,M_{\odot}\,{\rm Mpc}^{-3}$ if the correction factor for
sources on the faint-end slope of the luminosity function is
reasonable, and higher still when post-starburst and dust-obscured
galaxies are also taken into account.  Our lower limit value is
consistent with the findings of Yan et al.\ (2006), who find a
stellar mass density of $\approx 1.6\times 10^{6}\,M_{\odot}\,{\rm
  Mpc}^{-3}$ at $z\approx 6$.
The lower limit we place on the stellar mass density at $z\sim 6$ is roughly 20\% that measured by Stark et al. (2006) at $z\simeq 5$, requiring the assembly of $\simeq 5\times10^6\,M_{\odot}\,{\rm Mpc}^{-3}$ between $z\simeq 6$ and $z\simeq 5$.  Assuming constant star formation over this period, this corresponds to a star formation rate density of $0.02\,M_{\odot}\,{\rm Mpc}^{-3}{\rm yr}^{-1}$, a factor of four larger than measured by Bunker et al.\ (2004) at $z\simeq 6$ in the UDF.  The shortfall could be explained by substantial star formation below the sensitivity limits of Bunker et al.\ or an extreme amount of dust extinction which obscures ongoing star formation, which would support the suggestion by Stark et al.\ (2006).

\begin{figure}
\resizebox{0.48\textwidth}{!}{\includegraphics{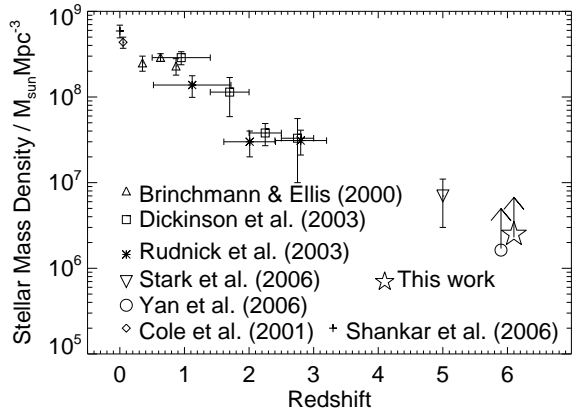}}
\caption{The evolution of the stellar mass density. Our work (star symbol)
and that of Yan et al.\ (2006, circle) are both at $z\approx 6$, but are
offset slightly in redshift for clarity.}
\label{fig:MASSDENSITYVZ}
\end{figure}

Galaxies  at intermediate redshifts, such as in Dickinson et al.\ (2003) and Rudnick et 
al.\ (2003), have fairly reliably placed constraints on the stellar mass 
already formed at redshifts $z\sim1-2$ (see Figure~\ref{fig:MASSDENSITYVZ}).
 These studies imply that roughly 
50-75\% of the stellar mass seen at the present day formed by $z\sim1$, i.e.\
$\sim 3 \times 10^8\,M_{\odot}\,{\rm Mpc}^{-3}$ out of the present day stellar mass 
of $5\times 10^8\,M_{\odot}\,{\rm Mpc}^{-3}$ (such as in Cole et al.\ 2001).  
This work, as well as that of Yan et al.\ (2006) and Stark 
et al.\ (2006), extend these measurements to $z\sim 5-6$,
and indicate that there may be more stellar mass 
already assembled at these early epochs than expected in some hierarchical scenarios
(see Section~\ref{sec:DETECTIONS}).  
If exceptional objects such as Mobasher et al.\ (2005) are considered 
real and possibly common -- at a stellar mass density another $10\times$ 
greater than those discussed in here  and in Yan et al.\ (2006) -- the shortfall of these
model scenarios is even greater.  
Different choices in the IMF slope and metallicity  may narrow the 
predicted versus observed stellar mass density.  However, with the 
conservative lower limits expressed here and in other studies, most 
corrections (such as reddening, surface brightness effects, and
limiting magnitude) tend to actually {\it 
increase} the mass. 

\subsection{SFRs and Star Formation Histories}
\label{sec:SFH}

The fitting of the B\&C spectral synthesis models to our photometric
data has allowed us to gain estimates of the current SFRs for our
selected $i'$-drop galaxies.  For our six sources which show evidence
of significant Balmer/4000\,\AA\ breaks, we find the inferred SFRs
span a substantial range, $\approx 2 -
140\,M_{\odot}\,{\rm yr}^{-1}$, involving between $\approx 0.1\% -
1.5\%$ of their total stellar mass (derived from the two-population composite model fits generally favoured for these objects).  In particular, $23\_6714$ \&
$31\_2185$ have, for their two population SED fits, inferred SFRs of
$\approx 80$ \& $140\,M_{\odot}\,{\rm yr}^{-1}$.  $23\_6714$
(SBM03\#1) is the brightest confirmed $i'$-drop in the UDF, with $z' =
25.35$ mag, and $31\_2185$ (SBM03\#3) is the brightest $i'$-drop in
the GOODS-South field, with $z' = 24.61$ mag; these two objects were
the focus of our previous study.  Their estimated SFRs are by far the
highest of our six Balmer/4000\,\AA\ break $z\sim 6$ sources; the
fainter four have inferred current SFRs between $\approx 2 -
25\,M_{\odot}\,{\rm yr}^{-1}$.  The remaining three IRAC-detected
objects have inferred ongoing SFRs of $\approx 0 - 20\,M_{\odot}\,{\rm
  yr}^{-1}$.  In fact, two of these sources nominally have SFRs of
$0\,M_{\odot}\,{\rm yr}^{-1}$ inferred by their SEDs, as the preferred
models for each of these are of young systems being viewed shortly
after a single starburst (i.e. no {\em current} star formation, but
rather very recent activity, within 10\,Myr so the rest-UV continuum
still persists).

In Eyles et al.\ (2005) we commented that SBM03\#1
\& SBM03\#3 may not be typical of the entire $i'$-drop population; our
analysis here suggests that in fact a substantial proportion of
$i'$-drops share similar masses and ages.  However, they are
significantly different from the subset of our selected objects which
are undetected at IRAC wavelengths, and which appear to be much
younger and less massive (see Section~\ref{sec:NONDETECTIONS}).  Hence 
these two objects are actually quite representative of the
older, more massive $i'$-drop population, but in terms of
current star formation rates, they do stand out alone as the most
currently active of $i'$-drop galaxies in the GOODS-South field.

\begin{figure}
\resizebox{0.48\textwidth}{!}{\includegraphics{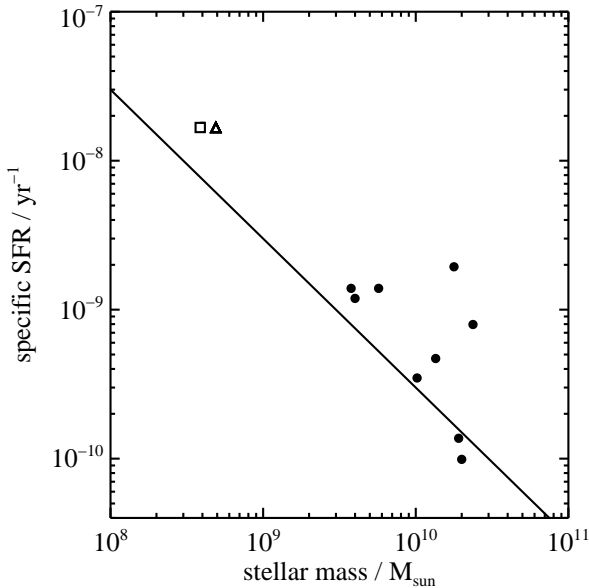}}
\caption{The specific SFR (the ratio of current star formation rate
to stellar mass) as a function of stellar mass for the IRAC-detected 
$i'$-drops in our analysis (circles). IRAC-undetected sources also included 
are $13\_3880$ (square), and the four IRAC-undetected sources (triangle) with the average star formation 
rate and mass for these galaxies derived from a stacking analysis. The line 
denotes the average of the lower-redshift work of Sajina et al.\ (2006), 
including the sample of Gil de Paz et al.\ (2000).}
\label{fig:bPARAM}
\end{figure}

Consideration of the specific star formation rate (SSFR; e.g., Brinchmann \& 
Ellis 2000)
is useful when investigating the SFRs of our $z\sim 6$ galaxies.  It
is defined as specific $SFR = SFR/M_{stellar}$, and is a measure of the 
fraction
of the total stellar mass that is currently being born as stars.
The SSFR is similar in concept to the dimensionless
`$b$-parameter' (e.g., Brinchmann et al.\ 2004), which is the ratio of the current
star formation rate to the past average.
  The
specific SFRs calculated from the results of the SED fitting are
tabulated in
Tables\,\ref{tab:MULTIGALTABLE},\,\ref{tab:MULTIDUSTTABLE}\,\&\,\ref{tab:MULTITWOPOPTABLE}.
For the best-fit models to our IRAC-detected sources, we
calculate values of specific $SFR = 1-50\times 10^{-10}\,{\rm yr}^{-1}$
(Figure~\ref{fig:bPARAM}), extending up to $\approx 200\times 10^{-10}\,{\rm yr}^{-1}$ for the composite SED of the IRAC-undetected galaxies.
These are comparable to the values inferred by Egami et al.\ (2005)
for a $z\sim 7$ lensed galaxy (Kneib et al.\ 2004), and are also
consistent with the findings of Yan et al.\ (2006).  These
specific SFR values are an indication of vigorous star formation
occurring in these nine IRAC-detected objects at the time of
observation, and these specific star formation rates are typically
higher than the lower-redshift results of Sajina et al.\ (2006) and Gil de
Paz et al.\ (2000).  However, we find that in most cases, the past-average
SFR must be comparable, if not greater than the current SFR. This is
the reason why the star formation histories that, for most of these
objects, return the best-fit results are decaying SFR models; the
exponential (tau) models, and for the younger galaxies, the burst
(SSP) models.  If this was not the case, then there would not have
been enough time, prior to the epoch of observation, for the stellar
masses of these objects to assemble, considering their inferred
formation redshifts.  Assuming the limiting case of a constant SFR,
most of our nine IRAC-detected objects require past average SFRs of
the order of SFR$\approx 50 - 60\,M_{\odot}\,{\rm yr}^{-1}$, with some
needing SFRs perhaps as high as $\approx 200\,M_{\odot}\,{\rm
  yr}^{-1}$.  The relative lack of $z\sim 7$ candidate galaxies and
the resultant continuing low luminosity density seen by Bouwens et
al.\ (2004b) in the {\em HST}/NICMOS UDF suggests that reionisation
may have been achieved by a considerably higher SFR density at even
earlier epochs (although see Bouwens et al.\ 2005), and our analysis
here supports this.

\begin{figure}
\resizebox{0.48\textwidth}{!}{\includegraphics{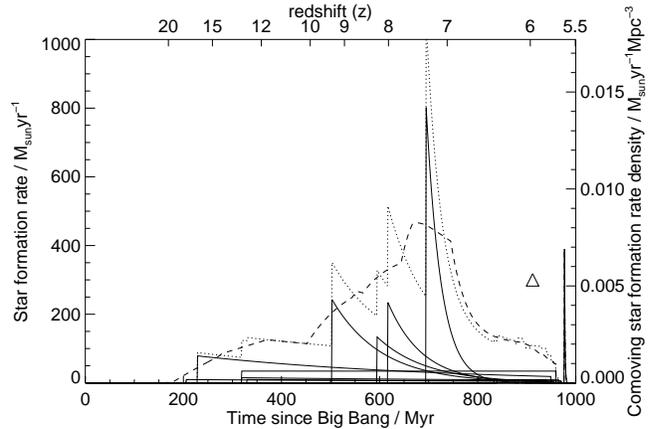}}
\caption{The best-fit star formation histories of individual $i'$-drops in our
sample, assuming an exponentially-decaying star formation rate and no
dust obscuration (solid lines). The dotted line is the sum of these star
formation histories to give the global average star formation rate density (right
axis, after correction for IRAC-confused sources and $i'$-drops
in our sample without GOODS-MUSIC photometric redshifts,
but {\em not} corrected for galaxies below our magnitude limit of $z'_{AB}<26.9$); the dashed line is this star formation rate density smoothed on
a timescale of 100\,Myr. The triangle is the star formation rate density at $z\approx 6$ from the rest-frame UV light of $i'$-drops in the Hubble Ultra Deep Field with $z'_{AB}<28.5$ (Bunker et al.\ 2004).}
\label{fig:SFH}
\end{figure}

Taking a different approach towards exploring the previous star
formation rates of our $i'$-drop sample, rather than assuming constant
SFRs prior to observation, we take into account the best-fit star
formation histories of each of our galaxies.  Figure~\ref{fig:SFH} is
a composite plot of the best-fit SFH (solid lines) for our nine
IRAC-detected objects.  The dashed line represents represents the
total SFR contribution due to these objects, smoothed using a time
interval of 100\,Myr, from the time of formation (obtained from their
inferred ages) until the epoch of observation (i.e. the redshift at
which they now reside).  The estimated star formation rate densities
extrapolated over each epoch shown in Figure~\ref{fig:SFH} represent
lower limits to the {\em actual} SFRD due to the populations of
galaxies not accounted for in our $z\sim 6$ galaxy selection (see
Section~\ref{sec:MASSDENSITY}).  However, our results seem to suggest
that the star formation density increases at $z>7$, relative to our
measured value at $z\sim 6$.

\subsection{Implications for Reionisation}
\label{sec:REIONISATION}

From SED fitting to {\em HST} and {\em Spitzer} photometry, we have
determined the best-fit preceding star formation histories of
individual $i'$-drop galaxies observed at $z\sim 6$. In
Section~\ref{sec:SFH} we summed these to recover the star formation
rate at $z>6$, necessarily a lower limit on this quantity as our
galaxy census in incomplete (magnitude-limited on the rest-frame UV).
However, our results do suggest a vigorous phase of star formation
activity, prior to $z\sim 6$. If the resultant ionizing photons are
able reach the IGM (i.e.\ if the escape fraction is high) then this
star formation may have played a key role in the reionisation of the
Universe.

Madau, Haardt \& Rees (1999) estimate the density of star
formation required for reionisation, and we use their equation 27
(updated for a more recent concordance cosmology estimate of the baryon
density from Spergel et al.\ 2003 of
$\Omega_b=0.0224\,h_{100}^{-2}=0.0457\,h_{70}^{-2}$):
\begin{equation}
{\dot{\rho}}_{\rm SFR}\approx \frac{0.026\,M_{\odot}\,{\rm yr}^{-1}\,{\rm Mpc}^{
-3}}{f_{\rm esc}}\,\left( \frac{1+z}{7}\right) ^{3}\,\left( \frac{\Omega_{b}\,h^
2_{70}}{0.0457}\right) ^{2}\,\left( \frac{C}{30}\right)
\end{equation}
This relation is based on the same Salpeter Initial Mass Function as
we have used throughout.  $C$ is the concentration factor of neutral
hydrogen, $C=\left< \rho^{2}_{\rm HI}\right> \left< \rho_{\rm
    HI}\right> ^{-2}$. Simulations suggest $C\approx 30$ (Gnedin \&
Ostriker 1997). The escape fraction of ionizing photons ($f_{\rm
  esc}$) for high-redshift galaxies is highly uncertain (e.g.,
Steidel, Pettini \& Adelberger 2001), but we consider here the
limiting case of a high escape fraction $f_{\rm esc}=1$ 
(i.e., no absorption by H{\scriptsize~I}/dust).

\begin{figure}
\resizebox{0.48\textwidth}{!}{\includegraphics{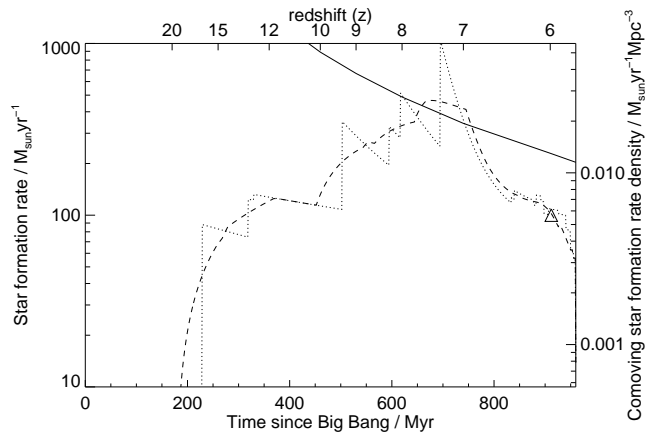}}
\caption{The solid curve shows the condition for reionisation from star formation,
as a function of time (bottom axis) and redshift (top axis),
assuming an escape fraction of unity for the Lyman continuum photons (from
Madau, Haardt  \& Rees 1999, updated for the effects of a warmer IGM temperature
by Stiavelli, Fall \& Panagia 2004). The dotted curve is the sum of the past
star formation rates for our $i'$-drop sample (left axis, with the corresponding
star formation rate density shown on the right axis, corrected for incompleteness
including a factor of 3.2 for galaxies below our flux threshold, Section~\ref{sec:MASSDENSITY}).
The dashed curve is this star formation history smoothed on a timescale of 100\,Myr.
The triangle is the estimate of the unobscured (rest-frame UV) star formation rate density
at $z\approx 6$ from $i'$-drops in the HUDF from Bunker et al.\ (2004).}
\label{fig:REION}
\end{figure}

It has recently been suggested by Stiavelli, Fall \& Panagia (2004)
that this star formation rate density requirement for reionisation
could be relaxed by a factor of $\sim 2$ at $z\sim 6$ since the IGM
temperature will be higher (perhaps 20,000\,K rather than the
10,000\,K assumed in Madau, Haardt \& Rees 1999). Figure~\ref{fig:REION} shows
this star formation rate requirement for reionisation (solid curve, for
a 20,000\,K IGM) overplotted with the star formation rate at $z>6$
inferred from the population synthesis fits to the $i'$-drops and
corrected for incompleteness (confused galaxies and those below our
magnitude limit; see Section~\ref{sec:SFH}). This past star formation
rate should still be regarded as a lower limit, as dormant
post-starburst galaxies are not found by the Lyman break
technique. The incompleteness is likely to be larger at higher
redshifts (earlier times) since by a redshift of 6 the light from
these fading populations will be swamped by any more recent star
formation activity in these galaxies.

What is apparent from Figure~\ref{fig:REION} is that our inferred 
lower limit on the past
star formation activity at $7<z<8$ is sufficient to achieve
reionisation, provided that the escape fraction is high. The star
formation history at higher redshift is more poorly constrained, and
could easily be sufficient given our incomplete census of galaxies (at
$z\sim 10$ our lower limit is only a factor of $\sim 3$ below the
reionisation requirement).  This interpretation may be consistent with
the CMB temperature-polarisation correlation measurement from WMAP
(Spergel et al.\ 2006) which indicates the Universe was 50\% reionized
at $z_r=9.3^{+2.8}_{-2.0}$.  However, we emphasize that the above
assumes that the UV ionizing photons escape galaxies; if star
formation is dust obscured even at these high redshifts, or
if the galaxy H{\scriptsize I} is optically thick to these photons, then our
inferred past star formation may be insufficient to provide the
necessary ionizing background. Yan et al.\ (2006) point out that
if the escape fraction is lower, $f_{esc}\approx 0.1$, then the previous
star formation inferred from {\em Spitzer} studies of $i'$-drops may
only be able to sustain reionisation for short periods.

\section{Conclusions}
\label{sec:CONCLUSIONS}

We have used multi-waveband imaging of the GOODS-South field ({\em
  HST}/ACS, VLT/ISAAC \& {\em Spitzer}/IRAC) to measure the stellar
mass density of the $z\sim 6$ $i'$-drop galaxy population.  From an
original catalog of 52 $i'$-drop candidates, we concentrated on 31
objects for which either spectroscopic or robust photometric redshifts
from GOODS-MUSIC were available.  A further 14 of these were eliminated from our
analysis due to IRAC-confusion with neighbouring sources, which could
not be satisfactorily removed.  One further source, singled out as
having peculiar photometry, had its SED matched very well to the
spectra of a T-dwarf object -- an interloper in our sample.  Fitting
of SEDs to B\&C spectral synthesis models was conducted on the
remaining 16 $z\sim 6$ galaxies, and properties including ages,
stellar masses and star formation histories were constrained.  Of
these, nine were detected at IRAC wavelengths, and six of these showed
evidence for significant Balmer/4000\,\AA\ spectral breaks,
brightening across the break by up to a factor of $2-3$ in $f_{\nu}$.
These indicate the presence of
old stellar populations that dominate the stellar masses of these galaxies, 
with inferred ages of $\sim 200 - 700$\,Myr, and stellar masses of $\sim 1.0 - 
3.0\times 10^{10}\,M_{\odot}$.  During the SED modelling process, we 
considered the
possibility of intrinsic dust reddening, and also the effects of
differing metallicities and IMF models.  We do not find evidence of
substantial dust reddening in our $i'$-drop galaxies, and using
differing metallicities did not have any significant effect on our
derived properties.  Use of a Chabrier, rather than Salpeter IMF had
little effect on our inferred galaxy ages, but did reduce our stellar
mass values by $\approx 30$\% due to a mass re-scaling.  The results
of the SED fitting of the three other fainter IRAC-detected sources were
inconclusive.  For the seven objects undetected in the IRAC wavebands, 
their SED fitting inferred much younger, less massive systems than their 
detected counterparts.

Using the constrained properties of our $i'$-drop sample, we were able
to calculate a value for the $z\sim 6$ stellar mass density of
$2.5\times 10^{6}M_{\odot}\,{\rm Mpc}^{-3}$, correcting for those
objects eliminated from our analysis due to their un-treatable
IRAC-confusion and those lacking GOODS-MUSIC photometric redshifts.  Using a somewhat uncertain correction in order to
account for the stellar mass in objects below our $z'$-band magnitude
selection limit, this value could perhaps be $5 - 8\times
10^{6}M_{\odot}\,{\rm Mpc}^{-3}$.  Any post-starburst and dust-obscured
$z\sim 6$ sources would not be found using the $i'$-drop selection
technique, and hence our $z\sim 6$ stellar mass density value is 
necessarily a lower limit, and is consistent with the estimates of Yan et al.\ 
(2006).

Exploring the previous star formation histories of our $i'$-drops, as inferred 
from their SED fitting, we suggest that the global star formation of these 
sources may have been substantially higher prior to the epoch of observation,
and the resultant UV flux at $z>7$ may have played an important role
in reionizing the Universe.

\subsection*{Acknowledgments}

This work is based on observations made with the {\em Spitzer Space 
Telescope}, which is operated by the Jet Propulsion Laboratory,
California Institute of Technology under NASA contract 1407.
Observations have been carried out using the Very Large Telescope at
the ESO Paranal Observatory under Program ID: LP168.A-0485.  This
paper is based in part on observations made with the NASA/ESA Hubble
Space Telescope, obtained from the Data Archive at the Space Telescope
Science Institute, which is operated by the Association of
Universities for Research in Astronomy, Inc., under NASA contract NAS
5-26555. These observations are associated with proposals
\#9425\,\&\,9583 (the GOODS public imaging survey). We are grateful to
the GOODS team for making their reduced images public. 
LPE acknowledges a Particle Physics and Astronomy Research
Council (PPARC) studentship supporting this study.  AJB is grateful
for financial support from a Leverhulme Prize.  ERS gratefully acknowledges 
support from NSF grant AST 02--39425.  The compilation of
T-dwarf spectra comes from Sandy Leggett. We acknowledge useful
discussions with Richard McMahon, Karl Glazebrook and Michelle
Doherty, and thank Carlton Baugh for providing electronic tables
of the GALFORM stellar mass function. LPE thanks Rob King for
useful discussions about low mass T-dwarf objects. We are grateful to Andrea Grazian
and Adriano Fontana
for providing us with the GOODS-MUSIC sample: a multicolour catalog of near-IR 
selected galaxies in the GOODS-South field. We thank the
referee for useful comments on this manuscript.

\bsp

\end{document}